\PassOptionsToPackage{unicode}{hyperref}
\PassOptionsToPackage{hyphens}{url}
\PassOptionsToPackage{dvipsnames,svgnames,x11names}{xcolor}
\documentclass[
  12pt]{article}

\usepackage{iftex}
\usepackage{amsmath, amssymb, amsthm, mathtools}
\usepackage{graphicx}
\usepackage{booktabs}
\usepackage{microtype}
\usepackage{hyperref}
\usepackage[nameinlink,noabbrev]{cleveref}
\usepackage{natbib}
\usepackage{enumitem}
\usepackage{authblk}
\usepackage{algorithm}
\usepackage{algpseudocode}
\usepackage{subcaption}
\usepackage{array}
\usepackage{xcolor}

\DeclareCaptionFont{captionspace}{\linespread{1.15}\selectfont}
\ifPDFTeX
  \usepackage[T1]{fontenc}
  \usepackage[utf8]{inputenc}
  \usepackage{textcomp} 
\else 
  \usepackage{unicode-math}
  \defaultfontfeatures{Scale=MatchLowercase}
  \defaultfontfeatures[\rmfamily]{Ligatures=TeX,Scale=1}
\fi
\usepackage{lmodern}
\ifPDFTeX\else  
\fi
\IfFileExists{upquote.sty}{\usepackage{upquote}}{}
\IfFileExists{microtype.sty}{
  \usepackage[]{microtype}
  \UseMicrotypeSet[protrusion]{basicmath} 
}{}
\makeatletter
\@ifundefined{KOMAClassName}{
  \IfFileExists{parskip.sty}{%
    \usepackage{parskip}
  }{
    \setlength{\parindent}{0pt}
    \setlength{\parskip}{6pt plus 2pt minus 1pt}}
}{
  \KOMAoptions{parskip=half}}
\makeatother
\usepackage{xcolor}
\makeatletter
\ifx\paragraph\undefined\else
  \let\oldparagraph\paragraph
  \renewcommand{\paragraph}{
    \@ifstar
      \xxxParagraphStar
      \xxxParagraphNoStar
  }
  \newcommand{\xxxParagraphStar}[1]{\oldparagraph*{#1}\mbox{}}
  \newcommand{\xxxParagraphNoStar}[1]{\oldparagraph{#1}\mbox{}}
\fi
\ifx\subparagraph\undefined\else
  \let\oldsubparagraph\subparagraph
  \renewcommand{\subparagraph}{
    \@ifstar
      \xxxSubParagraphStar
      \xxxSubParagraphNoStar
  }
  \newcommand{\xxxSubParagraphStar}[1]{\oldsubparagraph*{#1}\mbox{}}
  \newcommand{\xxxSubParagraphNoStar}[1]{\oldsubparagraph{#1}\mbox{}}
\fi
\makeatother

\usepackage{longtable,booktabs,array}
\usepackage{calc} 
\usepackage{etoolbox}
\makeatletter
\patchcmd\longtable{\par}{\if@noskipsec\mbox{}\fi\par}{}{}
\makeatother
\IfFileExists{footnotehyper.sty}{\usepackage{footnotehyper}}{\usepackage{footnote}}
\makesavenoteenv{longtable}
\usepackage{graphicx}
\makeatletter
\def\maxwidth{\ifdim\Gin@nat@width>\linewidth\linewidth\else\Gin@nat@width\fi}
\def\maxheight{\ifdim\Gin@nat@height>\textheight\textheight\else\Gin@nat@height\fi}
\makeatother
\setkeys{Gin}{width=\maxwidth,height=\maxheight,keepaspectratio}
\makeatletter
\def\fps@figure{!t}
\makeatother

\makeatletter
\@ifpackageloaded{caption}{}{\usepackage{caption}}
\AtBeginDocument{%
\ifdefined\contentsname
  \renewcommand*\contentsname{Table of contents}
\else
  \newcommand\contentsname{Table of contents}
\fi
\ifdefined\listfigurename
  \renewcommand*\listfigurename{List of Figures}
\else
  \newcommand\listfigurename{List of Figures}
\fi
\ifdefined\listtablename
  \renewcommand*\listtablename{List of Tables}
\else
  \newcommand\listtablename{List of Tables}
\fi
\ifdefined\figurename
  \renewcommand*\figurename{Figure}
\else
  \newcommand\figurename{Figure}
\fi
\ifdefined\tablename
  \renewcommand*\tablename{Table}
\else
  \newcommand\tablename{Table}
\fi
}
\@ifpackageloaded{float}{}{\usepackage{float}}
\floatstyle{ruled}
\@ifundefined{c@chapter}{\newfloat{codelisting}{h}{lop}}{\newfloat{codelisting}{h}{lop}[chapter]}
\floatname{codelisting}{Listing}

\makeatother
\makeatletter
\@ifpackageloaded{caption}{}{\usepackage{caption}}
\@ifpackageloaded{subcaption}{}{\usepackage{subcaption}}
\makeatother

\ifLuaTeX
  \usepackage{selnolig}  
\fi
\usepackage[]{natbib}
\usepackage{bookmark}

\IfFileExists{xurl.sty}{\usepackage{xurl}}{} 
\hypersetup{
  pdftitle={Title},
  pdfauthor={Author 1; Author 2},
  pdfkeywords={3 to 6 keywords, that do not appear in the title},
  colorlinks=true,
  linkcolor={blue},
  filecolor={Maroon},
  citecolor={Blue},
  urlcolor={Blue},
  pdfcreator={LaTeX via pandoc}}

\newcommand{\anon}{1}

\newtheorem{theorem}{Theorem}
\newtheorem{lemma}{Lemma}
\newtheorem{proposition}{Proposition}

\theoremstyle{remark}
\newtheorem{remark}{Remark}

\def\cA{\mathcal{A}}    \def\cB{\mathcal{B}}            
\def\cF{\mathcal{F}}                
            \def\cN{\mathcal{N}}    \def\cO{\mathcal{O}}
\def\cP{\mathcal{P}}        \def\cR{\mathcal{R}}        
\def\cU{\mathcal{U}}            \def\cX{\mathcal{X}}    \def\cY{\mathcal{Y}}

\newcommand{\RR}{\mathbb{R}}
\newcommand{\NN}{\mathbb{N}}

\newcommand{\EE}{\mathbb{E}}
\newcommand{\PP}{\mathbb{P}}

\newcommand{\dd}{\mathrm{d}}

\DeclareMathOperator{\Var}{Var}

\begin{document}

\def\spacingset#1{\renewcommand{\baselinestretch}%
{#1}\small\normalsize} \spacingset{1}


\if1\anon
{
  \title{\bf Free-support Wasserstein Resampling in Sequential Monte Carlo}
  \author{Mengyu Li$^1$, Cheng Meng$^{2*}$, Jun S. Liu$^1$\thanks{Corresponding author: Cheng Meng (chengmeng@ruc.edu.cn), Jun S. Liu (junsliu@tsinghua.edu.cn)}\\
    $^1$Department of Statistics and Data Science, Tsinghua University\\
    $^2$Center for Applied Statistics, Institute of Statistics and Big Data, Renmin University of China}
  \maketitle
} \fi

\if0\anon
{
  \bigskip
  \bigskip
  \bigskip
  \begin{center}
    {\LARGE\bf Free-support Wasserstein Resampling \\
    \smallskip
    in Sequential Monte Carlo}
\end{center}
  \medskip
} \fi

\bigskip
\begin{abstract}
In modern sequential Monte Carlo (SMC) applications, such as diffusion model samplers, propagating a particle is expensive, so only a few particles can be afforded.
In this regime, each resampling step decides which regions of the state space survive, so a better resampling scheme is worth extra computation.
However, classical resampling methods only replicate existing particles and suffer from sample impoverishment, especially when the particle size is small relative to the dimension.
To create new particles, we relax the unbiasedness of resampling and control the mean squared error via a Wasserstein criterion.
Specifically, we propose Wasserstein-Dirichlet resampling (WDR), which constructs an equal-weight measure that optimally approximates the weighted particle distribution by solving a free-support Wasserstein barycenter problem.
We then randomize the resulting optimal coupling by Dirichlet mixing.
Theoretically, we provide non-asymptotic mean squared error bounds that quantify the bias-variance trade-off and show that the rate depends on the intrinsic dimension.
A truncated variant preserves high-weight particles while applying WDR to the remaining mass, targeting mild weight degeneracy.
Experiments on state-space models and conditional image generation demonstrate that WDR consistently improves estimation accuracy, stability, and sample diversity.
The improvements persist at matched wall-clock cost when the propagation cost dominates.
\end{abstract}

\noindent%
\vfill

\newpage
\spacingset{1.75} 

\section{Introduction}\label{sec:intro}
Resampling is a fundamental operation that transforms a weighted sample $\{(X_i, W_i)\}_{i=1}^n$ with $\sum_{i=1}^n W_i = 1$ into an equally weighted collection $\{\tilde X_j\}_{j=1}^m$. The goal is to ensure that the two discrete distributions are close in some sense, i.e.,
\begin{equation*}
\sum_{i=1}^n W_i \delta_{X_i} \approx \sum_{j=1}^m m^{-1} \delta_{\tilde X_j},
\end{equation*}
where $\delta_x$ denotes the Dirac measure at $x$. 
Resampling plays an important role in a wide range of statistical applications, such as survey sampling~\citep{rao1988resampling}, causal inference~\citep{morgan2012rerandomization}, and missing data imputation~\citep{rubin1996multiple}. This work focuses on resampling within sequential Monte Carlo (SMC) methods, also known as particle filters, which have emerged as a powerful computational framework for Bayesian inference in dynamical systems~\citep{liu1998sequential}.

\subsection{Background and problem formulation}\label{subsec:background}
Sequential Monte Carlo algorithms recursively approximate a sequence of target distributions by maintaining a particle system and updating it through propagation, weighting, and resampling~\citep{doucet2011tutorial}. Resampling mitigates the progressive concentration of importance weights, known as weight degeneracy~\citep{liu1995blind}. Without resampling, the weights become increasingly concentrated on a few particles as iterations proceed, rendering most particles ineffective for downstream estimation and wasting computational resources.

In modern SMC applications, propagating each particle can require substantial time or memory.
For example, in diffusion model samplers, every propagation step requires a neural network forward pass for each particle~\citep{wu2023practical}; in scientific filtering problems, the latent dynamics must be numerically integrated between observation times~\citep{vanleeuwen2019particle}.
The number of particles $m$ that can be used is thus severely limited, while the number required to avoid weight collapse grows exponentially with the effective dimension~\citep{snyder2008obstacles, bengtsson2008curse}.
With so few particles, it is critical to handle resampling properly.

Various resampling schemes have been developed in the SMC literature. The most straightforward is multinomial resampling~\citep{gordon1993novel}, which draws particle indices independently from the categorical distribution with probabilities $(W_i)_{i=1}^n$. To reduce the variance introduced by resampling, refined schemes such as residual~\citep{liu1998sequential}, stratified~\citep{kitagawa1996monte}, and systematic resampling~\citep{carpenter1999improved} have been proposed. 
These methods are simple to implement, unbiased for arbitrary test functions, and consequently remain widely used in practice.

\begin{figure}[!t]
\centering
\includegraphics[width=0.7\linewidth]{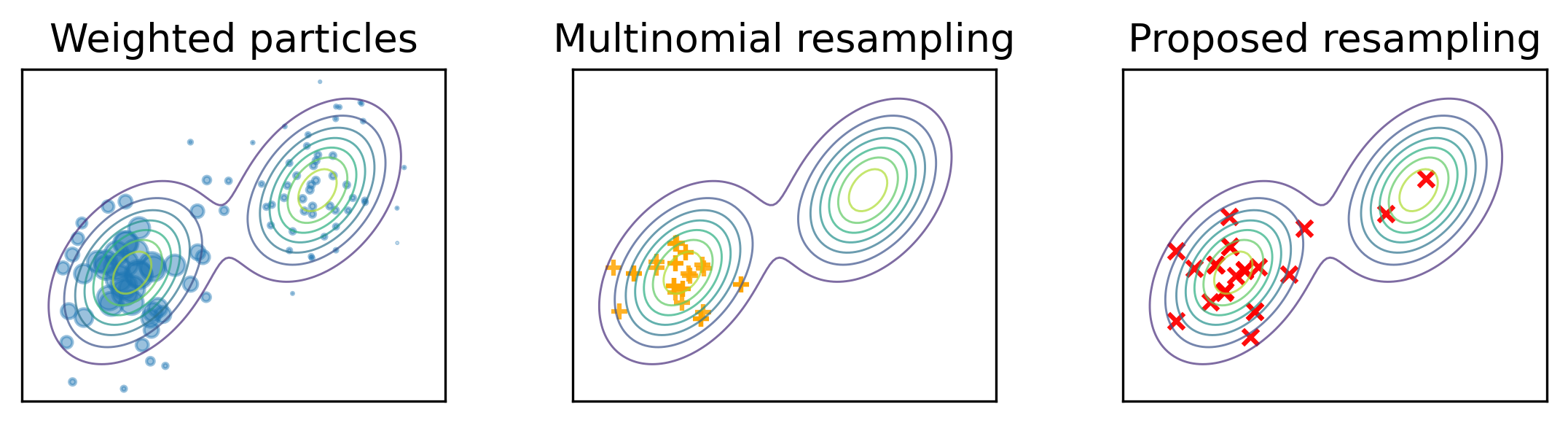}
\caption{\textbf{Left:} weighted particles ($n=100$) before resampling; marker size is proportional to importance weight, and contour lines depict the target density.
\textbf{Middle:} equally weighted particles ($m=30$) produced by multinomial resampling happen to be all located around the dominant high-weight region.
\textbf{Right:} particles ($m=30$) produced by the proposed free-support resampling are allowed to relocate, yielding a more spatially dispersed set.}
\label{fig:toy-impoverishment}
\end{figure}

However, these classical schemes only select among existing particles, i.e., they replicate or discard elements from $\{X_i\}_{i=1}^n$ but never relocate them or create new support points.
This limited flexibility leads to sample impoverishment: when a small number of particles carry most of the weight, traditional resampling replicates these particles many times while discarding all others. The resulting particle system concentrates in a narrow region of the state space and fails to represent the global shape of the target distribution; see Fig.~\ref{fig:toy-impoverishment} (left and middle panels) for a two-dimensional illustration.

To mitigate impoverishment, classical resampling can be augmented with additional randomization, such as adding jitter to resampled particles~\citep{gordon1993novel} or performing resample-move MCMC rejuvenation steps~\citep{gilks2001following}. While such mechanisms can improve diversity, they do so heuristically and locally. The resampling step itself still depends solely on the weights $\{W_i\}_{i=1}^n$ while leaving particle locations unchanged; any subsequent jitter or MCMC move is applied post-resampling and lacks an explicit global criterion for comparing the weighted empirical measure and its equal-weight approximation within the underlying geometry.

\subsection{Related work}
The aforementioned limitations have motivated a more recent line of work that directly incorporates geometric information into the resampling step. Existing spatially aware methods fall into three classes: sorting-based, clustering-based, and optimal transport resampling.

Sorting-based methods first order the particles and then apply stratified or systematic resampling to this sequence. In multiple dimensions, the ordering is typically induced by a Hilbert space-filling curve mapping that approximately preserves locality~\citep{gerber2015sequential}. 
\citet{li2022stratification} established an equivalence between sorted stratified resampling and an optimal transport formulation in one dimension. 
In higher dimensions, however, the performance depends on the quality of embedding, and the projection may introduce artifacts in highly anisotropic or multimodal settings.

Clustering-based resampling partitions particles into groups using algorithms such as $k$-means and performs resampling within each cluster~\citep{yang2019unsupervised}. 
This prevents clusters with small total weight from being eliminated entirely and encourages local diversity. 
However, such procedures typically lack theoretical guarantees and may be sensitive to clustering hyperparameters.

The third direction incorporates optimal transport (OT) theory into resampling \citep{villani2009optimal, peyre2019computational, zhang2021review}. \citet{reich2013nonparametric} 
proposed the ensemble transform particle filter (ETPF), which computes an OT plan between the weighted empirical measure and an equal-weight empirical measure supported on the original particles, and then uses the associated barycentric map to transform the particles deterministically.
Subsequent work developed second-order accurate ETPF \citep{acevedo2017second} with deterministic corrections to match covariance for bias reduction, and entropy-regularized variants \citep{kang2018particles,corenflos2021differentiable} to reduce the computational cost. 
While ETPF preserves geometry better than classical schemes, it optimizes transport only over couplings with fixed supports rather than optimizing the locations of resampled particles. Consequently, the resulting equal-weight empirical measure is not optimal in the Wasserstein sense. 
Moreover, because both marginals are supported on the same particle set, ETPF is naturally formulated only for $m=n$ and does not accommodate  multi-descendant SMC frameworks \citep{fearnhead2003line,li2022stratification}, where $m$ and $n$ may differ significantly. Finally, this line of work lacks non-asymptotic theoretical results.

\subsection{Contributions}
Our main idea is to relax the exact unbiasedness imposed by most classical resampling methods and instead control the overall mean squared error (MSE).
To this end, we first  consider resampling from the point of view of a free-support Wasserstein barycenter problem: given a weighted empirical measure $\mu_n = \sum_{i=1}^n W_i \delta_{X_i}$, we seek an equal-weight measure $\nu_m = m^{-1}\sum_{j=1}^m \delta_{\tilde X_j}$ that minimizes the Wasserstein distance $W(\mu_n,\nu_m)$.
Unlike fixed-support OT schemes such as ETPF, which constrain both marginals $\mu_n$ and $\nu_m$ to the original particle support and optimize only over the coupling $\Pi \in \mathbb{R}^{n\times m}_+$, our formulation  optimizes jointly over $\Pi \in \mathbb{R}^{n\times m}_+$ and the new support locations $\{\tilde X_j\}_{j=1}^m$.
Consequently, the resulting equal-weight empirical measure attains the minimum Wasserstein distance to $\mu_n$ among all $m$-atom equal-weight measures, distributing particles more evenly across the support (see Fig.~\ref{fig:toy-impoverishment} for illustration).

We next turn this deterministic free-support approximation into a randomized resampling operator.
A purely deterministic Wasserstein approximation may underestimate posterior spread, so we propose a Dirichlet mixing mechanism applied to the columns of the optimal coupling.
In contrast to a global Dirichlet reweighting on the original support~\citep{rubin1981bayesian, lyddon2019general}, our construction uses the optimal transport plan to build many local Dirichlet mixtures, one per output particle.
The resulting method, termed Wasserstein-Dirichlet resampling (WDR), produces a family of randomized resampling schemes interpolating between a deterministic transform and fully randomized ancestor selection, providing a tunable bias-variance trade-off.

Third, we propose some variants and algorithms for SMC implementation.
For mild weight-degeneracy regimes, we propose a truncated variant, T-WDR, which preserves a high-weight subset of particles and applies WDR to the remaining mass, thereby allocating more resampling budget to rejuvenate the lower-weight region.
Algorithmically, we develop an efficient block coordinate descent method to solve the underlying optimization problem.

Theoretically, we derive non-asymptotic error bounds for the WDR estimator.
Under mild regularity conditions, WDR achieves a rate of $\cO(m^{-1/s})$ for Lipschitz functions, where $s$ is the intrinsic dimension of the underlying distribution and can be substantially smaller than the ambient dimension $d$.
We also establish the convergence of the proposed block coordinate descent algorithm.

Empirically, we evaluate WDR on linear Gaussian state-space models simulated across a range of dimensions, including a spatial model that requires a block particle filter. Since all of these admit an exact Kalman filter, the Kalman solution provides a clean reference against which the effectiveness of each method can be assessed. Additional nonlinear studies are reported in the Supplementary Materials.
WDR consistently improves filtering accuracy and stability at matched particle sizes, and remains advantageous at matched wall-clock time in propagation-dominating regimes.
We further integrate WDR into the twisted diffusion sampler (TDS)~\citep{wu2023practical}, a state-of-the-art SMC framework for diffusion models.
In conditional image generation tasks, WDR consistently produces more diverse and higher-quality samples than baseline resampling strategies.

The rest of the article is organized as follows. Section 2 provides preliminary introductions to the general SMC framework and the Wasserstein barycenter; Section 3 describes our proposed WDR method; Section 4 studies theoretical properties of the proposed method; Section 5 details our simulation evaluations; Section 6 applies WDR to conditional image generation; and Section 7 concludes with a short discussion.

\section{Preliminaries}\label{sec:preliminaries}

\textbf{Notations.}
Superscripts $(t)$ and $[r]$ index timesteps and iterations, respectively; subscripts index particles. Temporal superscripts are omitted when clear from context.
We use $n$ for the number of weighted particles and $m$ for the resampled size. For vectors, $\|\cdot\|$ denotes the Euclidean norm. 
For a measurable space $\cX$, $\cP(\cX)$ denotes the set of Borel probability measures on $\cX$, and $\delta_x\in\cP(\cX)$ denotes the Dirac measure at $x\in\cX$. We write $1_m$ for the $m$-vector of ones, and $\Delta^{n-1}:=\{p\in\RR_+^n:\sum_{i=1}^n p_i=1\}$ for the probability simplex.
For two nonnegative sequences $(x_{n})_{n}$ and $(y_{n})_{n}$, we write $x_n=\cO(y_n)$ if there exists a constant $c>0$ such that $x_n\le cy_n$, and $x_{n}=\tilde\cO(y_{n})$ if there exist constants $c, c^\prime>0$ such that $x_{n} \leq c^\prime y_{n}(\log (n))^{c}$.
Sections numbered with the prefix S refer to the Supplementary Materials.

\subsection{Sequential Monte Carlo}\label{subsec:smc}

Consider a state-space model observed at $T$ timesteps, with a latent process $(X^{(t)})_{0\le t\le T}$ taking values in $\cX$ and observations $(Y^{(t)})_{1\le t\le T}$ taking values in $\cY$.
The latent process evolves from an initial probability density $\mu_0\in\cP(\cX)$ according to Markov transition kernels $f_t(x\mid x')$, $1\le t\le T$, where $f_t(x\mid x')$ denotes the probability density associated with moving from $x'$ to $x$.
Conditional on $(X^{(t)})_{0\le t\le T}$, the observations $(Y^{(t)})_{1\le t\le T}$ are statistically independent and their marginal densities are given by $g_t(y\mid x)$. That is, 
\begin{equation}\label{eq:state-space-model}
    \begin{aligned}
        X^{(0)} &\sim \mu_0, \\
        X^{(t)} \mid (X^{(t-1)} =x^{(t-1)}) &\sim f_t(\cdot \mid x^{(t-1)}),\quad 1\le t\le T, \\
        Y^{(t)} \mid (X^{(t)}=x^{(t)}) &\sim g_t(\cdot\mid x^{(t)}),\quad 1\le t\le T.
    \end{aligned}
\end{equation}
Given observations $y^{(1:t)}=(y^{(1)},\ldots,y^{(t)})$, the filtering problem primarily focuses on the target distribution $\pi_t(x) = \PP(X^{(t)} \in \dd x \mid Y^{(1:t)}=y^{(1:t)})$ and expectations $\pi_t(\varphi) = \EE [\varphi(X^{(t)})\mid Y^{(1:t)}=y^{(1:t)}]$ for some suitable test function $\varphi$.

Sequential Monte Carlo methods approximate $\pi_t$ by a random empirical measure
$$
  \pi_n^{(t)} = \sum_{i=1}^n W_i^{(t)}\,\delta_{X_i^{(t)}},
$$
where $\{X_i^{(t)}\}_{i=1}^n$ are particles and $\{W_i^{(t)}\}_{i=1}^n$ are normalized importance weights.
Given a particle set $\{(\tilde X_i^{(t-1)},\tilde W_i^{(t-1)})\}_{i=1}^m$, an SMC update at time $t$ proceeds as follows~\citep{doucet2011tutorial}:

\begin{enumerate}[leftmargin=*,itemsep=4pt]
  \item \textbf{Propagation.}  
        For each ancestor particle $i=1,\ldots,m$, simulate $K_t\ge1$ descendant particle(s) 
        $$
          X_{i,k}^{(t)} \sim q_t(\cdot\mid \tilde X_i^{(t-1)}), \quad k=1,\ldots,K_t,
        $$
        where $q_t$ is a proposal kernel; for example, $q_t=f_t$ in the bootstrap filter.
        The total number of candidate particles at time $t$ is $m K_t$.

  \item \textbf{Weighting.}  
        Update the importance weights according to the observation at time $t$: 
        $$
        W_{i,k}^{(t)} = \frac{\bar W_{i,k}^{(t)}}{\sum_{i=1}^m\sum_{k=1}^{K_t} \bar W_{i,k}^{(t)}},
        \quad \text{where}\quad
          \bar W_{i,k}^{(t)}
          \propto
          W_{i,k}^{(t-1)} \,
          \frac{g_t(Y^{(t)}\mid X_{i,k}^{(t)})\,f_t(X_{i,k}^{(t)} \mid \tilde X_{i,k}^{(t-1)})}
               {q_t(X_{i,k}^{(t)} \mid \tilde X_{i,k}^{(t-1)})}.
        $$
        For notational convenience, we relabel the subscripts $\{(i,k)\}_{i=1,\ldots,m, \, k=1,\ldots,K_t}$ as $\{i\}_{i=1,\ldots,n}$ with $n=mK_t$.

  \item \textbf{Resampling.}
        From $\{(X_i^{(t)},W_i^{(t)})\}_{i=1}^n$, generate 
        a set $\{\tilde X_i^{(t)}\}_{i=1}^m$ with equal weight, $\tilde W_i^{(t)}=1/m$, using some resampling scheme.
        We denote this operation by
        $$
          \{\tilde X_i^{(t)}\}_{i=1}^m
          \sim \cR_m\bigl(\{(X_i^{(t)},W_i^{(t)})\}_{i=1}^n\bigr),
        $$
        where $\cR_m$ represents a resampling operator. We keep $m$ fixed over time to ensure that the particle size does not grow uncontrollably over time.

\end{enumerate}

In practice, the multi-descendant proposal and resampling steps can be adapted based on the coefficient of variation (CV) of the weights. 
The CV of a weight vector $W=(W_1,\ldots,W_n)$ is defined as the ratio of its standard deviation to its mean, i.e., $\mathrm{cv}^2(W) = n^{-1}\sum_{i=1}^n (W_i-\bar W)^2/\bar W^2$, which is equal to $n\sum_{i=1}^n W_i^2 - 1$ if $W$ is normalized. In the survey literature, researchers often compute the effective sample size (ESS) as $\mathrm{ess}(W) =  n/(1+\mathrm{cv}^2(W))$ for inference with weighted samples, which was also suggested by \cite{kong1994sequential} for the SMC framework \citep{liu1995blind,liu1998sequential}. As such, CV can be used to provide a general measure of the degeneracy of a particle set, decoupled from the sample size $n$.

At timestep $t$, we compute $\mathrm{cv}(W^{(t-1)})$ and compare it with a predefined threshold $c_{\mathrm{thres}}\ge0$. If $\mathrm{cv}(W^{(t-1)})>c_{\mathrm{thres}}$, we regard the particle system as strongly degenerate and activate the multi-descendant proposal with $K_t = K > 1$; otherwise we set $K_t=1$ to save computational cost. 
After the weighting step at time $t$, we compute $\mathrm{cv}(W^{(t)})$ and perform resampling if either $K_t>1$ or $\mathrm{cv}(W^{(t)})>c_{\mathrm{thres}}$; otherwise we skip resampling and carry the weighted particles forward to the next timestep.

In what follows, we regard resampling as a stand-alone transformation that takes a weighted empirical measure $\mu_n$ and returns a random equal-weight measure $\nu_m$.
We say that a resampling operator $\cR_m$ is {\it unbiased}  if
$$
  \EE\big[ \nu_m(\varphi) \mid \mu_n \big]
  = \mu_n(\varphi),
  \quad \forall\,\varphi\in\cF,
$$
where $\cF$ denotes the set of all finite functions with respect to $\mu_n$.
Classical resampling schemes (e.g., multinomial, residual, stratified) are unbiased, while the method developed in this study trades off unbiasedness for improving geometric properties quantified via Wasserstein distance.

\subsection{Wasserstein distance and barycenter}

For $\mu,\nu\in\cP(\RR^d)$, a coupling (or transport plan) between $\mu$ and $\nu$ is a probability measure $\gamma$ on $\RR^d\times\RR^d$ with marginals
$\mu$ and $\nu$, i.e.,
$$
  \gamma(A\times\RR^d)=\mu(A),
  \quad
  \gamma(\RR^d\times B)=\nu(B),
  \quad
  \forall\,A,B\in\cB(\RR^d),
$$
where $\cB(\RR^d)$ denotes the Borel $\sigma$-algebra.
The set of all couplings between $\mu$ and $\nu$ is denoted by
$\Gamma(\mu,\nu)$.
For $p\ge1$, the $p$-Wasserstein distance~\citep{villani2009optimal, peyre2019computational} between $\mu$ and $\nu$ is defined as
$$
  W_p(\mu,\nu)
  = \Big(\inf_{\gamma\in\Gamma(\mu,\nu)}
    \int_{\RR^d\times\RR^d} \|x-y\|^p \,\gamma(\dd x,\dd y)\Big)^{1/p}.
$$

When $\mu$ and $\nu$ are discrete measures, the OT problem reduces to a finite-dimensional linear program.
For weighted empirical measures $\mu_n = \sum_{i=1}^n W_i\,\delta_{X_i}$ and $\nu_m = \sum_{j=1}^m V_j\,\delta_{Y_j}$ with weights $W\in\Delta^{n-1}$ and $V\in\Delta^{m-1}$, any coupling $\gamma\in\Gamma(\mu_n,\nu_m)$ can be represented by a nonnegative matrix $\Pi\in \cU(W,V)$, where
$$
  \cU(W,V) = \Big\{\Pi\in\RR_+^{n\times m}: \sum_{j=1}^m \Pi_{ij} = W_i, \quad \sum_{i=1}^n \Pi_{ij} = V_j\Big\}.
$$
The $p$-Wasserstein distance is then given by
$$
  W_p^p(\mu_n,\nu_m)
  = \min_{\Pi\in\cU(W,V)}
    \sum_{i=1}^n\sum_{j=1}^m \Pi_{ij}\,\|X_i-Y_j\|^p.
$$

\textbf{Wasserstein barycenters.}
Given a collection of probability measures $\mu^{(1)},\ldots,\mu^{(L)}$ on $\RR^d$ and nonnegative weights $\lambda_1,\ldots,\lambda_L$ with $\sum_{l=1}^L \lambda_l = 1$, a Wasserstein barycenter of $\{\mu^{(l)}\}_{l=1}^L$ is the minimizer
$$
  \nu^\star
  =
  \arg\min_{\nu\in\cP(\RR^d)}
  \sum_{l=1}^L \lambda_l \, W^p_p (\nu,\mu^{(l)}).
$$
This construction generalizes the notion of an average (or Fréchet mean) to the Wasserstein space.
In the \emph{free-support} setting, $\nu$ is allowed to be any probability measure in $\cP(\RR^d)$, so both its support and weights are determined by the optimization problem~\citep{claici2018stochastic, cuturi2014fast}.
In the \emph{fixed-support} setting, both $\nu$ and the input measures $\mu^{(l)}$ are discrete measures supported on a prescribed common support set.
In this case, the support locations are fixed and only the weights on that common support are optimized~\citep{benamou2015iterative, lin2020fixed}.
Our resampling formulation corresponds to a constrained free-support barycenter in which $\nu$ is required to be an equal-weight discrete measure with flexible support locations.

\section{Method}\label{sec:method}

\subsection{Free-support Wasserstein resampling with Dirichlet mixing}\label{subsec:wdr}

Recall that $\mu_n = \sum_{i=1}^n W_i \delta_{X_i}$ is a weighted empirical measure obtained after the weighting step of SMC, and $\nu_m = m^{-1} \sum_{i=1}^m \delta_{\tilde X_i}$ is the equal-weight measure returned by a resampling operator. For a test function $\varphi:\cX \to\RR$, the resampling estimator of $\mu_n(\varphi)=\sum_{i=1}^n W_i \varphi (X_i)$ is $\nu_m(\varphi)=m^{-1}\sum_{i=1}^m \varphi (\tilde X_i)$. Classical schemes are designed to satisfy $\EE[\nu_m(\varphi)\mid \mu_n]=\mu_n(\varphi)$ while reducing the conditional variance $\Var\{\nu_m(\varphi)\mid \mu_n\}$~\citep{liu1995blind, kitagawa1996monte, li2022stratification}.

However, exact unbiasedness for all test functions constrains resampled particles to the original support. If new locations were created with positive probability, one could always construct a function $\varphi$ separating the new support from the old, yielding $\EE[\nu_m(\varphi)\mid\mu_n]\neq\mu_n(\varphi)$.
To enable free supports, we relax exact unbiasedness and instead control the conditional MSE
$
  \EE [(\nu_m(\varphi)-\mu_n(\varphi))^2 \mid \mu_n].
$
For Lipschitz test functions, this MSE is controlled by the $p$-Wasserstein distance between $\mu_n$ and $\nu_m$, as established in Proposition~\ref{prop:mse-w}.

\begin{proposition}\label{prop:mse-w}
Let $\mu,\nu\in\cP(\RR^d)$ and let $\varphi:\RR^d\to\RR$ be $L$-Lipschitz, i.e.\ $|\varphi(x)-\varphi(y)|\le L\|x-y\|$ for all $x,y\in\RR^d$. Then, for any $p\ge1$,
\begin{equation}\label{eq:mse-W2}
  \EE\bigl[(\nu(\varphi)-\mu(\varphi))^2 \mid \mu\bigr]
  \;\le\;
  L^2\,\EE\bigl[W_p^2(\mu,\nu)\mid \mu\bigr].
\end{equation}
\end{proposition}

Inequality~\eqref{eq:mse-W2} motivates bringing $\nu_m$ as close as possible to $\mu_n$ in the $W_p$ distance. Throughout the remainder of the paper, we focus on $p=2$ and write $W(\mu,\nu)=W_2(\mu,\nu)$.

\textbf{Free-support Wasserstein approximation.}
Define the set of equal-weight discrete measures with $m$ atoms as $\cA_m = \{m^{-1}\sum_{j=1}^m \delta_{Y_j}: Y_j\in\RR^d \}$.
Motivated by Proposition~\ref{prop:mse-w}, our target resampled measure is the best $W_2$-approximation of $\mu_n$ in $\cA_m$:
\begin{equation}\label{eq:fs-barycenter}
  \nu_m^\star
  =
  \arg\min_{\nu_m\in\cA_m} W^2(\mu_n,\nu_m).
\end{equation}
This is a constrained free-support Wasserstein barycenter of the single input measure $\mu_n$, restricted to $\cA_m$. 
Using the discrete formulation in Section~\ref{sec:preliminaries},  problem~\eqref{eq:fs-barycenter} is equivalent to
\begin{equation}\label{eq:fs-barycenter-discrete}
  \min_{Y\in(\RR^d)^m}\,
  \min_{\Pi\in\cU(W,m^{-1}1_m)}\,
  \sum_{i=1}^n \sum_{j=1}^m \Pi_{ij}\,\|X_i - Y_j\|^2,
\end{equation}
where $W=(W_1,\ldots,W_n)$ are the input weights and $\cU(W,m^{-1}1_m)$ is the associated transportation polytope. The minimizer $(\Pi^\star,Y^\star)$ of~\eqref{eq:fs-barycenter-discrete} yields $\nu_m^\star = m^{-1}\sum_{j=1}^m \delta_{Y_j^\star}$, which is the closest distribution to $\mu_n$ in the $W_2$ metric within the class $\cA_m$.

Furthermore, $Y^\star$ can be expressed in terms of the columns of $\Pi^\star$.
For each $j=1,\ldots,m$, define $p^{(j)}_i := m\,\Pi^\star_{ij},\, i=1,\ldots,n$, so that $\sum_i p^{(j)}_i = 1$ and $p^{(j)}=(p^{(j)}_1,\ldots,p^{(j)}_n)$ is a probability vector. 
Then,
\begin{equation}\label{eq:barycentric-update}
Y^\star_j
=
\frac{\sum_{i=1}^n \Pi^\star_{ij} X_i}{\sum_{i=1}^n \Pi^\star_{ij}}
=
\sum_{i=1}^n p^{(j)}_i X_i,
\quad j = 1, \ldots, m.
\end{equation}
That is, each $Y^\star_j$ is the barycenter of the input particles under the column weights $p^{(j)}$.

The transformation $\mu_n\mapsto\nu_m^\star$ enjoys optimal geometric properties in the Wasserstein sense, but it is fully deterministic given $\mu_n$ and tends to underestimate particle spread~\citep{acevedo2017second}. To address this limitation, we introduce controlled stochasticity to $\nu_m^\star$ via a Dirichlet mixing mechanism.

\textbf{Dirichlet mixing randomization.}
Given $(\Pi^\star,Y^\star)$, we now define the Wasserstein-Dirichlet resampling (WDR) operator.
For $j=1,\ldots,m$, recall the probability vector $p^{(j)}\in\Delta^{n-1}$ and the barycentric projection $Y_j^\star = \sum_{i=1}^n p^{(j)}_i X_i$. Given a concentration parameter $\tau\ge 0$, we define the Dirichlet-mixed resampled particles as
\begin{equation}\label{eq:zj-tau}
  Z_j(\tau)
  := \sum_{i=1}^n V_i^{(j)} X_i,
  \quad \text{where}\quad
  V^{(j)} \sim \mathrm{Dirichlet}(\tau p^{(j)}),
  \quad j=1,\ldots,m,
\end{equation}
independently across $j$ conditional on $(\Pi^\star,X)$. Collecting $\tilde X_j := Z_j(\tau)$, $j=1,\ldots,m$, and assigning equal weights $1/m$ yields the equal-weight empirical measure $\nu_m^{\mathrm{WDR}}$.

This construction interpolates continuously between deterministic and fully randomized behavior. As $\tau\to\infty$, the Dirichlet weights $V^{(j)}$ concentrate around their mean $p^{(j)}$, and $Z_j(\tau)\to Y_j^\star$ in probability; in this limit WDR recovers the deterministic map $\mu_n\mapsto\nu_m^\star$ defined by~\eqref{eq:fs-barycenter}. 
At the other extreme, as $\tau\to 0^+$, the Dirichlet distribution with parameter $\tau p^{(j)}$ concentrates on the vertices of the probability simplex. In the limit case of $\tau=0$, we obtain $I_j\sim\mathrm{Categorical}(p^{(j)})$ and $Z_j(0)=X_{I_j}$, i.e., randomized resampling based on the optimal plan $\Pi^\star$, which is unbiased. For $d=1$, the optimal plan $\Pi^\star$ corresponds to a monotone transport map and $Z_j(0)$ reduces to sorted stratified resampling \citep{li2022stratification}.

For any fixed  $\tau>0$, the weights $V^{(j)}$ have full support on the simplex, and $Z_j(\tau)$ lies in the convex hull of the particles $\{X_i : p^{(j)}_i>0\}$. The parameter $\tau$ thus trades off Monte Carlo variability and shrinkage toward the barycentric projections $Y_j^\star$.

\medskip

\begin{remark}
    The discrete OT plan $\Pi^\star$ is sparse with at most $n+m-1$ nonzero entries~\citep{gerber2017multiscale}, so each column $p^{(j)}$ involves only $\cO(1)$ input particles on average. The Dirichlet mixing in~\eqref{eq:zj-tau} is therefore inherently local: each resampled particle combines only a small number of nearby ancestors rather than aggregating mass from the entire set. This locality preserves local structures and avoids excessive collapse of particles into a single location.
\end{remark}

\begin{remark}
The Dirichlet mixing~\eqref{eq:zj-tau} is related in spirit to the Bayesian bootstrap \citep{rubin1981bayesian,lyddon2019general}, which also uses Dirichlet weights to randomize an empirical measure. In the Bayesian bootstrap, one draws $V\sim\mathrm{Dirichlet}(1_n)$ on the simplex $\Delta^{n-1}$ and forms $\sum_{i=1}^n V_i\delta_{X_i}$ as a random empirical measure. The key difference is that, instead of employing a global Dirichlet weighting on the original support, WDR uses the optimal transport plan to construct many local Dirichlet mixtures, one for each output particle.
\end{remark}

\textbf{Truncated Wasserstein-Dirichlet resampling.}
In multi-descendant propagation, each ancestor generates multiple candidate descendants that probe the future independently. Hence, when weight degeneracy is mild, an additional resampling step to the entire population can waste budget by allocating multiple equal-weight output particles to essentially the same high-weight region.
We instead keep high-weight particles fixed and devote more budget to rejuvenate medium to low-weight regions; related thresholding ideas have appeared in, e.g.,  \cite{liu1998rejection} and \cite{fearnhead2003line}.
This leads to truncated WDR (T-WDR), a variant of WDR that preserves high-weight particles as anchors and reallocates only the residual mass, thereby retaining the dominant structure while providing sufficient opportunity for tail rejuvenation.
T-WDR shares the same objective~\eqref{eq:fs-barycenter} as WDR, and differs only in the feasible class $\cA_m$.
Given a threshold $\kappa>0$, define
$S_\kappa=\{i: W_i\ge\kappa\}$,
$r_\kappa=\sum_{i\notin S_\kappa}W_i$, and
$m_{\mathrm{tail}}=m-|S_\kappa|$.
We then restrict the optimization to
$
  \mathcal A_m^{\mathrm{T\text{-}WDR}}
  =
  \{
  \sum_{i\in S_\kappa}W_i\delta_{X_i}
  +r_\kappa m_{\mathrm{tail}}^{-1}\sum_{j=1}^{m_{\mathrm{tail}}}\delta_{Y_j}
  : Y_j\in\RR^d
  \}.
$
Under a continuity condition, the threshold $\kappa$ is characterized as the solution to $m = \sum_{j=1}^n \min(W_j/\kappa, 1)$. 
Further details are provided in Section~S1.
We recommend T-WDR as an efficient surrogate for WDR when weight degeneracy is mild; under severe degeneracy, full WDR is generally preferable.

\subsection{Block coordinate descent algorithm}\label{subsec:bcd}

In this section, we develop a numerical framework for efficient computation of the problem~\eqref{eq:fs-barycenter-discrete}.  
Although many numerical solvers for Wasserstein barycenters have been proposed, most focus on fixed-support formulations or regularized variants~\citep{benamou2015iterative, lin2020fixed}. For the free-support setting, existing algorithms are either not directly compatible with the equal-weight constraint in resampling~\citep{cuturi2014fast}, or are general-purpose solvers~\citep{claici2018stochastic} that can be prohibitively slow in our regime because the barycenter subproblem must be solved repeatedly within SMC.

Therefore, we propose an efficient algorithm tailored to the structure of~\eqref{eq:fs-barycenter-discrete}.
The objective in~\eqref{eq:fs-barycenter-discrete} is convex in $\Pi$ for fixed $Y$ and convex in $Y$ for fixed $\Pi$, yielding a biconvex structure.  
We exploit this with a block coordinate descent (BCD) procedure~\citep{bertsekas2016nonlinear} that alternates between updating the transport plan and the support locations. 
In particular, at iteration $r$, the BCD scheme performs the following two steps.

\begin{enumerate}[leftmargin=*,itemsep=4pt]
  \item \textbf{Transport step.}
        Given $Y^{[r]}$, compute the optimal transport plan
        \begin{equation}\label{eq:update-Pi}
          \Pi^{[r+1]}
          = \arg\min_{\Pi\in\cU(W,m^{-1}1_m)}
          \sum_{i=1}^n \sum_{j=1}^m \Pi_{ij}\,
          \|X_i - Y_j^{[r]}\|^2.
        \end{equation}
        This is a standard discrete OT problem with fixed supports $X$ and $Y^{[r]}$.

  \item \textbf{Support step.}
        Given $\Pi^{[r+1]}$, update the support locations $Y^{[r+1]}$ by minimizing the objective with respect to $Y$:
        $$
          Y^{[r+1]}
          = \arg\min_{Y\in(\RR^d)^m}
          \sum_{i=1}^n \sum_{j=1}^m \Pi_{ij}^{[r+1]}\,\|X_i - Y_j\|^2.
        $$
        This separates across columns and admits the closed-form update as in~\eqref{eq:barycentric-update}, that is,
        \begin{equation}\label{eq:update-Y}
          Y_j^{[r+1]}
          =
          \frac{\sum_{i=1}^n \Pi^{[r+1]}_{ij} X_i}
               {\sum_{i=1}^n \Pi^{[r+1]}_{ij}},
          \qquad j=1,\ldots,m,
        \end{equation}
        so each $Y_j^{[r+1]}$ is the barycenter of $X$ under the $j$th column of $\Pi^{[r+1]}$.
\end{enumerate}

Each iteration decreases the objective in~\eqref{eq:fs-barycenter-discrete}, and the sequence $\{(\Pi^{[r]},Y^{[r]})\}$ has convergence guarantees (see Section~\ref{subsec:algo-conv}).  
In practice, we terminate the iterations when the relative change in $Y$ falls below a small tolerance, or after a fixed maximum number of iterations. In our experiments, a small number of iterations (typically fewer than ten) is sufficient to reach numerical convergence.  

Equipped with the BCD solver, the complete Wasserstein-Dirichlet resampling procedure is summarized in Algorithm~\ref{alg:wdr}. In the SMC context of Section~\ref{sec:preliminaries}, we use Algorithm~\ref{alg:wdr} as a drop-in replacement for the generic resampling operator $\cR_m$, with $m$ chosen according to the desired number of descendants.

{\spacingset{1.25} 
\begin{algorithm}[!t]
  \caption{Wasserstein-Dirichlet resampling (WDR)}
  \label{alg:wdr}
  \begin{algorithmic}[1] 
    \State \textbf{Input:} weighted particles $\{(X_i,W_i)\}_{i=1}^n$, 
           target size $m$, concentration parameter $\tau \ge 0$
    \State Compute $(\Pi^\star, Y^\star)$ by solving
           \eqref{eq:fs-barycenter-discrete} via block coordinate descent:
    \Statex \hspace{\algorithmicindent}
            Initialize $Y^{[0]} = (Y_1^{[0]},\ldots,Y_m^{[0]})$, set $r \gets 0$
    \Statex \hspace{\algorithmicindent} \textbf{repeat}
    \Statex \hspace{\algorithmicindent}~~~
            Update $\Pi^{[r+1]}$ by solving~\eqref{eq:update-Pi} with $Y^{[r]}$
    \Statex \hspace{\algorithmicindent}~~~
            Update $Y^{[r+1]}$ using~\eqref{eq:update-Y} with $\Pi^{[r+1]}$
    \Statex \hspace{\algorithmicindent}~~~ $r \gets r+1$
    \Statex \hspace{\algorithmicindent} \textbf{until} convergence criterion is met
    \Statex \hspace{\algorithmicindent}
            $(\Pi^\star, Y^\star) \gets (\Pi^{[r]}, Y^{[r]})$
    \For{$j = 1,\dots,m$}
      \State Set $p^{(j)}\gets m\,\Pi^\star_{\cdot j}$
      \State Draw $V^{(j)} \sim \mathrm{Dirichlet}(\tau p^{(j)})$
               and set $\tilde X_j \gets \sum_{i=1}^n V^{(j)}_i X_i$
    \EndFor
    \State \textbf{Output:} equal-weight particles $\{\tilde X_j\}_{j=1}^m$
  \end{algorithmic}
\end{algorithm}
}

In Algorithm~\ref{alg:wdr}, we initialize $Y^{[0]}$ by subsampling $m$ particles at random.
Regarding the subproblems, when $n$ is moderate, we solve the transport subproblem~\eqref{eq:update-Pi} exactly using the network flow method~\citep{flamary2021pot}.  
For larger $n$, we use the sparsified approximation solver proposed by \citet{li2023importance}, which achieves near-linear time complexity.  
Due to the sparsity of the OT plan $\Pi$ with $\tilde\cO(n)$ nonzero entries, the barycentric update~\eqref{eq:update-Y} can also be implemented in $\tilde\cO(n)$ time.  
Thus, each BCD iteration has near-linear complexity $\tilde\cO(n)$. Since the BCD iterations converge within a small number of iterations (as shown in Section~S3.1), the overall cost of WDR scales approximately linearly in $n$, comparable to that of classical resampling methods, while exploiting richer geometric structure.

\section{Theoretical results}
\subsection{Statistical properties}\label{subsec:stat-prop}

We first quantify the approximation error of the optimal equal-weight atomic approximation to a given measure, then analyze the bias-variance trade-off of the Wasserstein-Dirichlet resampling.

\begin{lemma}
\label{lem:wasserstein-error}
Let $s\in\NN$ be the intrinsic dimension of the support of $\mu$ in the following sense: after a rigid change of coordinates in $\RR^d$, the support of $\mu$ is contained in an axis-aligned hyperrectangle $Q_s = \prod_{k=1}^s [a_k,b_k] \times \{0\}^{d-s}\subset \RR^d$ with side lengths $\ell_k := b_k-a_k > 0$ for $k=1,\dots,s$.
Let $D := (\sum_{k=1}^s \ell_k^2)^{1/2}$ be the Euclidean diameter of the active block $\prod_{k=1}^s[a_k,b_k]$.
Then for any $m\in\NN$ and any $p < s$, there exists an equal-weight atomic measure $\nu_m := m^{-1}\sum_{j=1}^m \delta_{y_j}$ with $y_1,\dots,y_m\in Q_s$ such that
\begin{equation*}
    W_p(\mu,\nu_m) \le C_{s,p}\,D\,m^{-1/s},
\end{equation*}
where $C_{s,p}$ depends only on $s$ and $p$.
\end{lemma}

This result extends standard quantization rates~\citep{gray2002quantization, kloeckner2012approximation} to the constrained setting of equal-weight discrete approximation, preserving the polynomial decay in the intrinsic dimension $s$. 
While Lemma~\ref{lem:wasserstein-error} assumes an $s$-dimensional axis-aligned support for brevity, the convergence rate $W_p(\mu,\nu_m) \lesssim m^{-1/s}$ can extend to more general structures, such as $s$-dimensional manifolds~\citep{weed2019sharp, hundrieser2024empirical}, up to a change in the constant factor. 

This viewpoint is motivated by the fact that, in many filtering and generative problems, the target distribution is far from isotropic in the ambient space. Strong dependence or latent geometric structure might cause the posterior to concentrate near a lower-dimensional set or along a small number of effective directions. For example, although image generation problems are formulated in high-dimensional pixel space, natural images are commonly believed to lie on a much lower-dimensional manifold. In such cases, geometry-aware resampling is expected to be advantageous.
We now state a non-asymptotic bound for the Wasserstein-Dirichlet resampling estimator of $L$-Lipschitz functions.

\begin{theorem}\label{thm:mse-error}
Let $\varphi:[0,1]^d \to \RR$ be $L$-Lipschitz and let $\mu_n(\varphi):= \sum_{i=1}^n W_i \varphi(X_i)$ be the target function.
Define the Wasserstein-Dirichlet resampling estimator $\nu_m^{\mathrm{WDR}}(\varphi):= m^{-1}\sum_{j=1}^m \varphi (Z_j(\tau))$,
where $\{Z_j(\tau)\}_{j=1}^m$ are the resampled particles constructed in Section~\ref{sec:method} with concentration parameter $\tau \ge 0$.
Then, under the assumption of Lemma~\ref{lem:wasserstein-error},
\begin{equation}\label{eq:mse-error}
  \EE\Bigl[\bigl(\nu_m^{\mathrm{WDR}}(\varphi) - \mu_n(\varphi)\bigr)^2 \mid \mu_n \Bigr]
  \le
  L^2 C_s^2
  \left[
    \frac{\tau}{\tau+1}\,m^{-2/s}
    +
    \frac{1}{\tau+1}\,m^{-1-2/s}
  \right],
\end{equation}
where $s$ is the intrinsic dimension under the structural condition of Lemma~\ref{lem:wasserstein-error}, and the constant $C_{s}$ depends only on $s$.
\end{theorem}

The right-hand side of \eqref{eq:mse-error} arises from a bias-variance decomposition of MSE.  
The first term $\cO(m^{-2/s})$, weighted by $\tau/(\tau+1)$, corresponds to the squared bias from the free-support Wasserstein approximation. As $\tau \to \infty$, WDR becomes the deterministic barycentric transform, so only this bias term remains.  
The second term $\cO(m^{-1-2/s})$, weighted by $1/(\tau+1)$, corresponds to the conditional variance from Dirichlet mixing. When $\tau = 0$, WDR reduces to ancestor sampling according to the optimal transport plan, and then the bias term vanishes, leaving only the variance term.
Especially, for linear test functions $\varphi$, the bias term vanishes, so the MSE reduces to the variance term $\cO(m^{-1-2/s})$ for all $\tau\ge0$.

It is also worth noting that the variance rate $\cO(m^{-1-2/s})$ matches that of Hilbert curve resampling~\citep{li2022stratification}, which samples ancestors according to a transport plan constructed via Hilbert curve projections. In this sense, WDR generalizes the Hilbert curve resampling by providing a continuous interpolation between deterministic and randomized schemes, thereby balancing estimation bias and variance.

\subsection{Algorithmic convergence}\label{subsec:algo-conv}

Let $\{(\Pi^{[r]},Y^{[r]})\}_{r\ge0}$ be the sequence produced by the BCD part of
Algorithm~\ref{alg:wdr}, i.e., alternating exact minimization over $\Pi$ and $Y$.
The next result establishes a well-behaved convergence guarantee.

\begin{theorem}\label{thm:algo-conv}
Denote $F(\Pi,Y)$ as the objective function in~\eqref{eq:fs-barycenter-discrete}.
Recall that, for each $r\ge0$,
$$
  \Pi^{[r+1]} \in \arg\min_{\Pi\in\cU(W,m^{-1}\mathbf 1_m)} F(\Pi,Y^{[r]}),
  \quad
  Y^{[r+1]} = \arg\min_{Y\in(\RR^d)^m} F(\Pi^{[r+1]},Y).
$$
Then, every limit point of $\{(\Pi^{[r]},Y^{[r]})\}$ is a stationary point of the problem~\eqref{eq:fs-barycenter-discrete}.
\end{theorem}

The monotone decrease property used in the convergence analysis is established in Lemma~S3 of the Supplementary Materials, while Theorem~\ref{thm:algo-conv} guarantees the stationarity of all limit points.
Since $F$ is nonconvex, the theorem does not assert global optimality without additional regularity.
Empirically, the BCD solver is numerically stable for the resampling problem and shows little sensitivity to initialization, as shown in Section~S3.1.

\section{Simulation study}\label{sec:simu}
This section evaluates the proposed resampling scheme on linear Gaussian state-space models.
Section~\ref{subsec:simu-lg} compares accuracy and stability across dimensions and condition numbers at matched particle sizes, and then identifies the propagation cost above which WDR is preferred at matched wall-clock time.
Section~\ref{subsec:simu-b3g} scales the comparison to a high-dimensional spatial model, which requires a block particle filter and has a high propagation cost.
Two nonlinear benchmarks, the Lorenz-63 system and a stochastic volatility model, are provided in Sections~S3.2 and~S3.4.

We compare the proposed (T-)WDR with classical multinomial, residual, and stratified resampling, and with the OT-based Hilbert curve resampling~\citep{li2022stratification}. Another OT-based approach, ETPF~\citep{reich2013nonparametric}, is not included in the comparison because it is incompatible with the multi-descendant SMC framework.
We set the Dirichlet concentration parameter $\tau$ to 30 as the default.
For the transport subproblem in Algorithm~\ref{alg:wdr}, we solve the OT problem exactly using a network flow solver~\citep{flamary2021pot} when $m \le 200$, covering the Lorenz-63 study; while for larger $m$, such as those considered in this section, we use a sparsified proximal point method~\citep{xie2020fast,li2023importance} to approximate the transport plan.

\textbf{Evaluation metrics.}
The primary goal is to estimate the posterior mean trajectory $I^\star=(I^{\star (1)},\ldots,I^{\star (T)})$, where $I^{\star(t)}$ denotes the oracle value of $\EE[X^{(t)} \mid Y^{(1:t)}]$ under the state-space model~\eqref{eq:state-space-model}.
For the linear Gaussian models, the oracle is available in closed form from the Kalman filter. 

We consider three metrics to evaluate each resampling method. At time $t$, a given method produces a weighted particle approximation $\{(X_i^{(t)},W_i^{(t)})\}_i$ to the filtering distribution. We compute the weighted particle mean $\hat I^{(t)}:=\sum_{i} W_i^{(t)} X_i^{(t)}$ and evaluate its discrepancy from the oracle posterior mean $I^{\star(t)}$ using $\mathrm{RMSE}_t = \|\hat I^{(t)} - I^{\star (t)}\|/\sqrt{d}$, and then report the time-averaged RMSE over the $T$ timesteps, $\mathrm{RMSE}=T^{-1}\sum_{t=1}^T \mathrm{RMSE}_t$.

We also evaluate the filtering distribution using the energy score (ES), a strictly proper scoring rule \citep{acevedo2017second}. At time $t$, given the weighted particle approximation $\sum_i W_i^{(t)} \delta_{X_i^{(t)}}$ and the oracle posterior mean $I^{\star(t)}$, the energy score is defined as
\begin{equation}\label{eq:energy-score}
\mathrm{ES}_t
=
\sum_{j=1}^n W_j^{(t)} \| X_j^{(t)} - I^{\star (t)} \|
-
\frac12 \sum_{j=1}^n\sum_{l=1}^n W_j^{(t)} W_l^{(t)}
\| X_j^{(t)} - X_l^{(t)} \|.
\end{equation}
We then compute its time-averaged version $\mathrm{ES} = T^{-1}\sum_{t=1}^T \mathrm{ES}_t$.
The first term in~\eqref{eq:energy-score} measures the proximity to the oracle state, and the second term accounts for the ensemble spread.
Thus, ES penalizes under-dispersion caused by the particle impoverishment.
For the linear Gaussian models, we additionally compare the particle approximation with the exact filtering distribution using the energy distance (ED); details are provided in Section~S3.7.
For all the three metrics, lower values indicate better performance.

\subsection{Linear Gaussian state-space model}\label{subsec:simu-lg}
We consider a linear Gaussian state-space model, which can be viewed as discrete observations at unit time intervals on the latent continuous-time Ornstein--Uhlenbeck (OU) diffusion, $\mathrm{d}X_u = B X_u\,\mathrm{d}u + C\,\mathrm{d}W_u$, with $W_u$ being a standard $d$-dimensional Brownian motion:
\begin{equation*}
\begin{aligned}
X^{(t)} &= AX^{(t-1)} + \xi^{(t)}, \quad \xi^{(t)}\sim\cN(0,\Sigma), \\
Y^{(t)} &= X^{(t)} + \varepsilon^{(t)}, \quad \varepsilon^{(t)}\sim\cN(0,I_d), \quad t=1,2,\ldots, T.
\end{aligned}
\end{equation*}
The process is initialized at $X^{(0)}\sim\cN(0,I_d)$, where $A=e^{B}$ and $\Sigma = \int_0^1 e^{Bs}\,CC^\top e^{B^\top s}\,\mathrm{d}s$.
We take $A=0.95\, I_d$ and $\Sigma = 0.5\,Q \,\mathrm{diag}(\lambda_1,\dots,\lambda_d)\, Q^\top$, where $Q$ is a fixed random orthogonal matrix and $\lambda_j = \kappa^{-(j-1)/(d-1)}$, so the eigenvalues decrease geometrically from $1$ to $1/\kappa$, where $\kappa$ is called the {\it condition number} of $\Sigma$.
To investigate the effects of dimensionality and anisotropy, we consider $d \in \{20, 50, 100\}$ and $\kappa \in \{1, 10, 50\}$, with particle sizes $m \in \{2^8, 2^9, \ldots, 2^{12}\}$.
The number of particles needed to avoid weight collapse grows exponentially with the effective dimension (Section~\ref{subsec:background}), so even the largest of these particle sizes is small at these dimensions.
Every setting thus lies in the small-particle regime that motivates this work.

In each of the twenty independent replications, we simulate a trajectory of $T=1000$ timesteps and run a bootstrap particle filter with the adaptive multi-descendant proposal and adaptive resampling of Section~\ref{subsec:smc}, setting $K=3$ and $c_{\mathrm{thres}}=2$.
We compare WDR with the baselines listed above, excluding T-WDR because the higher-dimensional settings exhibit severe weight degeneracy, for which it is not designed.

\subsubsection{Filtering results}
\begin{figure}[!t]
    \centering
    \includegraphics[width=\linewidth]{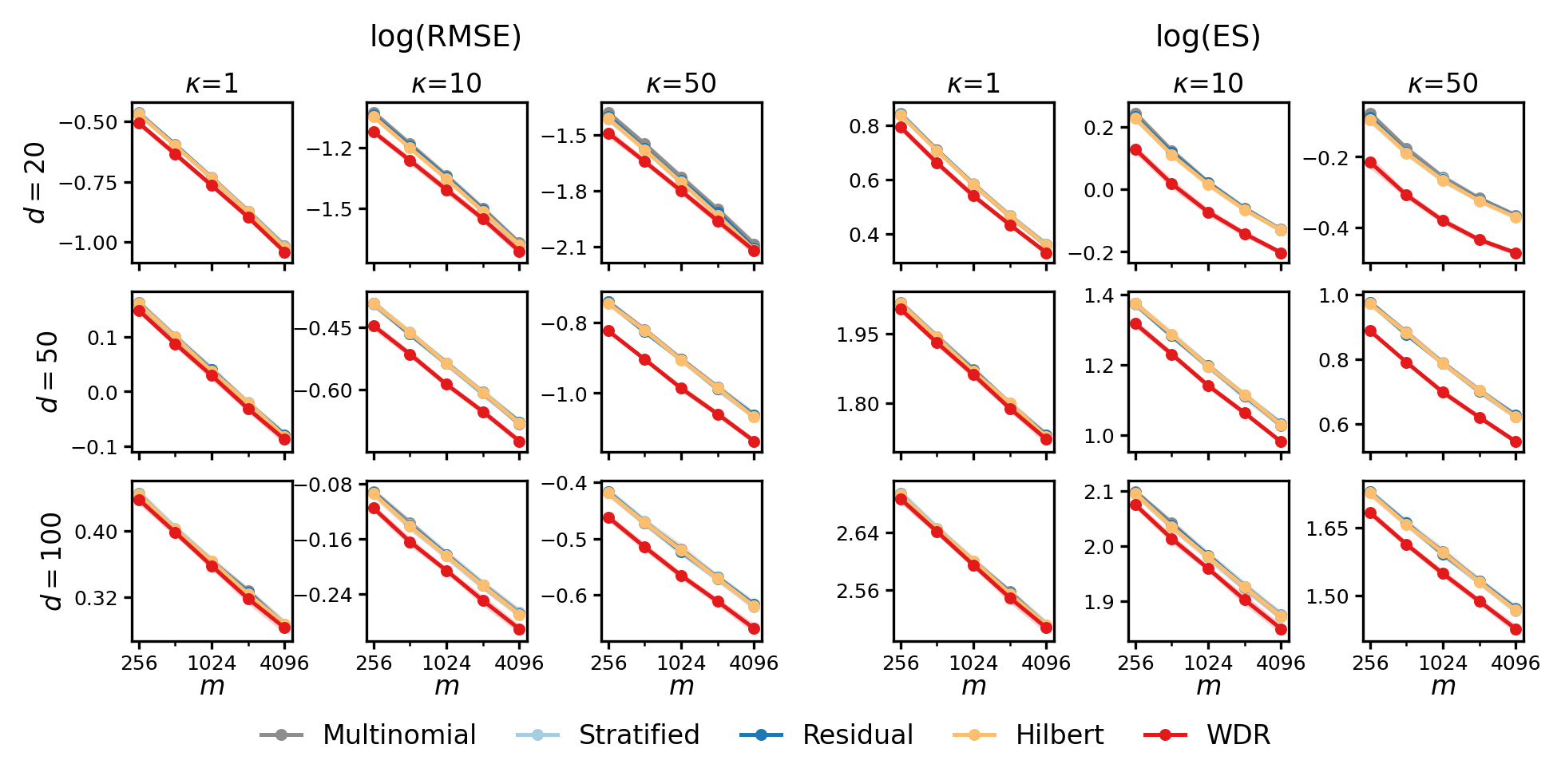}
    \caption{Time-averaged RMSE (left) and energy score (right) versus the particle size $m$ on the log-log scale for the linear Gaussian model. Rows correspond to the dimension $d \in \{20, 50, 100\}$ and columns to the condition number $\kappa \in \{1, 10, 50\}$. Shaded bands represent standard deviations based on twenty replications.}
    \label{fig:simu-lg-grid}
\end{figure}

Figure~\ref{fig:simu-lg-grid} reports the time-averaged RMSE and energy score against the Kalman reference $I^{\star(t)}$, and shows that WDR performs the best across all $(d,\kappa)$, while the four baselines are nearly indistinguishable.
For both metrics, the gap widens with $\kappa$, because classical schemes only replicate existing particles and thus suffer from weight degeneracy and sample impoverishment in ill-conditioned settings, whereas WDR's OT-informed coupling creates new support points that preserve particle dispersion and posterior coverage.
This is consistent with the intrinsic dimension dependence in Theorem~\ref{thm:mse-error}: stronger anisotropy concentrates the posterior along fewer effective directions, bringing it closer to a lower-dimensional structure, where geometry-aware resampling is beneficial.
RMSE depends on the particle distribution only through its mean, where WDR's margin over the baselines becomes smaller as $m$ grows when $d=20$, although its curve remains lowest throughout.
The energy score instead depends on the whole particle distribution, which is what WDR's free-support placement improves, and there its advantage persists at the largest particle size.
The energy distance results in Section~S3.7 yield rankings consistent with those based on RMSE and energy score.

Per-time trajectories further confirm that WDR's advantage holds over the whole time horizon for all the considered particle sizes; see Section~S3.3.1 for details.

\subsubsection{Stability analysis}
Besides accuracy, it is also important to assess the stability of resampling methods across repeated runs. To this end, we fix one dataset from the linear Gaussian model with $(d,\kappa)=(50,50)$ and rerun the particle filter twenty times with different random seeds. We consider three stability metrics, moving from the particle level to the estimator level and then to the evaluation level, all averaged over timesteps: (i) the between-run energy distance between the full particle distributions; (ii) the between-run standard deviation of the posterior mean estimate; and (iii) the between-run standard deviation of the RMSE relative to the Kalman oracle. Detailed definitions of these metrics are provided in Section~S3.3.2. For all three metrics, smaller values indicate greater stability.

\begin{figure}[!t]
    \centering
    \includegraphics[width=0.805\linewidth]{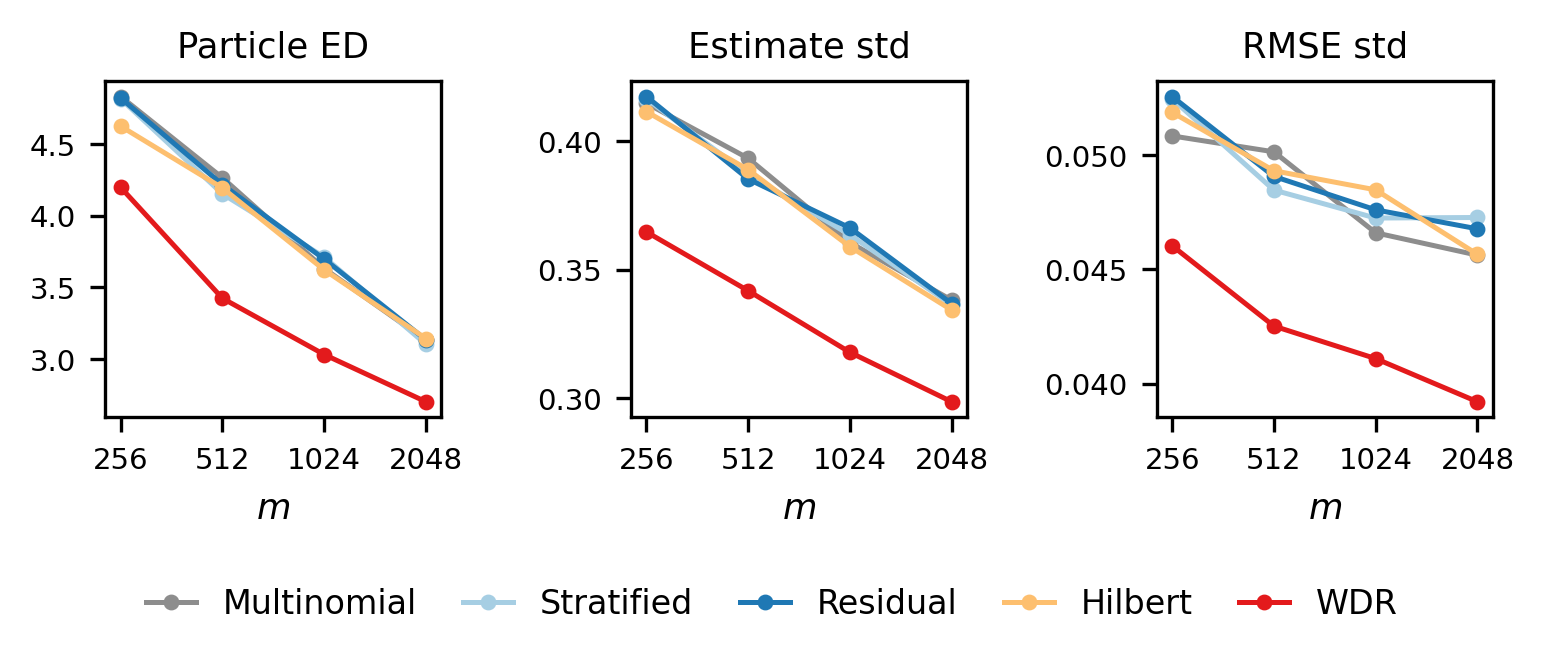}
    \caption{Between-run stability metrics versus particle size $m$ for the linear Gaussian model. From left to right: between-run particle energy distance, between-run standard deviation of the estimate, and between-run standard deviation of RMSE. Lower values are better.}
    \label{fig:simu-lg-stability}
\end{figure}

Figure~\ref{fig:simu-lg-stability} plots the stability metrics against particle size $m$. Although stratified and residual resampling are known to have lower conditional variance than multinomial resampling at a single step~\citep{chopin2020introduction}, the between-run variability accumulates with time and aggregates propagation randomness in the particle filter. Consequently, differences among these classical resampling methods are largely masked in the final run-to-run metrics, and their curves remain close. In contrast, WDR lies below all baseline methods across the full range of $m$ on all three metrics. This can be attributed to the fact that WDR constructs a structured and geometrically coherent transport of the particle cloud, rather than randomly selecting existing particles. This leads to a more stable approximation of the filtering distribution, demonstrating that WDR improves not only estimation accuracy but also between-run stability.

\subsubsection{Performance-cost analysis}\label{subsec:simu-cost}
The above experiments compared methods at equal particle size. Considering that WDR incurs additional resampling overhead, a more practical criterion is performance at equal wall-clock time. We design an experiment that varies the propagation cost to identify when WDR is the preferred choice.

We continue to use the linear Gaussian model with $(d,\kappa)=(50,50)$, and now make the cost of simulating the latent diffusion explicit. For a general nonlinear diffusion, the transition kernel has no closed form, so each propagation step must integrate the dynamics over many small sub-steps between observations, with cost proportional to the integration resolution. We emulate this cost by simulating the OU dynamics on a grid of $\gamma$ internal sub-steps per observation interval. Specifically, we set $A=P\,\mathrm{diag}(q_1,\dots,q_d)\,P^\top$, where $P$ is a fixed random orthogonal matrix and $q_j = 0.95^{\,1+s(2(j-1)/(d-1)-1)}$ with spread $s=0.5$, and each sub-step applies the transition $A_{\mathrm{sub}}=P\,\mathrm{diag}(q_1^{1/\gamma},\dots,q_d^{1/\gamma})\,P^\top$ with a matching sub-step noise covariance $\Sigma_{\mathrm{sub}}$, so that $\gamma$ successive sub-transitions under $(A_{\mathrm{sub}},\Sigma_{\mathrm{sub}})$ exactly reproduce the original one-step transition under $(A,\Sigma)$.
This construction leaves the target filtering distribution and the Kalman oracle unchanged, while making the propagation cost scale approximately linearly with $\gamma$.

\begin{figure}[!t]
    \centering
    \includegraphics[width=\linewidth]{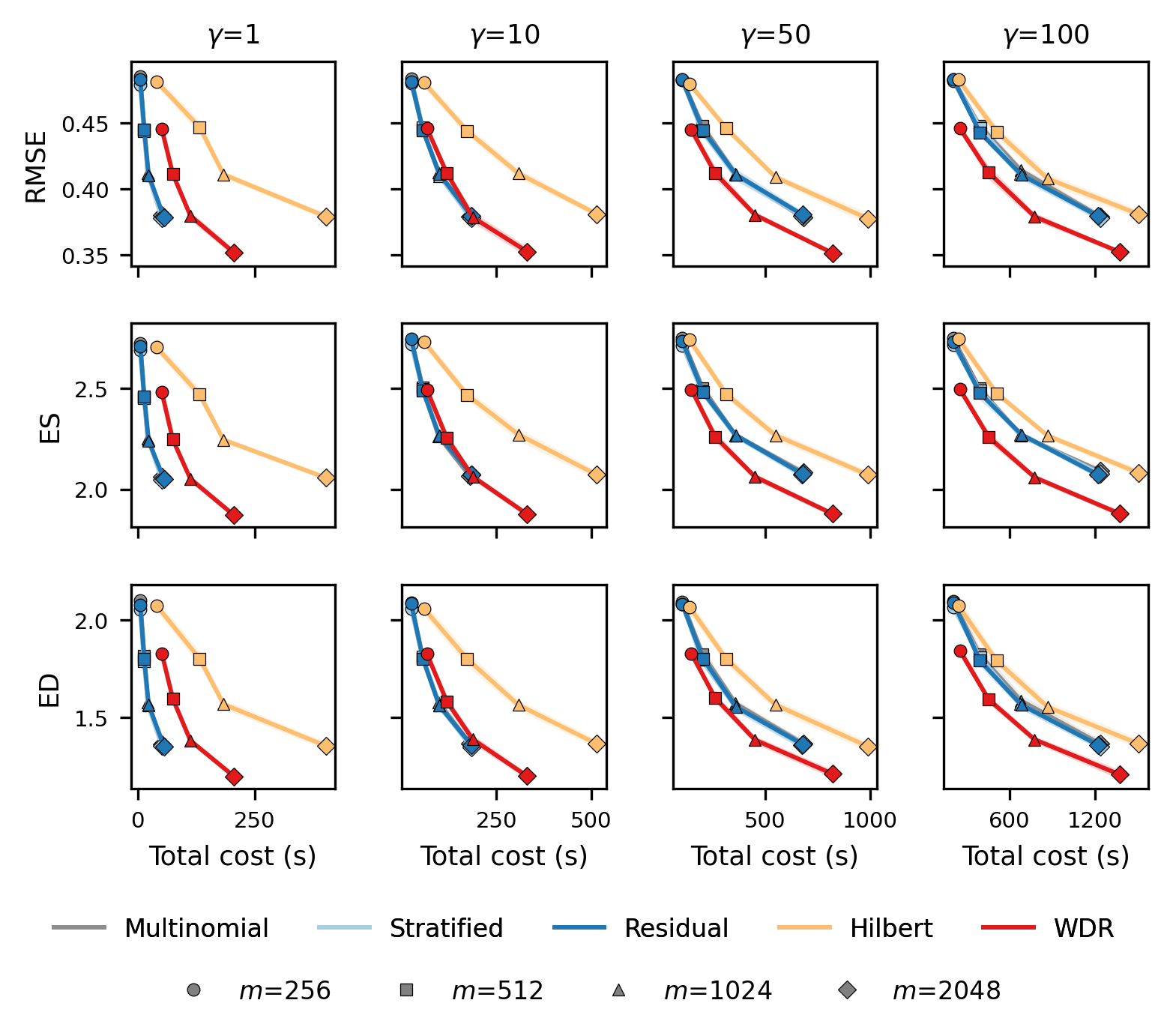}
    \caption{Time-averaged RMSE (top row), energy score (middle row), and energy distance (bottom row) versus total wall-clock cost on the linear Gaussian model with $(d,\kappa)=(50,50)$.
    Each panel corresponds to a sub-step multiplier $\gamma$.
    Shaded bands represent standard deviations based on twenty replications.}
    \label{fig:simu-lg-cost}
\end{figure}

Figure~\ref{fig:simu-lg-cost} plots the time-averaged RMSE, energy score, and energy distance against the total wall-clock time for each $\gamma$. 
As $\gamma$ grows, propagation gradually comes to dominate the computational budget, while resampling accounts for an ever-smaller fraction of total runtime.
In particular, the ratio of WDR resampling time to propagation time, averaged across $m$, falls from $5.1\times$ at $\gamma=1$ to $1.1\times$ at $\gamma=10$, $0.30\times$ at $\gamma=50$, and $0.16\times$ at $\gamma=100$.
Consistent with this shift in the computational budget, the practical benefit of WDR becomes increasingly clear as $\gamma$ rises.
For $\gamma\ge10$, WDR lies on the Pareto frontier for each of the three metrics: no alternative method simultaneously achieves a lower value on the corresponding metric and a lower total runtime.

Translated to absolute time, $\gamma=10$ corresponds to a per-particle-per-step propagation cost of roughly $0.1$\,ms, giving a practical threshold above which WDR is the recommended choice. Although this threshold is a representative order-of-magnitude reference in our experiments, the qualitative claim that WDR is preferable when propagation dominates is robust across settings. This regime is directly relevant to modern SMC applications, such as the diffusion model of Section~\ref{sec:application}, where the propagation cost is far above this threshold. We also conduct sensitivity analyses over the Dirichlet concentration parameter $\tau$ (defined in~\eqref{eq:zj-tau}), the multi-descendant factor $K$, and the CV threshold $c_{\mathrm{thres}}$ (see Section~\ref{subsec:smc}), and observe that WDR performs robustly, with intermediate values of $\tau$ performing the best. Details are provided in Section~S3.5.

\subsection{High-dimensional Gaussian spatial model}\label{subsec:simu-b3g}
To test a problem in which the model itself makes propagation expensive, we consider a linear Gaussian spatial model on a $45 \times 45$ lattice with $d = 2025$ grid sites:
\begin{equation*}
\begin{aligned}
X^{(t)} &= 0.95\, X^{(t-1)} + \xi^{(t)}, \quad \xi^{(t)}\sim\cN(0,\Sigma), \\
Y^{(t)} &= H X^{(t)} + \varepsilon^{(t)}, \quad \varepsilon^{(t)}\sim\cN(0,I_{d_y}),
\end{aligned}
\end{equation*}
where each $X^{(t)}$ is $d$-dimensional; $\Sigma_{ij} = 0.5(1+\sqrt{3}r_{ij}/\rho)\exp(-\sqrt{3}r_{ij}/\rho)$ is a Mat\'ern covariance with $\rho = 6$ and $r_{ij}$ the distance between sites $i$ and $j$ on the unit-spacing lattice; and $H\in\{0,1\}^{d_y\times d}$ has $d_y = 202$ rows, each selecting one site (only one ``1'' in each row). 
Generating the spatially correlated process noise $\xi^{(t)}$ requires a dense $d\times d$ matrix-vector product per particle per step, incurring substantial computational cost when \(d = 2025\).
At this dimension, the generic bootstrap filter collapses with any resampling scheme, so we employ the block particle filtering strategy~\citep{rebeschini2015can}, detailed in Section~S3.6 of the Supplementary Materials.
We simulate a trajectory of $T = 200$ timesteps and compare all the resampling methods discussed in Section~\ref{subsec:simu-lg}. 
The oracle reference is again available from the Kalman filter.

\begin{figure}[!t]
    \centering
    
        \includegraphics[width=\linewidth]{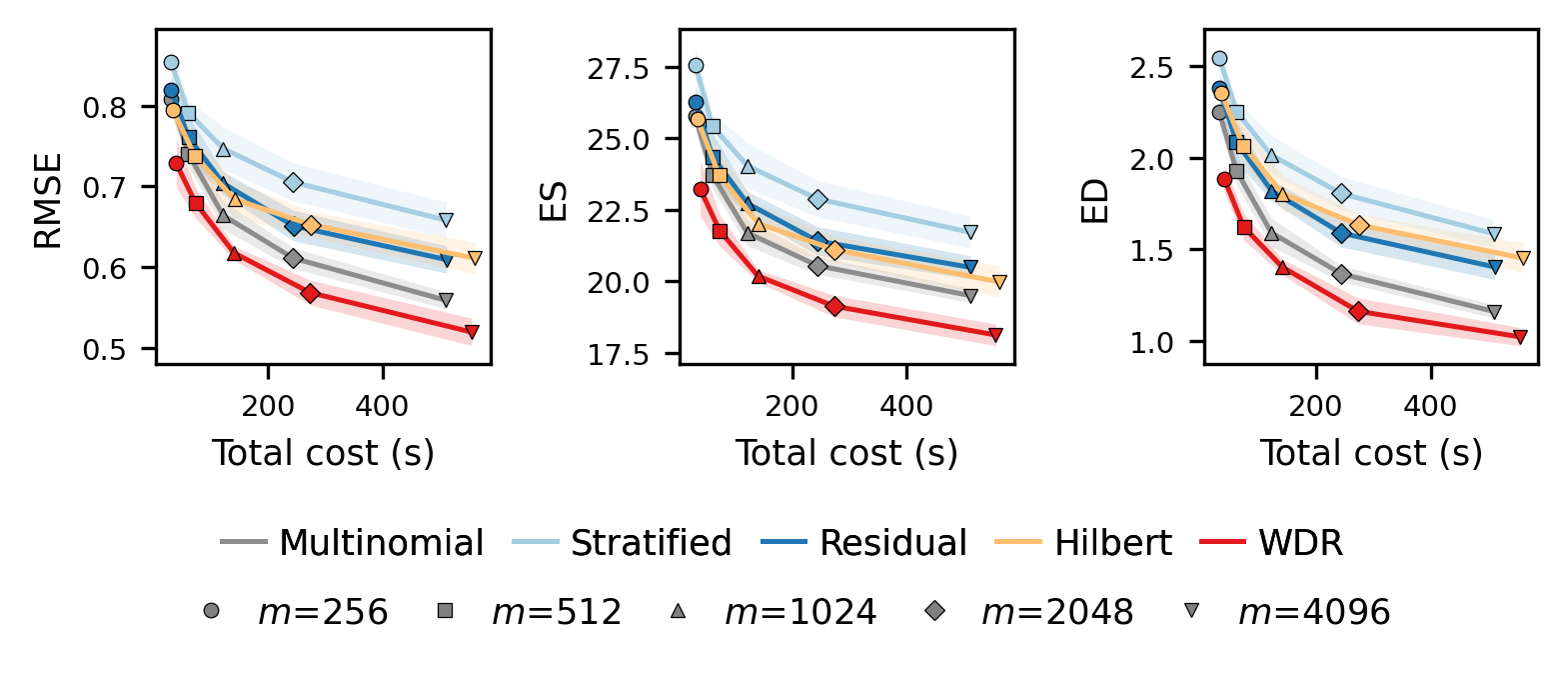}
    
    \caption{Time-averaged RMSE (left), energy score (middle), and energy distance (right) versus total wall-clock cost on the $d = 2025$ spatial model.
    Marker shape indicates the particle size $m$.
    Shaded bands represent standard deviations based on twenty replications.}
    \label{fig:simu-b3g-cost}
\end{figure}

Figure~\ref{fig:simu-b3g-cost} reports the evaluation metrics against total wall-clock cost, with the particle size $m$ represented by the marker.
WDR outperforms all baselines on all the three metrics at every $m$, and its advantage persists as $m$ increases.
The propagation step costs about $0.6$\,ms per particle per step, six times the threshold identified in Section~\ref{subsec:simu-cost}, and accounts for over $80\%$ of the per-step cost when WDR is used for resampling.
Among the classical schemes, multinomial resampling performs the best, reversing its usual ranking. 
We attribute this reversal to the i.i.d. output indices generated by multinomial resampling, which preserve the blockwise independence that the block particle filter is based on.

\section{Conditional image generation}\label{sec:application}

We next apply Wasserstein-Dirichlet resampling to conditional image generation with diffusion models, using the twisted diffusion sampler (TDS) of \citet{wu2023practical} as the baseline SMC framework.
Let $x^{(0:T)}=(x^{(0)},\ldots,x^{(T)})$ denote the latent diffusion trajectory (with $t$ decreasing in the reverse-time sampling procedure) and let $y$ be a conditioning variable, such as a class label.
A pretrained diffusion model specifies a joint density $p_\theta(x^{(0:T)})$ over trajectories, and the task is to draw samples from the conditional distribution $\pi_0(x^{(0)}\mid y)$, which is the marginal of $\pi(x^{(0:T)}\mid y) \propto p_\theta(x^{(0:T)})\,p(y\mid x^{(0)})$,
where $p(y\mid x^{(0)})$ is an externally supplied likelihood (e.g., a classifier on clean images).

\textbf{Twisted diffusion sampler with WDR.}
TDS views $\pi(x^{(0:T)}\mid y)$ as the extended target of an SMC sampler and runs particles backward in diffusion time.
In particular, TDS augments an unconditional score network with a likelihood gradient term as the twisted proposal and corrects this proposal using importance weights, so that the resulting particle approximation converges to $\pi(x^{(0:T)}\mid y)$ as the number of particles grows. Algorithmic details of TDS are provided in Section~S4.1.
At each reverse timestep, TDS (i) propagates descendants via the twisted kernel, (ii) updates the particle weights by the associated incremental weight, and (iii) resamples particles according to these weights.
In the original TDS,  systematic sampling (i.e., stratified with a single shared random offset) is used at each resampling step.
As shown in Section~\ref{subsec:simu-lg}, such classical schemes are inefficient in high-dimensional state spaces, whereas Wasserstein-Dirichlet resampling better preserves geometric structure and diversity.
We therefore replace the classical resampling in TDS with (T-)WDR and incorporate multi-descendant propagation with $K=3$.

\subsection{Experimental setup}
We follow the experimental design of \citet{wu2023practical} and consider class-conditional image generation on MNIST and CIFAR-10 datasets, each consisting of ten classes. In both experiments, we use the same pretrained score-based diffusion models and ResNet classifiers as in the original TDS. Experiments were conducted on one NVIDIA H100 GPU (80~GB), using PyTorch 2.0.1 and CUDA 11.8.

For MNIST, the conditioning variable is the digit label, and the goal is to generate $28\times 28$ grayscale images. For CIFAR-10, the conditioning variable is the class label, and the goal is to generate $32\times 32$ RGB images. In all runs, we simulate the reverse diffusion trajectory for $T=100$ discretization steps and consider particle sizes $m\in\{2^4,2^5,2^6,2^7\}$. Following \citet{wu2023practical}, we trigger resampling adaptively when the effective sample size falls below $0.5m$. 
Considering that the squared Euclidean distances in high-dimensional pixel space are poorly aligned with perceptual similarity, we implement Wasserstein-Dirichlet resampling in a learned low-dimensional representation that better reflects image semantics. Implementation details are provided in Section~S4.2.

For each method, we run 100 independent replications per class, with each run using $m$ particles (i.e., $m$ generated images). We evaluate the generated images using four complementary metrics: 
(i) \textit{Fr\'echet inception distance (FID)}~\citep{heusel2017gans}, lower indicates better distributional fidelity; 
(ii) \textit{classification accuracy}~\citep{wu2023practical}, higher indicates better label-conditional correctness;
(iii) \textit{within-run LPIPS diversity}~\citep{zhang2018unreasonable}, higher indicates more within-class sample diversity; and
(iv) \textit{between-run over within-run LPIPS ratio}, smaller (i.e., closer to one) indicates that different random seeds explore the same correct posterior region rather than collapsing onto different modes.
Detailed definitions of these metrics are provided in Section~S4.3.

\subsection{Experimental results}
\begin{figure}[!t]
    \centering
    \includegraphics[width=\linewidth]{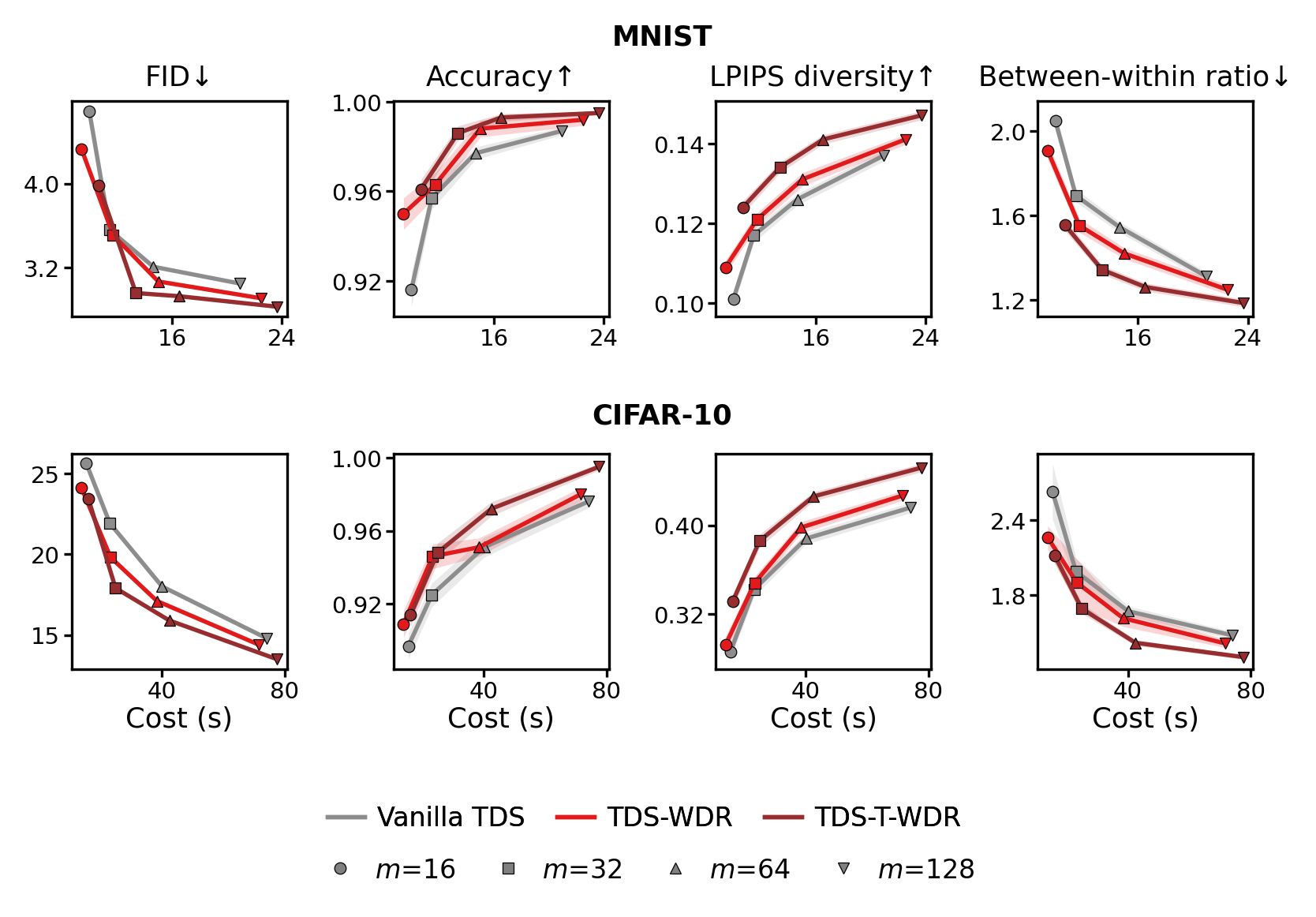}
    \caption{Conditional image generation metrics versus wall-clock time (in seconds). Lower is better for FID and the between-within ratio; higher is better for classification accuracy and within-run LPIPS diversity. Shaded bands represent standard errors based on one hundred replications.}
    \label{fig:tds-cost}
\end{figure}

Figure~\ref{fig:tds-cost} compares vanilla TDS with its (T-)WDR variants on MNIST and CIFAR-10. To evaluate practical benefit, we plot each performance metric against computational costs directly. As expected, all methods improve as the particle size $m$ increases, reflecting a more accurate approximation of the twisted path measure. Within this overall pattern, TDS-T-WDR performs the best, followed by TDS-WDR, and both consistently outperform vanilla TDS on both datasets.

This result is consistent with the geometric structure of the problem. Although they lie in a high-dimensional pixel space, natural images are concentrated near a much lower-dimensional manifold. Thus, in our conditional generation tasks, weight degeneracy is often mild to moderate when employing the twisted SMC. This is precisely the regime where T-WDR is most effective, leading to improved generation qualities.

In terms of computational cost, TDS-WDR introduces negligible additional cost compared to vanilla TDS, whereas TDS-T-WDR incurs a modest overhead, which is consistent with the wall-clock times reported in Section~S3.2.2 and can be attributed to more resampling triggers. Overall, (T-)WDR remains computationally competitive across the particle sizes considered. Averaged over $m$, the per-particle-per-step propagation cost is about $4$\,ms on MNIST and $8$\,ms on CIFAR-10, both far above the practical threshold identified in Section~\ref{subsec:simu-cost}. Hence, diffusion-based conditional generation fits the propagation-dominating regime, where the overhead of WDR is negligible, while its performance gains remain practically meaningful.

\begin{figure}[!t]
    \centering
    \includegraphics[width=\linewidth]{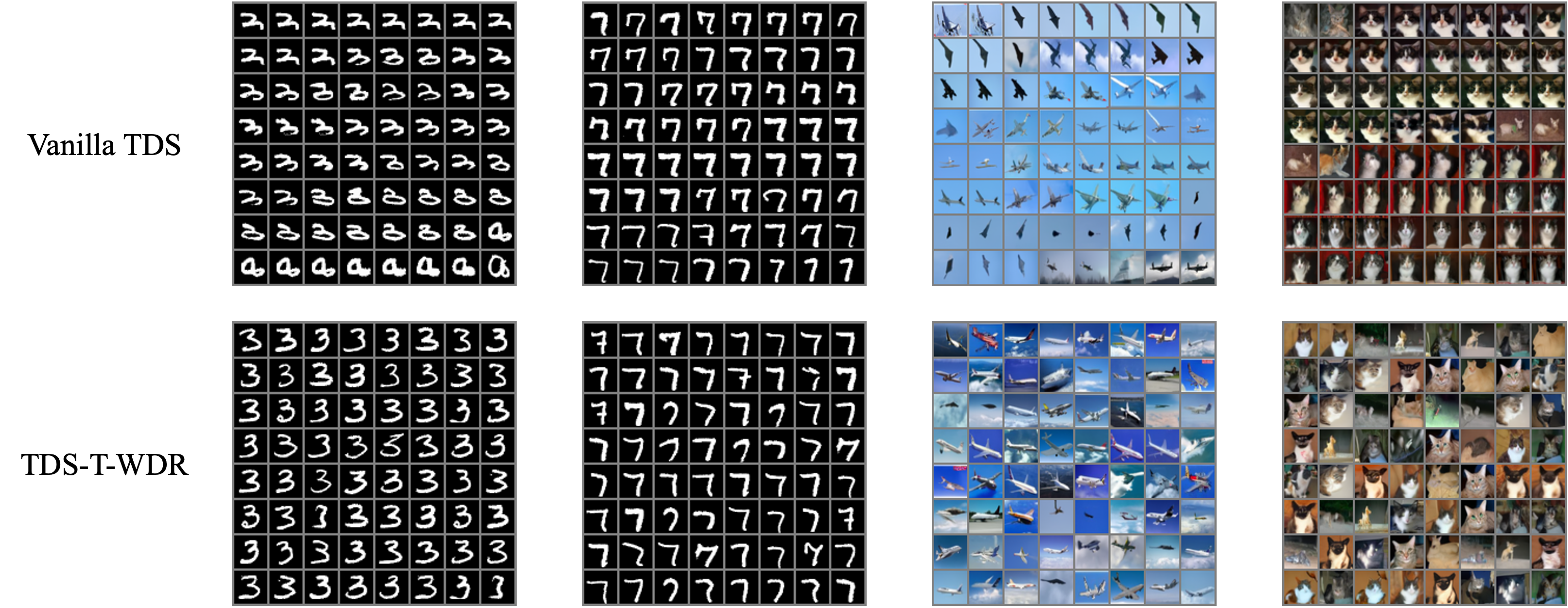}
    \caption{Samples generated by vanilla TDS (top row) versus those by the proposed TDS-T-WDR (bottom row) with $m=64$.
    The image batches are randomly selected.
    From left to right: digit 3 (MNIST), digit 7 (MNIST), airplane (CIFAR-10), and cat (CIFAR-10).}
    \label{fig:tds-samples}
\end{figure}

Figure~\ref{fig:tds-samples} provides examples of the generated images with $m=64$. Compared with vanilla TDS, our TDS-T-WDR produces samples that are both more class-consistent and more diverse within each class. For example, the MNIST digits generated by TDS-T-WDR show richer variation in stroke thickness, curvature, and writing style. In contrast, the vanilla TDS occasionally shows low-quality artifacts and exhibits apparent within-class repetition, indicating reduced effective diversity.
Taken together, these results demonstrate that incorporating Wasserstein-Dirichlet resampling into TDS improves fidelity, conditional correctness, and sample coverage without compromising practical efficiency.

\section{Concluding remarks}\label{sec:conclude}
We propose a new resampling paradigm, termed WDR, for sequential Monte Carlo from the perspective of free-support Wasserstein barycenters. Unlike conventional replication-only approaches, our scheme employs an optimal transport formulation that jointly optimizes the coupling and support locations, yielding an equal-weight particle system that best approximates the weighted empirical distribution in the Wasserstein sense.  A Dirichlet mixing mechanism injects controlled stochasticity, interpolating between a deterministic transform and OT-informed ancestor sampling. 
Additionally, when weight degeneracy is mild, a truncated version of the method offers further computational efficiency.
Algorithmically, we develop a block coordinate descent solver that exploits the problem's biconvex structure with provable convergence guarantees. Theoretically, we derive a non-asymptotic MSE bound that makes the bias-variance trade-off explicit and reveals that the convergence rate depends only on the intrinsic dimension of the target distribution.

Empirically, WDR consistently improves filtering accuracy and run-to-run stability, particularly in higher-dimensional and anisotropic settings. At per-particle propagation costs above roughly $0.1$\,ms, the method occupies the Pareto frontier, trading off accuracy and runtime optimally. 
Moreover, WDR integrates seamlessly into modern SMC-based diffusion samplers, boosting conditional image generation across fidelity, coverage, and diversity metrics—all while remaining computationally competitive.

Several directions are worthy of further investigation. First, it would be valuable to propagate our single-step theoretical guarantee through the full SMC recursion to quantify the overall filtering error. Second, an adaptive selection strategy for the Dirichlet concentration parameter $\tau$ would be useful in practice. Third, it is of interest to generalize the truncated WDR to the severely degenerate high-dimensional scenarios. Finally, the proposed resampling paradigm can extend to broader particle filters, such as the SMC sampler~\citep{dai2022invitation} and particle MCMC~\citep{andrieu2009particle}, where geometry-aware rejuvenation may help reduce path degeneracy and improve mixing.

\textbf{Supplementary materials}\quad
The supplementary materials include: a PDF containing proofs of the theoretical results, additional details, and further numerical results; and a ZIP implementing the proposed methods and reproducing the experiments. The reproducibility materials are also available at \url{https://github.com/Mengyu8042/wdr-repro}.

\textbf{Disclosure statement}\quad
The authors declare no competing interests.

\textbf{Data availability statement}\quad
The datasets used in this study are publicly available at \url{http://yann.lecun.com/exdb/mnist/} and \url{https://www.cs.toronto.edu/~kriz/cifar.html}.

\textbf{Declaration of generative AI use}\quad
The authors used ChatGPT (OpenAI, GPT-5 and GPT-6 series) for language polishing. They reviewed and edited the output and take full responsibility for the manuscript.

\if1\anon
{
\textbf{Acknowledgments}\quad
The authors thank the Editor, the Associate Editor, and three anonymous referees for their constructive comments that helped improve the work.
This work was supported by the National Natural Science Foundation of China (Grant No. 12601554), the China Postdoctoral Science Foundation (Grant No. 2025M783111), and the Shenzhen Science and Technology Program (Grant No. AI2026019).
} \fi

\begingroup
\spacingset{1.0}
\setlength{\bibsep}{4pt}
\bibliography{refs}
\endgroup

\end{document}


\maketitle

This supplementary material is structured as follows.
Section~\ref{sec:twdr} provides additional details on truncated Wasserstein-Dirichlet resampling.
Section~\ref{sec:proof} provides technical derivations of the theoretical results stated within the manuscript.
Section~\ref{sec:additional-results} includes additional experimental details and results for simulation studies.
Section~\ref{sec:tds-supp} provides additional details for the twisted diffusion sampler (TDS) application.

\section{Truncated Wasserstein-Dirichlet resampling (T-WDR)}\label{sec:twdr}

Recall that the resampling step takes a weighted empirical measure $\mu_n = \sum_{i=1}^n W_i\delta_{X_i}$ as input, with $W_i\ge 0$ and $\sum_{i=1}^n W_i=1$.
WDR solves the free-support Wasserstein barycenter problem
$$
\nu_m^\star = \arg\min_{\nu\in\cA_m} W_2^2(\mu_n,\nu), \qquad\cA_m = \Bigl\{\nu=\frac1m\sum_{j=1}^m \delta_{Y_j}:\ Y_j\in\RR^d\Bigr\}.
$$

The truncated variant, T-WDR, introduces a weight threshold $\kappa>0$ to decompose $\mu_n$ into an anchor component containing high-weight particles and a residual tail component. Specifically, define
$$
S_\kappa := \{i\in[n] : W_i\ge \kappa\},
\quad
r_\kappa := \sum\nolimits_{i\notin S_\kappa} W_i,
\quad
m_{\mathrm{tail}} := m-|S_\kappa|.
$$
The key idea of T-WDR is to preserve the anchor particles $\{X_i : i\in S_\kappa\}$ with their original weights, while reallocating only the residual mass $r_\kappa$ among $m_{\mathrm{tail}}$ equal-weight atoms. This yields a restricted feasible class
$$
\cA_m^{\mathrm{T\text{-}WDR}}(\kappa)
:=
\Bigl\{
\sum_{i\in S_\kappa}W_i\delta_{X_i}
+
\frac{r_\kappa}{m_{\mathrm{tail}}}
\sum_{j=1}^{m_{\mathrm{tail}}}\delta_{Y_j}
:\ Y_j\in\RR^d
\Bigr\}.
$$
Accordingly, T-WDR solves
\begin{equation}\label{eq:supp-twdr-ot}
\nu_{m,\kappa}^\star
=
\arg\min_{\nu\in\cA_m^{\mathrm{T\text{-}WDR}}(\kappa)} W_2^2(\mu_n,\nu).
\end{equation}
When $\kappa>\max_i W_i$, we have $S_\kappa=\varnothing$, $r_\kappa=1$, and $m_{\mathrm{tail}}=m$, so $\cA_m^{\mathrm{T\text{-}WDR}}(\kappa)=\cA_m$, recovering the original WDR formulation as a special case.

In practice, we seek an output measure with ``continuous'' mass distribution. This leads to the constraint $\kappa = r_\kappa/m_{\mathrm{tail}}$, which can be rewritten as
\begin{equation}\label{eq:supp-kappa-equation}
    m = \sum_{j=1}^n \min(W_j/\kappa, 1).
\end{equation}
Note that this threshold criterion is consistent with the optimality condition derived in~\cite{fearnhead2003line}, though we reach it from a distinct perspective.

For any $1\le m<n$, equation~\eqref{eq:supp-kappa-equation} admits at least one solution. Without ties among the weights, the solution is unique. Numerically, \eqref{eq:supp-kappa-equation} can be solved efficiently via bisection search over $\kappa$.

\textbf{Algorithmic implementation.}
Algorithm~\ref{alg:supp-twdr} provides complete procedures of T-WDR. It mirrors the standard WDR algorithm but optimizes only on the tail component, thereby preserving the anchor particles exactly while reducing computational cost.

\begin{algorithm}[!t]
  \caption{Truncated Wasserstein-Dirichlet resampling (T-WDR)}\label{alg:supp-twdr}
  \begin{algorithmic}[1]
    \State \textbf{Input:} weighted particles $\{(X_i,W_i)\}_{i=1}^n$, target size $m$, concentration $\tau\ge 0$
    \State Choose threshold $\kappa$ by solving \eqref{eq:supp-kappa-equation}
    \State Set $S_\kappa=\{i:W_i\ge\kappa\}$, $r_\kappa=\sum_{i\notin S_\kappa}W_i$, $m_{\mathrm{tail}}=m-|S_\kappa|$
    \State Define the normalized tail measure
    $$
      \mu_{\mathrm{tail}}
      :=
      \sum_{i\notin S_\kappa} \bar W_i\delta_{X_i},
      \qquad
      \bar W_i := W_i/r_\kappa .
    $$
    \State Compute $(\Pi^\star_{\mathrm{tail}},Y^\star_{\mathrm{tail}})$ by solving the free-support OT problem
    $$
      \arg\min_{\nu\in\cA_{m_{\mathrm{tail}}}} W_2^2(\mu_{\mathrm{tail}},\nu),
    $$
    via the same BCD solver as in the main WDR algorithm
    \For{$j=1,\dots,m_{\mathrm{tail}}$}
      \State Set $p^{(j)}\gets m_{\mathrm{tail}}\Pi^\star_{\mathrm{tail},\cdot j}$
      \State Draw $V^{(j)} \sim \mathrm{Dirichlet}(\tau p^{(j)})$ and set $\tilde Y_j \gets \sum_{i\notin S_\kappa} V_i^{(j)} X_i$
    \EndFor
    \State \textbf{Output:} the mixed measure
    $$
      \nu^{\mathrm{T\text{-}WDR}}
      :=
      \sum_{i\in S_\kappa} W_i\delta_{X_i}
      + \frac{r_\kappa}{m_{\mathrm{tail}}}\sum_{j=1}^{m_{\mathrm{tail}}}\delta_{\tilde Y_j}
    $$
  \end{algorithmic}
\end{algorithm}

Empirically, T-WDR is more effective under mild-to-moderate degeneracy, where $r_\kappa$ is non-negligible, so that the tail resampling step acts on a substantial residual mass. In severely degenerate cases (i.e., $\sum_{i\in S_\kappa}W_i\approx 1$), the tail mass $r_\kappa$ becomes very small and T-WDR reduces to no resampling, in which case full WDR is generally preferable.

\section{Technical details}\label{sec:proof}

\subsection{Proof of Proposition~1}

\begin{proof}
By the Kantorovich--Rubinstein duality, we have $|\nu(\varphi)-\mu(\varphi)| \le L W_1(\nu,\mu)$.
For any $p\ge 1$, Jensen's inequality implies $W_1(\nu,\mu) \le W_p(\nu,\mu)$.
Taking the conditional expectation with respect to the randomness of $\nu$ yields the desired result.
\end{proof}

\subsection{Proof of Lemma~1}

The proof follows the dyadic partitioning argument of \citet[Section~2.2]{chewi2025statistical}, using the idea of coupling $\mu$ and $\nu_m$ by recursively constructing candidate couplings at multiple scales.

We first partition the active block $Q=\prod_{k=1}^s[a_k,b_k]\subset\RR^s$ into smaller axis-aligned cells.
We then allocate the $m$ equal-weight support points so that every cell receives almost the correct mass $m\mu(U)$, and choose one representative point in each finest-level cell.
Finally, a multiscale transport bound says that the mass discrepancy visible at level $\ell+1$ needs to be moved only inside cells of level $\ell$, whose diameter is $\Delta_\ell$.
Summing these scale-by-scale costs yields the rate $m^{-1/s}$.

\begin{figure}[ht]
\centering
\begin{tikzpicture}[x=0.9cm,y=0.9cm,every node/.style={text=black,font=\footnotesize},line cap=round,line join=round]
  \draw[black,thick] (0,0) rectangle (2,2);
  \node at (1,2.4) {Level $0$: $\mathcal P_0$};
  \node at (1,1) {$Q$};

  \node[align=center] at (3.5,1.5) {split along\\coordinate $1$};
  \draw[black,->,thick] (2.2,0.8) -- (4.8,0.8);

  \draw[black,thick] (5,0) rectangle (7,2);
  \draw[black,thick] (6,0) -- (6,2);
  \node at (6,2.4) {Level $1$: $\mathcal P_1$};

  \node[align=center] at (8.5,1.5) {split along\\coordinate $2$};
  \draw[black,->,thick] (7.2,0.8) -- (9.8,0.8);

  \draw[black,thick] (10,0) rectangle (12,2);
  \draw[black,thick] (11,0) -- (11,2);
  \draw[black,thick] (10,1) -- (11,1);
  \draw[black,thick] (11,1) -- (12,1);
  \filldraw[draw=black,fill=red!10,thick] (10,1) rectangle (11,2);
  \node at (11,2.4) {Level $2$: $\mathcal P_2$};
  \node at (10.5,1.5) {\textcolor{red}{$U$}};

  \draw[blue,dashed,thick] (10.05,1.95) -- (10.95,1.05);
  \node[anchor=west,text=blue] at (10.6,1.55) {$\operatorname{diam}(U)$};
  \draw[blue,<->,thick] (10,-0.35) -- (11,-0.35);
  \node[text=blue] at (10.5,-0.7) {$\operatorname{side}_1(U)$};
  \draw[blue,<->,thick] (12.4,1) -- (12.4,2);
  \node[rotate=-90,text=blue] at (12.7,1.5) {$\operatorname{side}_2(U)$};

  \node[align=center] at (14.15,1.5) {split along\\coordinate $1$};
  \draw[black,->,thick] (13,0.8) -- (15.3,0.8);

  \draw[black,thick] (15.5,0) rectangle (17.5,2);
  \fill[red!10] (15.5,1) rectangle (16.5,2);
  \draw[black,thick] (16.5,0) -- (16.5,2);
  \draw[black,thick] (15.5,1) -- (17.5,1);
  \draw[black,thick] (16,0) -- (16,2);
  \draw[black,thick] (17,0) -- (17,2);
  \node at (16.5,2.4) {Level $3$: $\mathcal P_3$};
  \node at (15.75,1.5) {\textcolor{red}{$U^-$}};
  \node at (16.25,1.5) {\textcolor{red}{$U^+$}};

\end{tikzpicture}
\caption{Schematic of the cyclic axis-aligned partition used in Lemma~1, illustrated for $s=2$. At level $\ell$, each cell is bisected along coordinate $\mathrm{axis}(\ell)=1+(\ell\bmod s)$, so the splitting axis cycles $1,2,1,2,\ldots$ between the two active coordinates. The shaded rectangle in $\mathcal P_2$ is a typical small cell $U$, included to indicate its side lengths and diameter; the shaded region in $\mathcal P_3$ shows how this $U$ is further bisected along coordinate $1$ into its two level-$3$ subcells $U^-$ and $U^+$, illustrating the recursion $N(U^-)+N(U^+)=N(U)$ used in Lemma~\ref{lem:integer-allocation-supp}.}
\label{fig:kdtree-supp}
\end{figure}
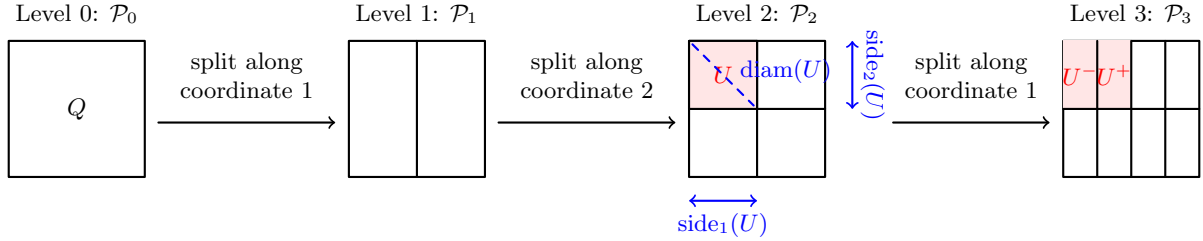

\begin{proof}
Since the support of $\mu$ is contained in $Q_s\times\{0\}^{d-s}$ and the Euclidean distance on this set depends only on the first $s$ coordinates, it suffices to work in $\RR^s$. Accordingly, we identify $Q_s$ with the active block
$$
Q := \prod_{k=1}^s [a_k,b_k] \subset \RR^s
$$
and identify $\mu$ with its pushforward to $\RR^s$.
Throughout, $C_{s,p}$ denotes a positive constant depending only on $(s,p)$, whose value may change from line to line.

Set $t_0 := \lfloor \log_2 m\rfloor$.\footnote{This choice makes $2^{t_0}\asymp m$, so the finest-level cells are meant to carry mass of order $1/m$.}
We construct nested partitions $\mathcal P_0,\ldots,\mathcal P_{t_0}$ of $Q$ by starting from $\mathcal P_0:=\{Q\}$ and, for each $\ell=0,\dots,t_0-1$, bisecting every cell in $\mathcal P_\ell$ at the midpoint of its $\mathrm{axis}(\ell)$th coordinate, where
$$
  \mathrm{axis}(\ell) := 1 + (\ell \bmod s).
$$
This is the usual balanced cyclic $k$-d partition; see Fig.~\ref{fig:kdtree-supp}. Here and below, the $k$th coordinate means the $k$th coordinate axis of the reduced space $\RR^s$, equivalently the $k$th active dimension.
Since every cell $U\in\mathcal P_\ell$ is an axis-aligned hyperrectangle, we write
$$
  U = \prod_{k=1}^s [\alpha_k(U),\beta_k(U)].
$$
We then define the side length of $U$ along the $k$th active coordinate by
$$
  \operatorname{side}_k(U) := \beta_k(U)-\alpha_k(U),
  \qquad k=1,\dots,s,
$$
and define its Euclidean diameter by
$$
  \operatorname{diam}(U)
  :=
  \Bigl(\sum_{k=1}^s \operatorname{side}_k(U)^2\Bigr)^{1/2}.
$$
Since each active coordinate has been split at least $\lfloor \ell/s\rfloor$ times by level $\ell$,
$$
  \operatorname{side}_k(U)
  \le
  (b_k-a_k)\,2^{-\lfloor \ell/s\rfloor},
  \qquad k=1,\dots,s,
$$
and therefore
$$
  \Delta_\ell
  :=
  \max_{U\in\mathcal P_\ell}\operatorname{diam}(U)
  \le
  D\,2^{-\lfloor \ell/s\rfloor},
  \qquad \ell=0,\dots,t_0.
$$
Because $t_0=\lfloor \log_2 m\rfloor \ge \log_2 m-1$, we also have
$$
  2^{\lfloor t_0/s\rfloor}
  \ge
  2^{(\log_2 m - 1)/s - 1}
  =
  2^{-1} 2^{-1/s} m^{1/s}
  \ge \tfrac14 m^{1/s},
$$
so
$$
  \Delta_{t_0} \le 4D\,m^{-1/s}.
$$

By Lemma~\ref{lem:integer-allocation-supp} below, there exist integers $N(U)\in\{0,1,\dots,m\}$ for all cells $U$ in the nested partition such that $N(Q)=m$, each level-$\ell$ cell count equals the sum of its two level-$(\ell+1)$ subcell counts, and
$$
  |N(U)-m\mu(U)| \le 1
  \qquad\text{for every cell } U\in\mathcal P_\ell,\ \ell=0,1,\dots,t_0.
$$
For each finest-level cell $L\in\mathcal P_{t_0}$, set $a(L):=N(L)$, choose a representative point $x_L\in L$, and define
$$
  \nu_m
  :=
  \frac1m \sum_{L\in\mathcal P_{t_0}} a(L)\delta_{x_L}.
$$
Then $\sum_{L\in\mathcal P_{t_0}} a(L)=m$, and at every level $\ell=0,1,\dots,t_0$, for every cell $U\in\mathcal P_\ell$,
$$
  \nu_m(U)
  =
  \frac1m \sum_{L\subset U} a(L)
  =
  \frac{N(U)}{m},
  \qquad
  |\nu_m(U)-\mu(U)| \le \frac1m.
$$
In words, $\nu_m(U)$ is simply the total allocated count of the finest-level cells contained in $U$, divided by $m$.

The key transport observation is that, if $\mu$ and $\nu_m$ already place the same total mass inside a level-$(\ell+1)$ cell, then that matched mass can be coupled within that finer cell and therefore does not contribute to the transport cost at level $\ell$. Only the remaining level-$(\ell+1)$ discrepancy needs to be transported across a level-$\ell$ cell of diameter at most $\Delta_\ell$. Lemma~\ref{lem:multiscale-bound-supp} therefore gives
$$
  W_p^p(\mu,\nu_m)
  \le
  \sum_{\ell=0}^{t_0-1}
    \Delta_\ell^p
    \sum_{U\in\mathcal P_{\ell+1}} |\mu(U)-\nu_m(U)|
  + \Delta_{t_0}^p.
$$
Since $|\mathcal P_{\ell+1}|=2^{\ell+1}$ and the per-cell discrepancy satisfies $|\mu(U)-\nu_m(U)|\le 1/m$ for every $U\in\mathcal P_{\ell+1}$, we have $\sum_{U\in\mathcal P_{\ell+1}}|\mu(U)-\nu_m(U)|\le 2^{\ell+1}/m$, and combining with $\Delta_\ell\le D\,2^{-\lfloor\ell/s\rfloor}$ and $\Delta_{t_0}\le 4D\,m^{-1/s}$ gives

$$
  W_p^p(\mu,\nu_m)
  \le
  \frac{D^p}{m}
  \sum_{\ell=0}^{t_0-1} 2^{\ell+1-p\lfloor \ell/s\rfloor}
  + (4D)^p m^{-p/s}.
$$
Using $\lfloor \ell/s\rfloor \ge \ell/s - 1$,
$$
  2^{\ell+1-p\lfloor \ell/s\rfloor}
  \le
  2^{p+1}\,2^{\ell(1-p/s)}.
$$
Because $p<s$, the ratio $2^{1-p/s}$ is greater than $1$, so the resulting geometric series is dominated by its last term:
$$
  \sum_{\ell=0}^{t_0-1} 2^{\ell(1-p/s)}
  \le
  C_{s,p}\,2^{(t_0-1)(1-p/s)}
  \le
  C_{s,p}\,m^{1-p/s}.
$$
Consequently,
$$
  W_p^p(\mu,\nu_m) \le C_{s,p}\,D^p\,m^{-p/s},
$$
and taking the $p$th root yields
$$
  W_p(\mu,\nu_m) \le C_{s,p}\,D\,m^{-1/s}.
$$
This proves Lemma~1.
\end{proof}

\medskip
The two technical ingredients used above are provided next.

\begin{lemma}[Integer allocation on a nested binary partition]
\label{lem:integer-allocation-supp}
Let $\mathcal P_0,\dots,\mathcal P_{t_0}$ be nested partitions of $Q$ such that $\mathcal P_0=\{Q\}$ and each cell $U\in\mathcal P_\ell$ is the disjoint union of two level-$(\ell+1)$ subcells, denoted $U^-$ and $U^+$. Let $\mu$ be a probability measure on $Q$, and let $m\in\NN$. Then one can assign integers $N(U)\in\{0,1,\dots,m\}$ to all cells in the nested partition so that:
\begin{itemize}
\item[(i)] $N(Q)=m$;
\item[(ii)] $N(U^-)+N(U^+)=N(U)$ for every cell $U$ that is split;
\item[(iii)] $|N(U)-m\mu(U)|\le 1$ for every cell $U$.
\end{itemize}
Consequently, if $a(L):=N(L)$ for each finest-level cell $L\in\mathcal P_{t_0}$, then
$$
  \sum_{L\in\mathcal P_{t_0}} a(L)=m,
  \qquad
  \Bigl|\sum_{L\subset U} a(L)-m\mu(U)\Bigr|\le 1
  \quad\text{for every cell } U.
$$
\end{lemma}

\begin{proof}
We argue recursively over the levels of the partition. Start from $N(Q):=m$, which already satisfies $|N(Q)-m\mu(Q)|=0$.

Assume that counts have been assigned to every cell of $\mathcal P_\ell$ and that they satisfy $|N(U)-m\mu(U)|\le 1$ for each $U\in\mathcal P_\ell$. Fix one such cell $U$, and let $U^-,U^+\in\mathcal P_{\ell+1}$ be its two subcells. Consider the interval
$$
  I_U
  :=
  [m\mu(U^-)-1,\; m\mu(U^-)+1]
  \cap
  [N(U)-m\mu(U^+)-1,\; N(U)-m\mu(U^+)+1].
$$
The centers of these two intervals differ by
$$
  N(U)-m\mu(U^-)-m\mu(U^+)
  =
  N(U)-m\mu(U),
$$
which lies in $[-1,1]$ by the induction hypothesis. Hence $I_U$ has length at least $1$.

Choose an integer $N(U^-)\in I_U\cap[0,N(U)]$. Such an integer always exists: if $I_U\subset[0,N(U)]$, its length is at least $1$, so it contains an integer; if $I_U$ extends to the left of $0$, then it contains $0$; if it extends to the right of $N(U)$, then it contains $N(U)$. Now set
$$
  N(U^+) := N(U)-N(U^-).
$$
Then $N(U^-),N(U^+)\in\{0,1,\dots,N(U)\}$, they add up to $N(U)$, and the definition of $I_U$ gives
$$
  |N(U^-)-m\mu(U^-)|\le 1,
  \qquad
  |N(U^+)-m\mu(U^+)|\le 1.
$$
Applying this construction independently to every cell of $\mathcal P_\ell$ produces counts on $\mathcal P_{\ell+1}$ with the same property. Iterating to level $t_0$ proves (i)--(iii).

For the final claim, let $L$ range over the finest-level cells. Repeatedly using (ii) shows that for every cell $U$,
$$
  N(U)=\sum_{L\subset U} a(L).
$$
Combining this identity with (iii) yields the displayed bound.
\end{proof}

The next lemma formalizes the scale-by-scale transport estimate. Intuitively, mass that already agrees inside a finer cell is postponed to the next level, and only the remaining cellwise discrepancy is charged at the current diameter.

\begin{lemma}[Multiscale transport bound on a nested partition]
\label{lem:multiscale-bound-supp}
Let $Q\subset\RR^s$, and let $\mathcal P_0,\dots,\mathcal P_{t_0}$ be nested finite partitions with $\mathcal P_0=\{Q\}$ such that each cell in $\mathcal P_\ell$ is a disjoint union of cells in $\mathcal P_{\ell+1}$. Define
$$
  \Delta_\ell := \max_{U\in\mathcal P_\ell}\operatorname{diam}(U),
  \qquad \ell=0,\dots,t_0.
$$
Then for any probability measures $\mu$ and $\nu$ supported on $Q$,
$$
  W_p^p(\mu,\nu)
  \le
  \sum_{\ell=0}^{t_0-1}
    \Delta_\ell^p
    \sum_{U\in\mathcal P_{\ell+1}} |\mu(U)-\nu(U)|
  + \Delta_{t_0}^p.
$$
\end{lemma}

\begin{proof}
This is the positive measure decomposition used in the proof of \citet[Section~2.2, Theorem~2.2]{chewi2025statistical}, specialized to the present nested partition. We show the result because it is exactly what Lemma~1 needs.

There exist nonnegative measures $\mu_0,\dots,\mu_{t_0}$ and $\nu_0,\dots,\nu_{t_0}$ such that:
\begin{itemize}
\item[(i)] $\sum_{\ell=0}^{t_0}\mu_\ell = \mu$ and $\sum_{\ell=0}^{t_0}\nu_\ell = \nu$;
\item[(ii)] for every level $\ell$ and every cell $U\in\mathcal P_\ell$, one has $\mu_\ell(U)=\nu_\ell(U)$;
\item[(iii)] for $\ell<t_0$,
$$
  \mu_\ell(Q)=\nu_\ell(Q)
  \le
  \sum_{U\in\mathcal P_{\ell+1}} |\mu(U)-\nu(U)|,
$$
and at the final level $\mu_{t_0}(Q)=\nu_{t_0}(Q)\le 1$.
\end{itemize}
Property (iii) is the scale-by-scale content of the construction: the mass paid at level $\ell$ is controlled by the mismatch seen one level finer.

By (ii), for each level $\ell$ and each cell $U\in\mathcal P_\ell$, the restricted measures $\mu_\ell|_U$ and $\nu_\ell|_U$, defined by $(\mu_\ell|_U)(A):=\mu_\ell(A\cap U)$ and $(\nu_\ell|_U)(A):=\nu_\ell(A\cap U)$ for any measurable $A$, have the same total mass. Hence they can be coupled inside $U$, i.e., there exists $\gamma_\ell^U\in\Pi(\mu_\ell|_U,\nu_\ell|_U)$ supported on $U\times U$. Gluing these cell-wise couplings $\gamma_\ell:=\sum_{U\in\mathcal P_\ell}\gamma_\ell^U$ gives a coupling $\gamma_\ell\in\Pi(\mu_\ell,\nu_\ell)$ such that
$$
  \operatorname{supp}(\gamma_\ell)
  \subset
  \bigcup_{U\in\mathcal P_\ell}(U\times U).
$$
Therefore, every $(x,y)\in\operatorname{supp}(\gamma_\ell)$ satisfies $\|x-y\|\le \Delta_\ell$, and so
$$
  \int \|x-y\|^p\,\gamma_\ell(\dd x,\dd y)
  \le
  \Delta_\ell^p\,\mu_\ell(Q).
$$

Now define a nonnegative measure on $Q\times Q$ by summing the level-wise couplings,
$$
  \gamma := \sum_{\ell=0}^{t_0}\gamma_\ell.
$$
Its first marginal is $\sum_\ell\mu_\ell=\mu$ and its second marginal is $\sum_\ell\nu_\ell=\nu$, both by (i); hence $\gamma\in\Pi(\mu,\nu)$ is a valid coupling of $\mu$ and $\nu$. By the definition of the Wasserstein distance,
$$
  W_p^p(\mu,\nu)
  \le
  \int \|x-y\|^p\,\gamma(\dd x,\dd y)
  \le
  \sum_{\ell=0}^{t_0} \Delta_\ell^p\,\mu_\ell(Q).
$$
Applying (iii) to the first $t_0$ terms and using $\mu_{t_0}(Q)\le 1$ for the last term gives
$$
  W_p^p(\mu,\nu)
  \le
  \sum_{\ell=0}^{t_0-1}
    \Delta_\ell^p
    \sum_{U\in\mathcal P_{\ell+1}} |\mu(U)-\nu(U)|
  + \Delta_{t_0}^p,
$$
which is the desired bound.
\end{proof}

\subsection{Proof of Theorem~1}

\begin{proof}
Recall from the free-support formulation that an optimal plan $\Pi^\star \in \cU(W,m^{-1}1_m)$ and support points $Y^\star = (Y_1^\star,\ldots,Y_m^\star)$ solve
$$
  W_2^2(\mu,\nu_m^\star)
  =
  \min_{\Pi\in\cU(W,m^{-1}1_m)}
  \sum_{i=1}^n\sum_{j=1}^m \Pi_{ij}\|X_i - Y_j\|^2,
  \quad 
  \nu_m^\star := \frac1m\sum_{j=1}^m \delta_{Y_j^\star}.
$$
For each column $j=1,\dots,m$, we set $p^{(j)}_i := m\Pi^\star_{ij},i=1,\dots,n,$ so $p^{(j)}=(p^{(j)}_1,\dots,p^{(j)}_n)$ is a probability vector and $Y_j^\star = \sum_{i=1}^n p^{(j)}_i X_i.$
Define the local covariance matrices
$
  \Sigma_j
  := \sum_{i=1}^n p^{(j)}_i (X_i - Y_j^\star)(X_i - Y_j^\star)^\top,
   j=1,\dots,m,
$
and the averaged local variance
$
  A_m
  := m^{-1}\sum_{j=1}^m \tr(\Sigma_j).
$
A direct computation shows
\begin{equation}\label{eq:Am-is-W2}
  A_m
  = \frac{1}{m}\sum_{j=1}^m \sum_{i=1}^n p^{(j)}_i \|X_i - Y_j^\star\|^2
  = \sum_{i=1}^n \sum_{j=1}^m \Pi^\star_{ij} \|X_i - Y_j^\star\|^2
  = W_2^2(\mu,\nu_m^\star).
\end{equation}

By definition,
$$
  \nu_m^{\mathrm{WDR}}(\varphi)
  = \frac{1}{m}\sum_{j=1}^m \varphi\bigl(Z_j(\tau)\bigr),
  \quad
  Z_j(\tau)
  = \sum_{i=1}^n V_i^{(j)} X_i,
  \quad
  V^{(j)} \sim \mathrm{Dirichlet}\bigl(\tau p^{(j)}\bigr),
$$
independently over $j$, conditional on $\mu_n$.
We decompose its conditional mean squared error into squared bias plus variance:
$$
  \EE\Bigl[
    \bigl(
      \nu_m^{\mathrm{WDR}}(\varphi) - \mu_n(\varphi)
    \bigr)^2
    \mid \mu_n
  \Bigr]
  =
  \Bigl(
    \EE[\nu_m^{\mathrm{WDR}}(\varphi)\mid \mu_n] - \mu_n(\varphi)
  \Bigr)^2
  +
  \Var\bigl(\nu_m^{\mathrm{WDR}}(\varphi)\mid \mu_n\bigr).
$$

\textbf{Step 1: Variance term.}
Conditional on $\mu_n$, the random variables $\{Z_j(\tau)\}_{j=1}^m$ are independent across $j$. Thus
$$
  \Var\bigl(\nu_m^{\mathrm{WDR}}(\varphi)\mid \mu_n\bigr)
  = \frac{1}{m^2}\sum_{j=1}^m
      \Var\bigl(\varphi(Z_j(\tau))\mid \mu_n\bigr).
$$
Since $\varphi$ is $L$-Lipschitz and $Y_j^\star$ is deterministic given $\mu_n$,
$$
  \Var\bigl(\varphi(Z_j(\tau))\mid \mu_n\bigr)
  \le
  \EE\bigl[(\varphi(Z_j(\tau))-\varphi(Y_j^\star))^2 \mid \mu_n\bigr]
  \le
  L^2\EE\bigl[\|Z_j(\tau)-Y_j^\star\|^2 \mid \mu_n\bigr].
$$
Note that
$$
  \EE[V^{(j)}] = p^{(j)}, 
  \quad
  \Cov\bigl(V^{(j)}\bigr)
  = \frac{1}{\tau+1}\bigl(\mathrm{diag}(p^{(j)}) - p^{(j)}p^{(j)\top}\bigr).
$$
It follows that
$$
  \EE[Z_j(\tau)\mid \mu_n] = Y_j^\star,
  \quad
  \Cov\bigl(Z_j(\tau)\mid \mu_n\bigr)
  = \frac{1}{\tau+1}\Sigma_j,
$$
and hence
$$
  \EE\bigl[\|Z_j(\tau)-Y_j^\star\|^2 \mid \mu_n\bigr]
  = \tr\bigl(\Cov\bigl(Z_j(\tau)\mid \mu_n\bigr)\bigr)
  = \frac{1}{\tau+1}\tr(\Sigma_j).
$$
Substituting into the variance bound gives
$$
  \Var\bigl(\nu_m^{\mathrm{WDR}}(\varphi)\mid \mu_n\bigr)
  \le
  \frac{L^2}{m^2}\sum_{j=1}^m
    \frac{1}{\tau+1}\tr(\Sigma_j)
  =
  \frac{L^2}{\tau+1}\frac{A_m}{m}.
$$

\textbf{Step 2: Bias term.}
We now bound
$$
  \bigl|
    \EE[\nu_m^{\mathrm{WDR}}(\varphi)\mid \mu_n] - \mu_n(\varphi)
  \bigr|.
$$
First rewrite $\mu_n(\varphi)$ in terms of the column probabilities.
Using $\sum_j \Pi^\star_{ij} = W_i$ and $p_i^{(j)} = m \Pi^\star_{ij}$,
$$
  \mu_n(\varphi)
  = \sum_{i=1}^n W_i \varphi(X_i)
  = \frac{1}{m}\sum_{j=1}^m \sum_{i=1}^n p_i^{(j)} \varphi(X_i).
$$
For each $j$, define the mixture measure
$$
  \mu_j
  := \sum_{i=1}^n p_i^{(j)}\delta_{X_i},
$$
and let $\nu_{j,\tau}$ denote the law of $Z_j(\tau)$ conditional on $\mu_n$.
Then
$$
  \EE[\nu_m^{\mathrm{WDR}}(\varphi)\mid \mu_n]
  = \frac{1}{m}\sum_{j=1}^m \int \varphi \,\dd \nu_{j,\tau},
  \quad
  \mu_n(\varphi)
  = \frac{1}{m}\sum_{j=1}^m \int \varphi \,\dd \mu_j.
$$
Hence
$$
  \Bigl|
    \EE[\nu_m^{\mathrm{WDR}}(\varphi)\mid \mu_n] - \mu_n(\varphi)
  \Bigr|
  \le
  \frac{1}{m}\sum_{j=1}^m
    \Bigl|
      \int \varphi \,\dd \nu_{j,\tau} - \int \varphi \,\dd\mu_j
    \Bigr|.
$$

By Kantorovich--Rubinstein duality, for each $j$,
$$
  \Bigl|
    \int \varphi \,\dd \nu_{j,\tau} - \int \varphi \,\dd\mu_j
  \Bigr|
  \le LW_1(\nu_{j,\tau},\mu_j).
$$
We now construct an explicit coupling to bound $W_1(\nu_{j,\tau},\mu_j)$.
Given $V^{(j)}\sim\mathrm{Dirichlet}(\tau p^{(j)})$, define
$$
  Z_j(\tau) = \sum_{i=1}^n V_i^{(j)} X_i,
  \quad
  I_j \mid V^{(j)} \sim \mathrm{Categorical}\bigl(V^{(j)}\bigr),
  \quad
  \widetilde X_j := X_{I_j}.
$$
Conditionally on $V^{(j)}$, we have
$
  \EE[\widetilde X_j\mid V^{(j)}] = Z_j(\tau)
$,
and marginally $\widetilde X_j$ has law $\mu_j$ (since $\EE[V^{(j)}]=p^{(j)}$).
Thus $(Z_j(\tau),\widetilde X_j)$ is a valid coupling between $\nu_{j,\tau}$ and $\mu_j$.
Under this coupling,
$$
  W_1(\nu_{j,\tau},\mu_j)
  \le
  \EE\bigl[\|\widetilde X_j - Z_j(\tau)\|\mid \mu_n\bigr]
  \le
  \Bigl(
    \EE\bigl[\|\widetilde X_j - Z_j(\tau)\|^2\mid \mu_n\bigr]
  \Bigr)^{1/2}.
$$

We now compute the second moment.
Conditionally on $V^{(j)}$, 
$$
  \EE\bigl[\|\widetilde X_j - Z_j(\tau)\|^2\mid V^{(j)},\mu_n\bigr]
  = \sum_{i=1}^n V_i^{(j)} \|X_i - Z_j(\tau)\|^2.
$$
Let $\Sigma_j(V^{(j)}) := \sum_i V_i^{(j)}(X_i - Z_j(\tau))(X_i - Z_j(\tau))^\top$ be the conditional covariance matrix of $\widetilde X_j$ given $V^{(j)}$.
Then
$$
  \EE\bigl[\|\widetilde X_j - Z_j(\tau)\|^2\mid V^{(j)},\mu_n\bigr]
  = \tr\bigl(\Sigma_j(V^{(j)})\bigr),
$$
and hence
$$
  \EE\bigl[\|\widetilde X_j - Z_j(\tau)\|^2\mid \mu_n\bigr]
  = \EE\bigl[\tr\bigl(\Sigma_j(V^{(j)})\bigr)\mid \mu_n\bigr].
$$

A direct calculation using the Dirichlet moments gives
$$
  \EE\bigl[\Sigma_j(V^{(j)})\mid \mu_n\bigr]
  = \frac{\tau}{\tau+1}\Sigma_j,
$$
and therefore
$$
  \EE\bigl[\|\widetilde X_j - Z_j(\tau)\|^2\mid \mu_n\bigr]
  = \frac{\tau}{\tau+1}\tr(\Sigma_j).
$$
Consequently,
$$
  W_1(\nu_{j,\tau},\mu_j)
  \le
  \sqrt{\frac{\tau}{\tau+1}\tr(\Sigma_j)}.
$$

Putting this into the bias bound and applying Cauchy--Schwarz,
$$
\begin{aligned}
  \Bigl|
    \EE[\nu_m^{\mathrm{WDR}}(\varphi)\mid \mu_n] - \mu_n(\varphi)
  \Bigr|
  &\le
  \frac{L}{m}\sum_{j=1}^m
    \sqrt{\frac{\tau}{\tau+1}\tr(\Sigma_j)}
  \\
  &\le
  L\sqrt{\frac{\tau}{\tau+1}}
  \Biggl(
    \frac{1}{m}\sum_{j=1}^m \tr(\Sigma_j)
  \Biggr)^{1/2}
  =
  L\sqrt{\frac{\tau}{\tau+1}A_m}.
\end{aligned}
$$
Thus the squared bias satisfies
$$
  \Bigl(
    \EE[\nu_m^{\mathrm{WDR}}(\varphi)\mid \mu_n] - \mu_n(\varphi)
  \Bigr)^2
  \le
  L^2\frac{\tau}{\tau+1}A_m.
$$

\textbf{Step 3: Combine.}
Combining the bias and variance bounds gives
$$
  \EE\Bigl[
    \bigl(
      \nu_m^{\mathrm{WDR}}(\varphi) - \mu_n(\varphi)
    \bigr)^2
    \Big| \mu_n
  \Bigr]
  \le
  L^2\frac{\tau}{\tau+1}A_m
  +
  \frac{L^2}{\tau+1}\frac{A_m}{m}.
$$
Using \eqref{eq:Am-is-W2} and Lemma~1, we have
$$
  A_m 
  = W_2^2(\mu,\nu_m^\star)
  \le C_s^2m^{-2/s},
$$
so that
$$
  \EE\Bigl[
    \bigl(
      \nu_m^{\mathrm{WDR}}(\varphi) - \mu_n(\varphi)
    \bigr)^2
    \Big| \mu_n
  \Bigr]
  \le
  L^2 C_s^2
  \left[
    \frac{\tau}{\tau+1}m^{-2/s}
    +
    \frac{1}{\tau+1}m^{-1-2/s}
  \right].
$$
\end{proof}

\subsection{Proof of Theorem~2}

Recall the free-support OT problem
\begin{equation}\label{eq:app-free-support}
  \min_{(\Pi,Y)\in\cD} F(\Pi,Y),
  \quad
  F(\Pi,Y) := \sum_{i=1}^n \sum_{j=1}^m \Pi_{ij}\|X_i - Y_j\|^2,
\end{equation}
where $\cD := \cU(W,m^{-1}1_m)\times(\RR^d)^m$ and
$$
  \cU(W,m^{-1}1_m)
  := \Bigl\{
         \Pi\in\RR_+^{n\times m}:
         \Pi1_m=W,
         \Pi^\top 1_n=m^{-1}1_m
     \Bigr\}.
$$
The BCD iterates satisfy
\begin{align}
  \Pi^{[r+1]} &\in \arg\min_{\Pi\in\cU(W,m^{-1}1_m)} F(\Pi,Y^{[r]}),
  \label{eq:app-bcd-Pi}\\
  Y^{[r+1]}   &= \arg\min_{Y\in(\RR^d)^m} F(\Pi^{[r+1]},Y).
  \label{eq:app-bcd-Y}
\end{align}

\begin{lemma}\label{lem:app-basic}
The sequence $\{(\Pi^{[r]},Y^{[r]})\}_{r\ge0}$ generated by~\eqref{eq:app-bcd-Pi}--\eqref{eq:app-bcd-Y} satisfies:
\begin{enumerate}[label=(\roman*),leftmargin=*,itemsep=2pt]
\item $F(\Pi^{[r]},Y^{[r]})$ is nonincreasing and converges to some $F^\infty\ge0$.
\item The iterates remain in a compact level set and hence admit limit points.
\item $\sum_{r=0}^\infty \|Y^{[r+1]}-Y^{[r]}\|^2<\infty$, in particular
$\|Y^{[r+1]}-Y^{[r]}\|\to0$.
\end{enumerate}
\end{lemma}

\begin{proof}
(i) Monotonicity follows from the two exact minimization steps:
$$
F(\Pi^{[r+1]},Y^{[r]}) \le F(\Pi^{[r]},Y^{[r]}),
\quad
F(\Pi^{[r+1]},Y^{[r+1]}) \le F(\Pi^{[r+1]},Y^{[r]}).
$$
Since $F\ge0$ on $\cD$, the values converge to a finite limit $F^\infty$.

(ii) The polytope $\cU(W,m^{-1}1_m)$ is defined by linear equality constraints together with the nonnegativity constraint $\Pi_{ij}\ge 0$; thus it is closed. It is also bounded: for every $\Pi\in\cU(W,m^{-1}1_m)$,
$$
0\le \Pi_{ij}\le \sum_{i'=1}^n \Pi_{i'j}=\frac{1}{m}, \quad 1\le i\le n,\ 1\le j\le m.
$$
Hence, $\cU(W,m^{-1}1_m)$ is a closed and bounded subset of the finite-dimensional space $\RR^{n\times m}$, and is therefore compact.

To bound $Y$, let $M_0:=F(\Pi^{[0]},Y^{[0]})$, and consider any $(\Pi,Y)\in\cD$ with $F(\Pi,Y)\le M_0$. Write
$$
F(\Pi,Y)=\sum_{j=1}^m \sum_{i=1}^n \Pi_{ij}\|X_i-Y_j\|^2.
$$
Since each outer summand is nonnegative and the total is at most $M_0$, for every fixed $j$ we have $\sum_{i=1}^n \Pi_{ij}\|X_i-Y_j\|^2 \le M_0$.
Using $\sum_{i=1}^n \Pi_{ij}=1/m$, it follows that
$$
\frac{1}{m}\min_{1\le i\le n}\|X_i-Y_j\|^2
\le
\sum_{i=1}^n \Pi_{ij}\|X_i-Y_j\|^2
\le M_0.
$$
Hence, $\min_{1\le i\le n}\|X_i-Y_j\|\le \sqrt{mM_0}$.
Choosing $i_j\in\arg\min_{1\le i\le n}\|X_i-Y_j\|$, we obtain
$$
\|Y_j\|
\le
\|X_{i_j}\|+\|Y_j-X_{i_j}\|
\le
\max_{1\le i\le n}\|X_i\|+\sqrt{mM_0}.
$$
Therefore each $Y_j$ lies in a fixed Euclidean ball, so the corresponding level set is bounded. Since it is also closed, it is compact.

(iii) For fixed $\Pi$, the map $Y\mapsto F(\Pi,Y)$ is a convex quadratic. A direct
calculation yields
$$
\nabla_{YY}^2 F(\Pi,Y)
= 2\mathrm{diag}(\Pi^\top 1_n)\otimes I_d.
$$
On $\cU(W,m^{-1}1_m)$ we have $\Pi^\top 1_n = m^{-1}1_m$, hence
$\nabla_{YY}^2 F(\Pi,Y) = \frac{2}{m}I_{md}$ and $F(\Pi,\cdot)$ is $(2/m)$-strongly convex.
Therefore, with $Y^{[r+1]}=\arg\min_Y F(\Pi^{[r+1]},Y)$,
$$
F(\Pi^{[r+1]},Y^{[r]}) - F(\Pi^{[r+1]},Y^{[r+1]})
\ge \frac{1}{m}\|Y^{[r]}-Y^{[r+1]}\|^2.
$$
Combining with $F(\Pi^{[r]},Y^{[r]})\ge F(\Pi^{[r+1]},Y^{[r]})$ gives
$$
F(\Pi^{[r]},Y^{[r]}) - F(\Pi^{[r+1]},Y^{[r+1]})
\ge \frac{1}{m}\|Y^{[r]}-Y^{[r+1]}\|^2.
$$
Summing over $r$ and using (i) implies $\sum_r \|Y^{[r+1]}-Y^{[r]}\|^2<\infty$.
\end{proof}

\begin{proof}[Proof of Theorem~2]
Let $(\bar\Pi,\bar Y)$ be any limit point; take a subsequence $r_k\to\infty$ such that
$$
(\Pi^{[r_k]},Y^{[r_k]})\to(\bar\Pi,\bar Y).
$$
By Lemma~\ref{lem:app-basic}(iii),
$$
\|Y^{[r_k]}-Y^{[r_k-1]}\|\to 0,
$$
hence $Y^{[r_k-1]}\to \bar Y$ as well.

Since $F(\Pi,Y)$ is a quadratic function of $(\Pi,Y)$, both $F$ and $\nabla_Y F$ are continuous on $\cD$.

\textbf{Step 1: stationarity in $Y$.}
For each $r$, $Y^{[r]}$ minimizes the map $Y\mapsto F(\Pi^{[r]},Y)$, so
$$
\nabla_Y F(\Pi^{[r]},Y^{[r]})=0.
$$
Passing to the limit along the subsequence $r_k$ and using continuity of $\nabla_YF$ yields
$$
\nabla_Y F(\bar\Pi,\bar Y)=0.
$$

\textbf{Step 2: stationarity in $\Pi$.}
For each $k$, by~\eqref{eq:app-bcd-Pi},
$$
F(\Pi^{[r_k]},Y^{[r_k-1]})
\le
F(\tilde\Pi,Y^{[r_k-1]}),
\qquad \forall\,\tilde\Pi\in\cU(W,m^{-1}1_m).
$$
Letting $k\to\infty$ and using continuity of $F$, together with $\Pi^{[r_k]}\to\bar\Pi$ and $Y^{[r_k-1]}\to\bar Y$ gives
$$
F(\bar\Pi,\bar Y)\le F(\tilde\Pi,\bar Y),
\qquad \forall\,\tilde\Pi\in\cU(W,m^{-1}1_m).
$$
Therefore,
$$
\bar\Pi\in\arg\min_{\Pi\in\cU(W,m^{-1}1_m)}F(\Pi,\bar Y).
$$

Since $\cU(W,m^{-1}1_m)$ is closed and convex, and $F(\cdot,\bar Y)$ is differentiable, the first-order optimality condition yields
$$
-\nabla_\Pi F(\bar\Pi,\bar Y)\in N_{\cU(W,m^{-1}1_m)}(\bar\Pi),
$$
where, for a closed convex set $C$ and a point $z\in C$, $N_C(z):=\{g: \langle g,\tilde z-z\rangle\le 0, \forall\,\tilde z\in C\}$ denotes the normal cone to $C$ at $z$, and $\langle\cdot,\cdot\rangle$ is the Frobenius inner product.

Combining Step~1 and Step~2 shows that $(\bar\Pi,\bar Y)$ is a stationary point of~\eqref{eq:app-free-support}.
\end{proof}

\section{Additional details and results for simulation studies}\label{sec:additional-results}

\subsection{Algorithmic convergence diagnostics}\label{subsec:simu-convergence}

We empirically assess the convergence behavior of the block coordinate descent (BCD) solver for the free-support Wasserstein problem. Let $J^{[r]}:=\sum_{i,j}\Pi^{[r]}_{ij}\|X_i-Y_j^{[r]}\|^2$ denote the objective value at BCD iteration $r$. Our diagnostics focus on two questions: whether $J^{[r]}$ decreases monotonically and converges within a small number of iterations, and whether the attained solution is stable with respect to the initialization of $Y^{[0]}$.

We consider one SMC run for each of the Lorenz-63 model (Section~S3.2, $m=128$) and the linear Gaussian model (Section~5.1 of the main text, $m=512$). At each timestep where WDR is applied, we initialize $Y^{[0]}$ by randomly subsampling from $X$, run BCD for ten iterations, and record both the objective value $J^{[r]}$ and the relative support change $\Delta^{[r]} := \|Y^{[r]} - Y^{[r-1]}\| / \|Y^{[r-1]}\|$. We use $\Delta^{[r]}<10^{-2}$ as the convergence criterion.

\begin{figure}[ht]
    \centering
    \includegraphics[width=0.95\linewidth]{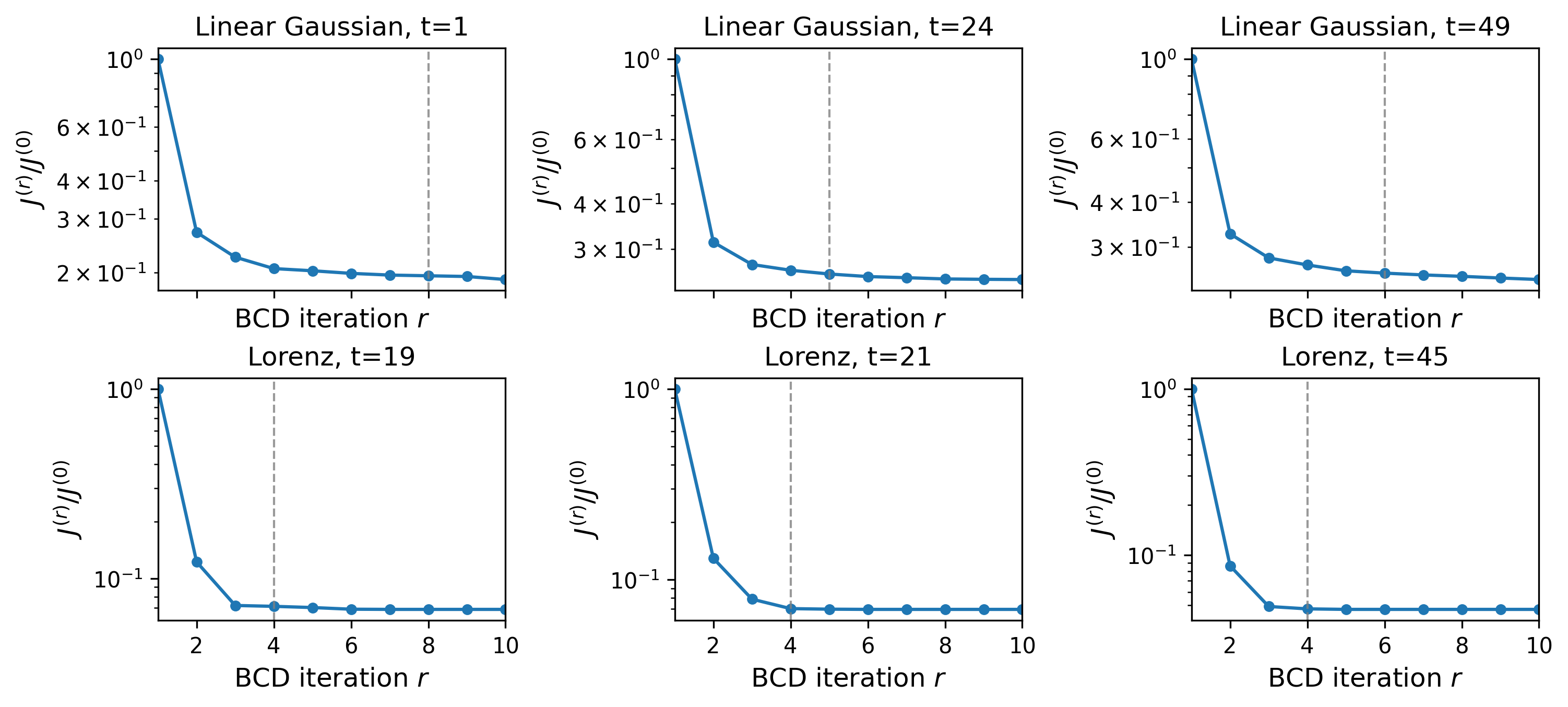}
    \caption{BCD objective trajectories $J^{[r]}/J^{[0]}$ at six randomly selected timesteps. Top row: linear Gaussian model. Bottom row: Lorenz-63 system. In each panel, the gray dashed vertical line marks the first iteration at which the relative support change $\|Y^{[r]}-Y^{[r-1]}\|/\|Y^{[r-1]}\|$ falls below $10^{-2}$.}
    \label{fig:bcd-conv}
\end{figure}

Figure~\ref{fig:bcd-conv} plots $J^{[r]}/J^{[0]}$ against the BCD iteration $r$ for three randomly selected timesteps from each model. In every panel, the objective decreases monotonically and reaches the convergence criterion within a few iterations. The convergence criterion is achieved within a median of 4.5 iterations.

\begin{figure}[ht]
    \centering
    \includegraphics[width=0.5\linewidth]{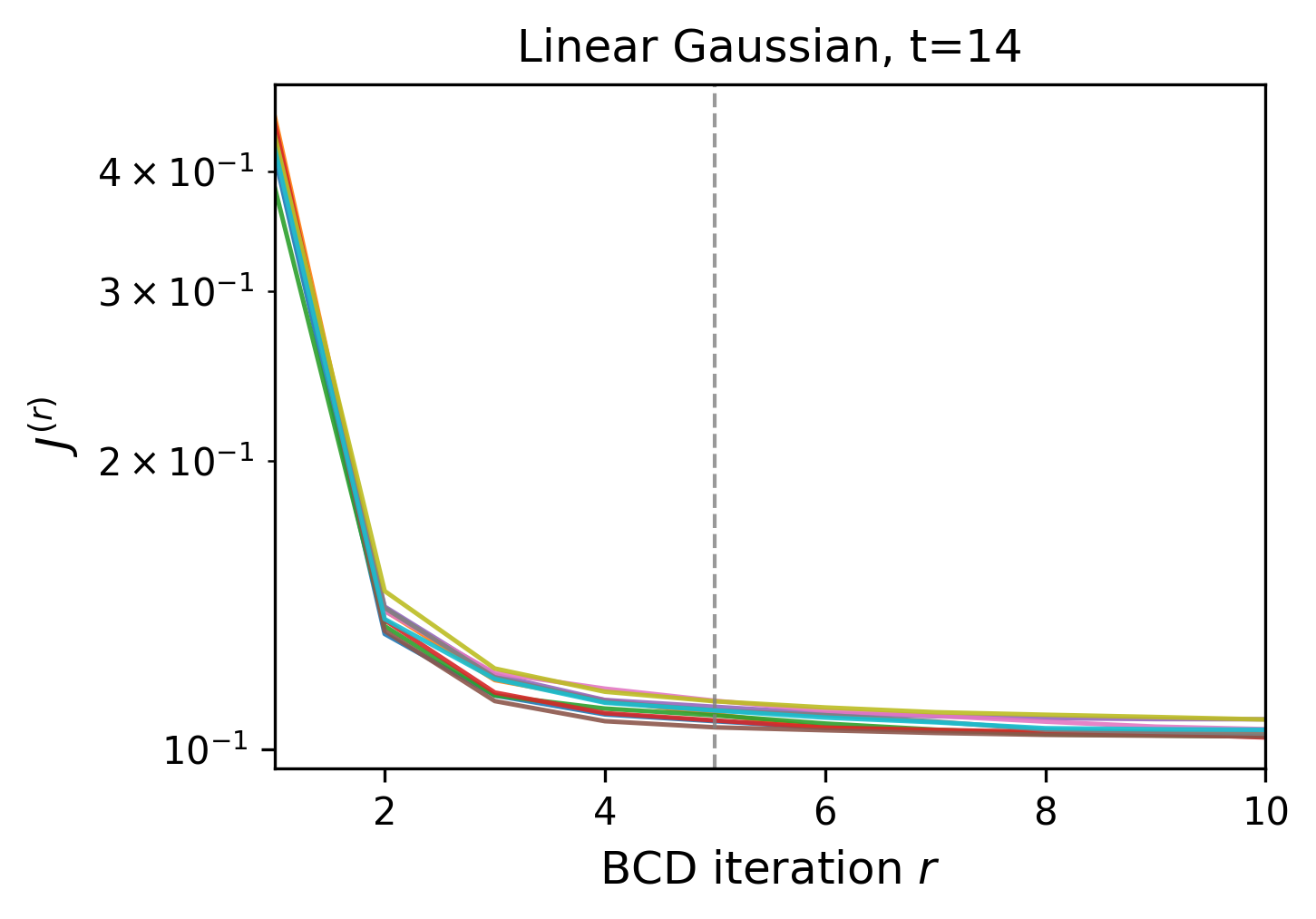}
    \caption{BCD objective trajectories $J^{[r]}$ from ten independent random initializations of $Y^{[0]}$ for one timestep in the linear Gaussian model. The gray dashed vertical line marks the first iteration at which $\|Y^{[r]}-Y^{[r-1]}\|/\|Y^{[r-1]}\|<10^{-2}$ holds simultaneously for all ten trajectories.}
    \label{fig:bcd-init}
\end{figure}

To assess sensitivity to initialization, we rerun BCD on one timestep from the linear Gaussian model using ten independent random initializations of $Y^{[0]}$. As shown in Fig.~\ref{fig:bcd-init}, the ten objective trajectories quickly come together and converge to very similar final values. All ten runs satisfy $\Delta^{[r]}<10^{-2}$ by iteration $r=5$.

Taken together, Figs.~\ref{fig:bcd-conv} and~\ref{fig:bcd-init} indicate that the BCD solver converges quickly and stably for our free-support Wasserstein resampling problem.

\subsection{Lorenz-63 system}\label{subsec:supp-lorenz}

This subsection evaluates on the classical three-dimensional Lorenz-63 system
\begin{equation}\label{eq:l63}
\begin{aligned}
\frac{\dd x_1}{\dd t} &= \sigma(x_2 - x_1),\\
\frac{\dd x_2}{\dd t} &= x_1(\rho - x_3) - x_2,\\
\frac{\dd x_3}{\dd t} &= x_1 x_2 - \beta x_3,
\end{aligned}
\end{equation}
with the standard parameters $(\sigma,\rho,\beta) = (10, 28, 8/3)$~\citep{reich2013nonparametric, acevedo2017second}.

The latent state $X^{(t)}\in\RR^3$ follows
$$
X^{(t+1)} = \Phi_{\Delta t_{\text{obs}}}(X^{(t)}) + \xi^{(t)},\quad
\xi^{(t)} \sim \cN(0, \, 0.5\,I_3),
$$
where $\Phi_{\Delta t_{\text{obs}}}$ denotes the flow of~\eqref{eq:l63} over one observation interval $\Delta t_{\text{obs}} = 0.12$. Numerically, we integrate the ordinary differential equation with an implicit midpoint scheme with step size $\Delta t = 0.01$ and perform $\Delta t_{\text{obs}} / \Delta t = 12$ internal steps between observations.

The observation is
$$
Y^{(t)} = H X^{(t)} + \varepsilon^{(t)},\quad
H = I_3,\quad
\varepsilon^{(t)} \sim \cN(0,\,8\,I_3).\quad 
$$
For each replication, we sample an initial state $X^{(0)} \sim \cN(m_0,\Sigma_0)$ with $m_0 = (1.5,-1.5,25)^\top$ and $\Sigma_0 = \mathrm{diag}(2,2,2)$, and then simulate ${(X^{(t)},Y^{(t)})}_{t=1}^T$ for $T=5000$ timesteps according to the above dynamics.
We run one hundred independent replications, and set the number of particles $m \in \{2^4, 2^5, 2^6, 2^7\}$. 
On each replication, we run SMC with the adaptive multi-descendant proposal and adaptive resampling introduced in Section~2.1 of the main text, where we set $K=3$ and $c_{\mathrm{thres}}=2$.

\textbf{Evaluation reference.}
For nonlinear models, we construct the oracle $I^{\star}$ of $\EE[X^{(t)} \mid Y^{(1:t)}]$ using a high-accuracy particle filter approximation, following standard practices \citep{li2022stratification,fearnhead2003line}. Specifically, for each replicated dataset, we run a bootstrap filter with $10^5$ particles and stratified resampling, and record its filtered mean trajectory. To reduce variability due to resampling randomness, we repeat this oracle run ten times and average the resulting filtered mean trajectories, yielding the final oracle reference $I^{\star}$.

To test the reliability of this large-$m$ oracle, given a Lorenz-63 dataset, we examine the cross-run variability of the posterior mean of above ten trajectories. The results are reported in Table~\ref{tab:oracle-test}. Since the ranges of $x_1$ and $x_2$ are roughly $[-20,20]$ and that of $x_3$ is roughly $[0,40]$, the observed variability is very small. We therefore believe that the large-$m$ particle filter is sufficiently reliable to serve as an oracle reference for the Lorenz-63 experiment.

\begin{table}[ht]
\centering
\caption{Cross-run variability of the oracle posterior means based on ten runs. Let $\hat I^{\star(t)}_r(j)$ denote the oracle posterior mean at time $t$, coordinate $j$, and run $r$. For each coordinate $j$, ``mean std'' is the time average of $\mathrm{sd}_r\{\hat I^{\star(t)}_r(j)\}$, and ``max std'' is the maximum over $t$ of $\mathrm{sd}_r\{\hat I^{\star(t)}_r(j)\}$. The row ``Overall'' averages these quantities over the three coordinates.}
\label{tab:oracle-test}
\begin{tabular}{@{}lrr@{}}
\toprule
Coordinate & mean std & max std \\
\midrule
$x_1$ & 0.0076 & 0.0952 \\
$x_2$ & 0.0109 & 0.1389 \\
$x_3$ & 0.0102 & 0.2362 \\
Overall & 0.0096 & 0.2362 \\
\bottomrule
\end{tabular}
\end{table}

\subsubsection{Filtering results}\label{subsec:supp-l63-results}

\begin{figure}[ht]

    \centering
    \includegraphics[width=\linewidth]{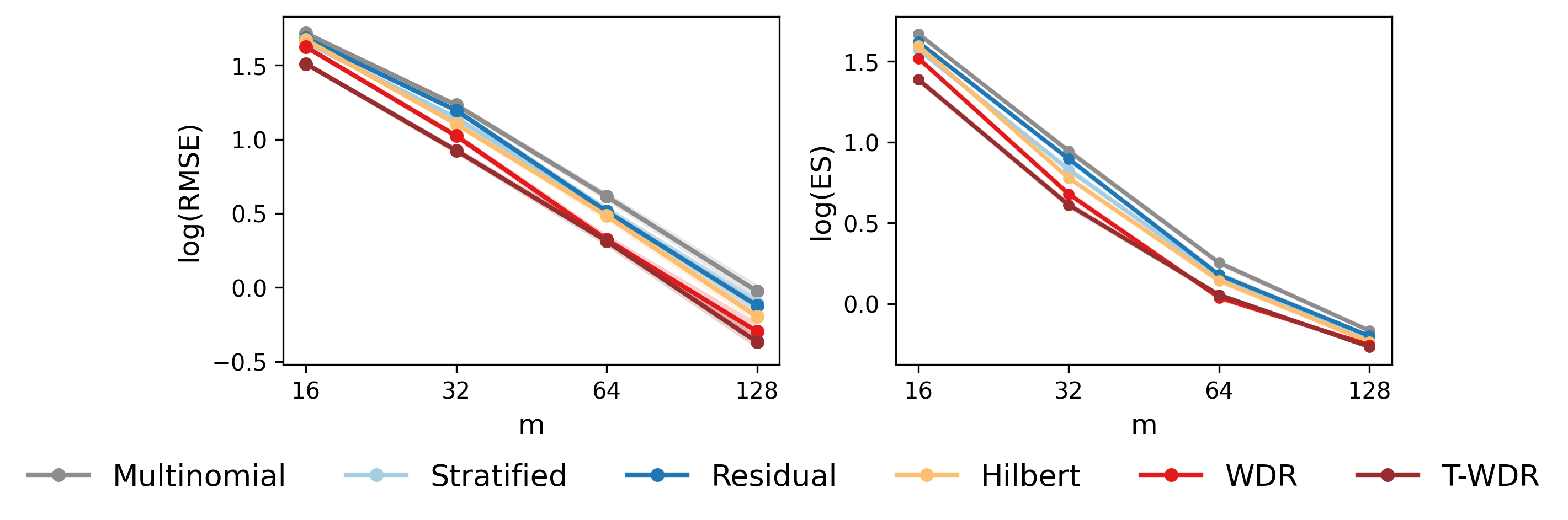}
    \caption{Time-averaged RMSE (left) and energy score (right) versus the particle size $m$ on the log-log scale for the Lorenz-63 model. Shaded bands represent standard errors based on one hundred replications.}
    \label{fig:simu-l63}
\end{figure}

Figure~\ref{fig:simu-l63} shows the average RMSE and energy score versus particle size on a log-log scale. For all resampling methods, both error metrics decrease as $m$ increases. Across all particle sizes, T-WDR consistently achieves the lowest RMSE and energy scores, with WDR a close second.

\begin{figure}[ht]

    \centering
    \includegraphics[width=0.982\linewidth]{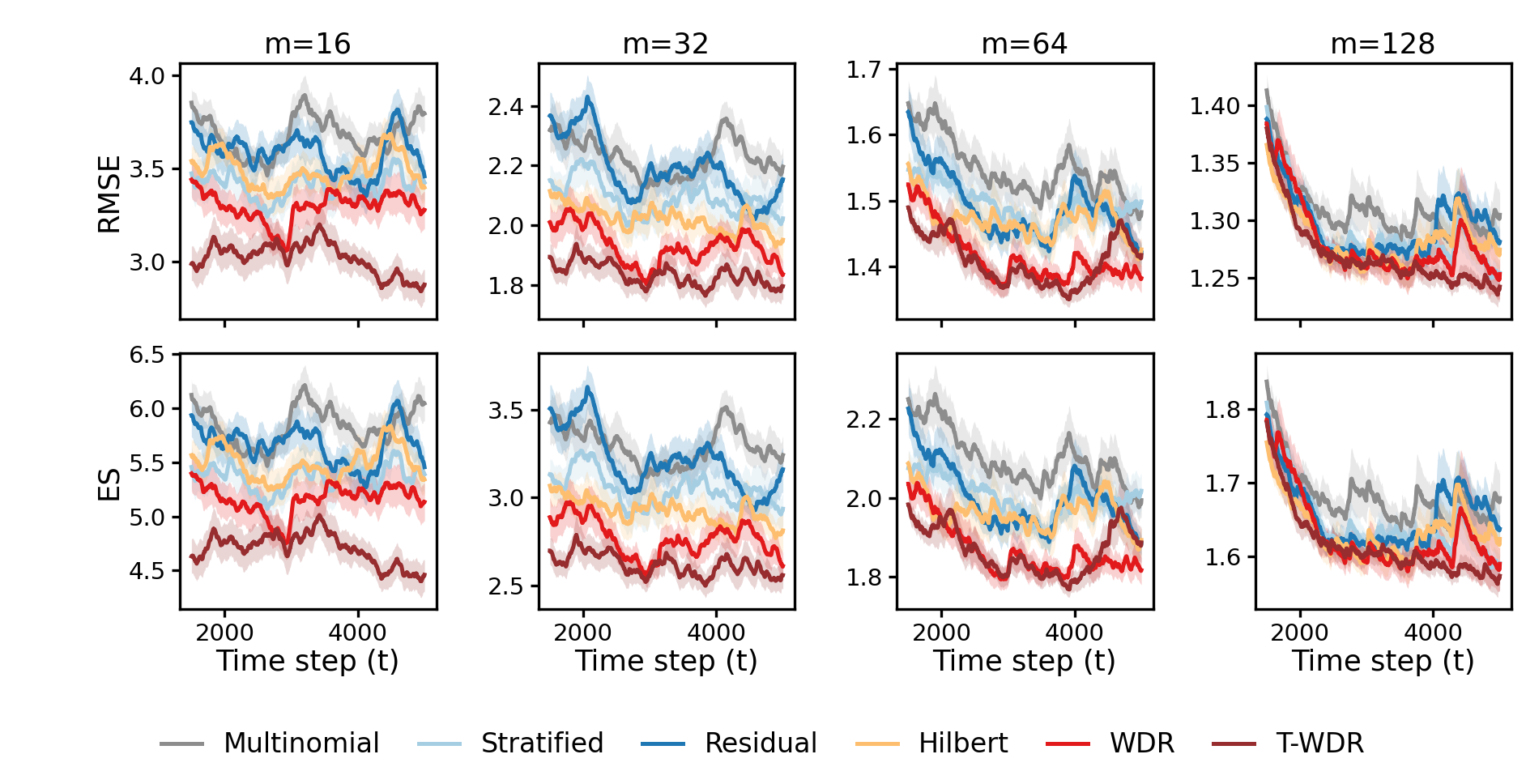}
    \caption{Per-time $\mathrm{RMSE}_t$ (top) and $\mathrm{ES}_t$ (bottom) versus the timestep $t$ for the Lorenz-63 model. Each curve is averaged over one hundred replications, and shaded bands represent standard errors.}
    \label{fig:simu-l63-time}
\end{figure}

Further, we examine how the estimation error evolves over time. Figure~\ref{fig:simu-l63-time} shows per-time $\mathrm{RMSE}_t$ and $\mathrm{ES}_t$ smoothed by an exponential moving average with an effective window of 500.
WDR and T-WDR achieve the leading performance throughout the entire time horizon. 
In particular, when $m$ is small (e.g., $m=16$), the advantage is more pronounced, especially that of T-WDR. As $m$ increases, the gap between methods narrows, since the Lorenz-63 system is only three-dimensional and leaves limited room for a more sophisticated resampling scheme. In the higher-dimensional settings (e.g., Section~5 of the main text), the gap becomes much more significant.

\subsubsection{Computational cost and scalability}\label{subsec:supp-time}
We next assess the computational cost of Wasserstein-Dirichlet resampling and its scalability compared to classical schemes.
From an algorithmic perspective, standard multinomial, stratified, and residual resampling have $\mathcal{O}(m)$ complexity, where $m$ denotes the number of equally weighted particles after resampling.
Hilbert curve resampling incurs an additional $\mathcal{O}(m \log m)$ cost from embedding and ordering.
For Wasserstein-Dirichlet resampling, as discussed in Section~3.2 of the main text, the computational complexity grows approximately linearly in $m$.

To quantify empirical behaviors, we record the wall-clock time of a complete filtering run under each resampling method on the Lorenz-63 model with $T=5000$ timesteps, and average over multiple replications. The results are reported in Table~\ref{tab:time-l63}.
As expected, the three classical resampling methods (multinomial, stratified, and residual) are the fastest.
Hilbert resampling and (T-)WDR incur additional overhead, but their runtimes exhibit similar growth trends as classical methods, consistent with the complexity analyses.
Additionally, T-WDR can be slightly slower than WDR. This is mainly because T-WDR does not enforce a globally equal-weight output, and the resulting weight variation may lead to more frequent activation of resampling in downstream steps, increasing the overall wall-clock time.

\begin{table}[ht]

\centering
\small
\setlength{\tabcolsep}{6pt}
\caption{Wall-clock time (seconds) per run for different resampling schemes and particle sizes $m$ on the Lorenz-63 model with $T=5000$ and $K=3$.}
\label{tab:time-l63}
\begingroup
\renewcommand{\arraystretch}{1}
\small
\begin{tabular}{lccccc}
\toprule
Method & $m=2^4$ & $m=2^6$ & $m=2^8$ & $m=2^{10}$ & $m=2^{12}$ \\
\midrule
Multinomial & 9.27 & 10.85 & 17.25 & 39.96 & 130.33 \\
Stratified & 9.54 & 10.96 & 17.50 & 39.80 & 130.18 \\
Residual & 9.82 & 10.89 & 17.35 & 40.99 & 135.50 \\
Hilbert & 9.99 & 13.02 & 24.90 & 72.29 & 269.14 \\
WDR & 17.39 & 29.37 & 45.85 & 71.11 & 161.21 \\
T-WDR & 16.22 & 26.61 & 46.93 & 100.97 & 227.52 \\
\bottomrule
\end{tabular}
\endgroup
\end{table}

In the Lorenz-63 setting, propagation and weighting are relatively inexpensive, so the cost of WDR appears comparatively large in Table~\ref{tab:time-l63}.
In modern SMC applications such as the image generation tasks and the protein design problems in \cite{wu2023practical}, the propagation and weighting steps often dominate the computational budget, making the additional overhead of Wasserstein-Dirichlet resampling negligible in the total runtime.
The performance-cost analysis of Section~5.1.3 of the main text quantifies this transition.

\subsection{Linear Gaussian state-space model}\label{subsec:supp-lg}

\subsubsection{Per-time results}

\begin{figure}[ht]

    \centering
    \begin{subfigure}{0.6\linewidth}
        \centering
        \includegraphics[width=\linewidth]{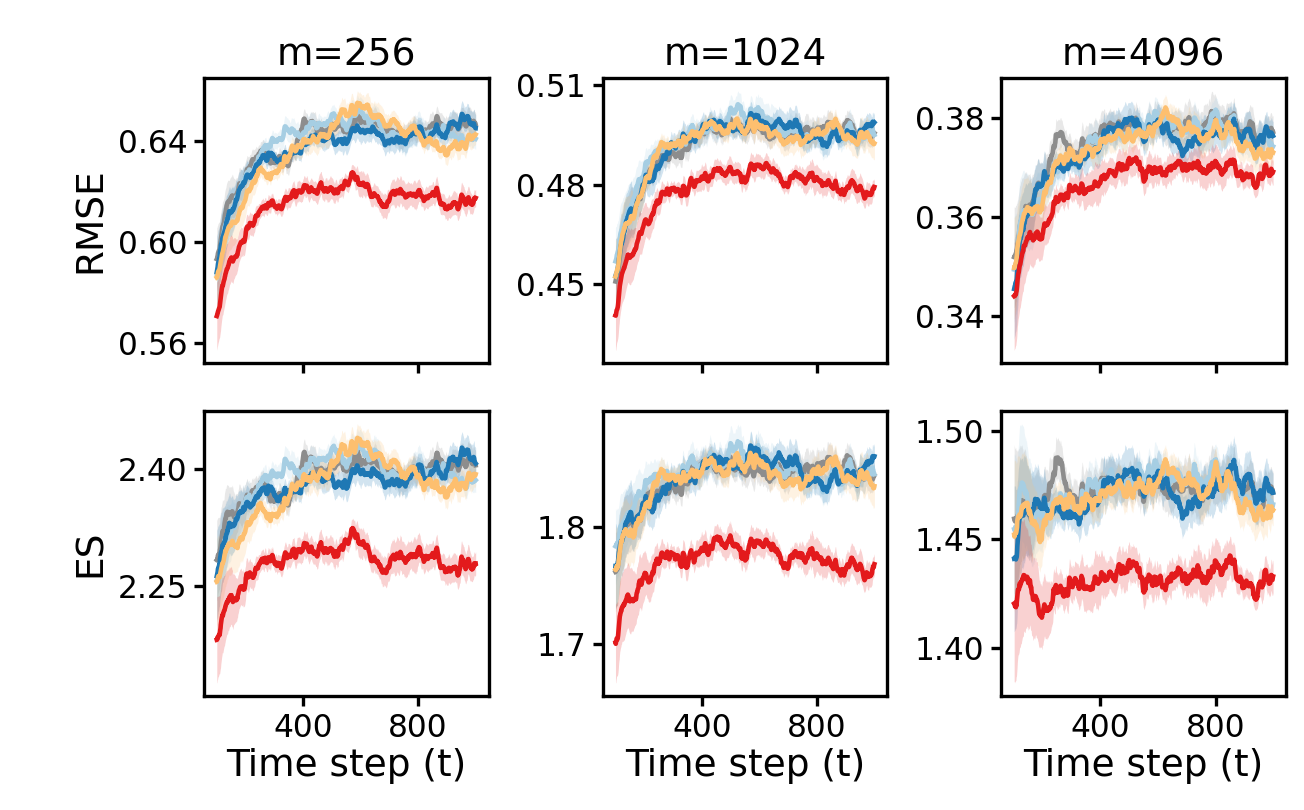}
        \caption{$d=20$, $\kappa=1$.}
    \end{subfigure}
    \\[6pt]
    \begin{subfigure}{0.6\linewidth}
        \centering
        \includegraphics[width=\linewidth]{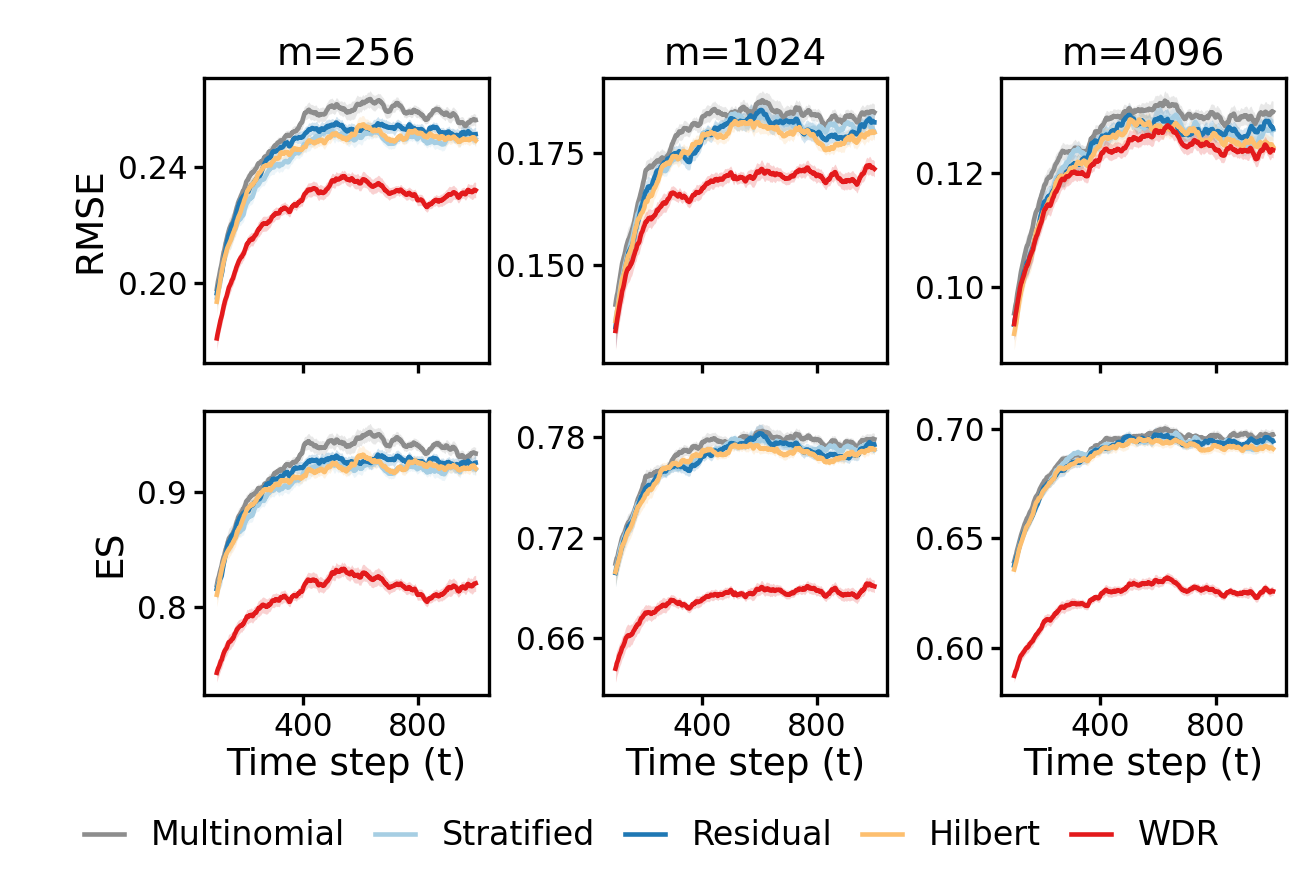}
        \caption{$d=20$, $\kappa=50$.}
    \end{subfigure}
    \caption{Per-time $\mathrm{RMSE}_t$ and $\mathrm{ES}_t$ versus the timestep $t$ for the linear Gaussian model. Shaded bands represent standard errors.}
    \label{fig:simu-lg-time}
\end{figure}

Figure~\ref{fig:simu-lg-time} reports the per-time RMSE and energy score versus the timestep $t$ for two representative settings. After a short transient phase, all methods reach a reasonably steady level. The advantage of WDR is more pronounced in the anisotropic than in the isotropic setting. In both settings, WDR stays below every baseline throughout the time horizon. Its margin narrows as $m$ grows in RMSE and persists in the energy score.

\subsubsection{Stability metrics}
Let $\hat P_r^{(t)}$ denote the weighted particle distribution at time $t$ from run $r$, and let $\hat I_r^{(t)}$ denote its posterior mean estimate. We report three complementary between-run stability metrics:
\begin{enumerate}[leftmargin=*,itemsep=2pt]
\item \textit{Between-run particle energy distance.} For each time $t$ and each pair of distinct runs $(r,r')$, we compute the energy distance between the corresponding particle distributions,
$$
\mathrm{ED}\big(\hat P_r^{(t)},\hat P_{r'}^{(t)}\big) = 2\,\EE\|X-Y\|-\EE\|X-X'\|-\EE\|Y-Y'\|,
$$
where $X,X'\sim \hat P_r^{(t)}$ and $Y,Y'\sim \hat P_{r'}^{(t)}$ are independent. We then average this quantity over all distinct run pairs $(r,r')$ and over all timesteps $t$.
\item \textit{Between-run standard deviation of the estimate.} For each time $t$, we compute the across-run variance of the posterior mean estimate in each coordinate, namely $\Var_r[\hat I_r^{(t)}[j]]$, average these variances over coordinates $j$ and then take the square root. We then average this quantity over $t$.
\item \textit{Between-run standard deviation of RMSE.} For each run $r$ and time $t$, let $\mathrm{RMSE}_{t,r}=\|\hat I_r^{(t)}-I^{\star(t)}\|/\sqrt{d}$, where $I^{\star(t)}$ is the Kalman oracle posterior mean. We compute the standard deviation of $\mathrm{RMSE}_{t,r}$ across runs $r$ at each $t$, and then average this standard deviation over $t$.
\end{enumerate}

\subsection{Stochastic volatility model}\label{subsec:simu-sv}

The multivariate stochastic volatility (SV) model is a nonlinear state-space model~\citep{li2022stratification}.
Let $X^{(t)} \in \RR^d$ denote the latent log-volatility at time $t$ and $Y^{(t)} \in \RR^d$ the corresponding observation (e.g., asset returns). The SV model takes the form
\begin{equation*}
\begin{aligned}
X^{(t)} \mid X^{(t-1)} & \sim \mathcal{N}(\alpha X^{(t-1)}, \Sigma), \\
Y^{(t)} \mid X^{(t)} & \sim \mathcal{N}(0, \beta^2 \operatorname{diag}(\exp (X^{(t)}))).
\end{aligned}
\end{equation*}
That is, $X^{(t)}$ evolves according to a Gaussian autoregressive model, and conditionally on $X^{(t)}$, the coordinates of $Y^{(t)}$ are independent with variance $\beta^2 \exp(X^{(t)}_j)$ in dimension $j$. We set $\alpha = 0.95$ and $\beta = 1.0$.
The covariance $\Sigma$ is defined the same as in the linear Gaussian model (Section~5.1 of the main text), with the condition number $\kappa$.
To investigate the effects of dimensionality and anisotropy, we consider $d \in \{20, 50, 100\}$ and the condition number $\kappa \in \{1, 10, 50\}$.

We run twenty independent replications. For each replication, we initialize the latent state as $X^{(0)} \sim \cN(0,\Sigma)$ and generate a trajectory $\{(X^{(t)},Y^{(t)})\}_{t=1}^T$ by iterating the state and observation equations for $T=1000$ steps.
We set $m \in \{2^7, 2^8, 2^9, 2^{10}\}$, $K=3$, and $c_{\mathrm{thres}}=2$.

\begin{figure}[ht]
    \centering
    \includegraphics[width=\linewidth]{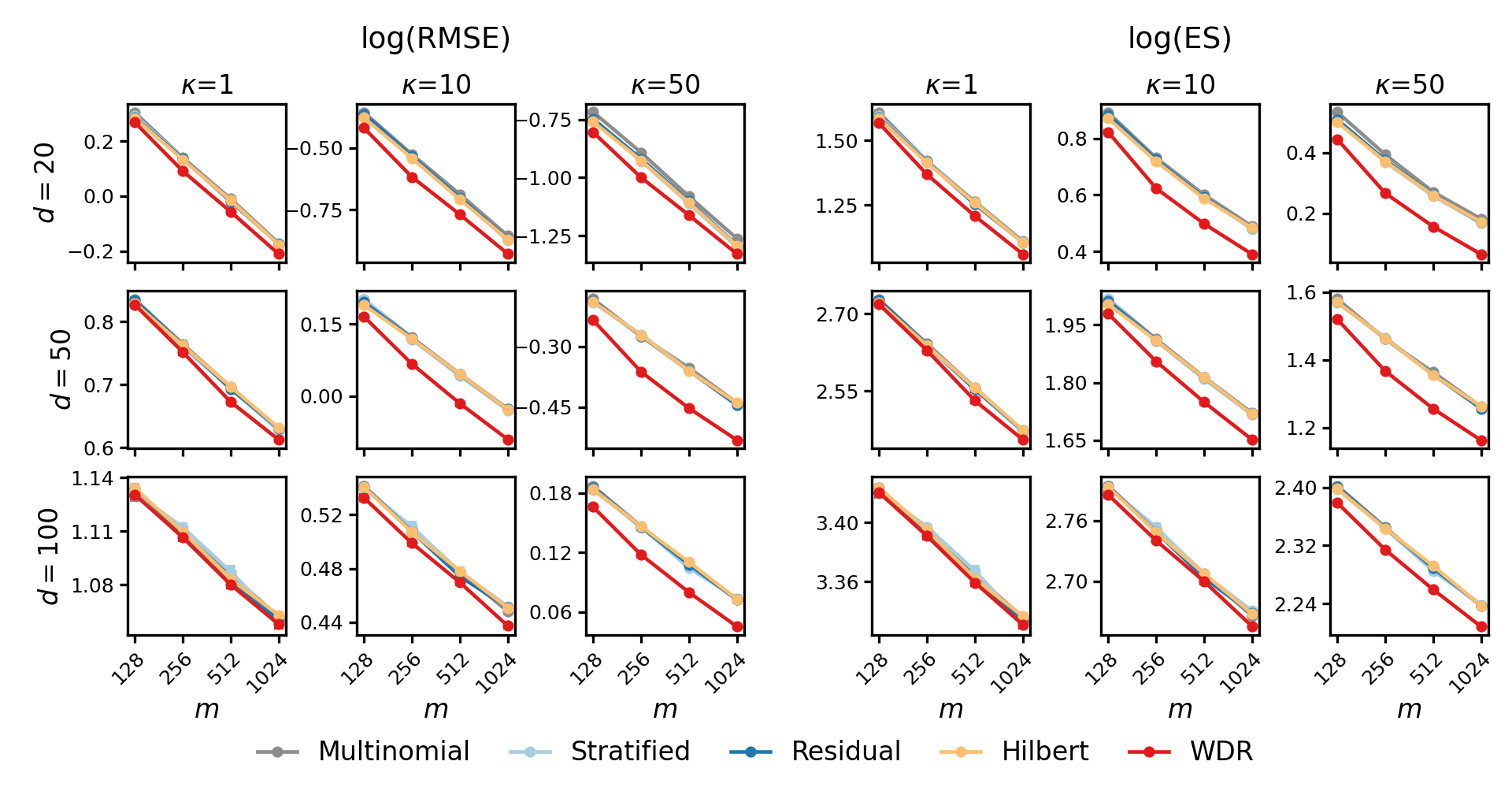}
    \caption{Time-averaged RMSE (left) and energy score (right) versus the particle size $m$ on the log-log scale for the stochastic volatility model. Each row corresponds to a specific dimension $d \in \{20, 50, 100\}$, and each column represents a specific condition number $\kappa \in \{1, 10, 50\}$.
    Vertical bars represent standard errors based on twenty replications.}
    \label{fig:simu-sv}
\end{figure}

Figure~\ref{fig:simu-sv} displays the time-averaged RMSE and energy score versus particle size $m$.
On both metrics, WDR is below every baseline in every case, and the gap grows with $\kappa$.
This reproduces the pattern of the linear Gaussian model (Fig.~2 in the main text) and is consistent with the theoretical benefit of WDR in low intrinsic dimensions (Theorem~1).

\subsection{Sensitivity analysis}\label{subsec:sensitivity}

The proposed Wasserstein-Dirichlet resampling involves a concentration parameter $\tau$ that controls the amount of randomization.  
Its performance in particle filters may also be influenced by the coefficient-of-variation threshold $c_{\mathrm{thres}}$ used to trigger adaptive resampling and by the multi-descendant factor $K$ in the proposal step.
In this subsection we investigate the impact of these hyperparameters on the Lorenz-63 model.

Unless otherwise stated, the experimental settings are the same as in Section~S3.2.
For ease of comparison, we fix a moderate particle size $m=2^6$ and focus on the time-averaged RMSE averaged over one hundred replications. We also include two most competing methods, stratified and Hilbert resampling, for reference.
When varying one tuning parameter, the others are kept at their default values $\tau=30$, $K=3$, and $c_{\mathrm{thres}}=2$. 

\begin{figure}[ht]
    \centering
    \begin{subfigure}{0.32\linewidth}
        \centering
        \includegraphics[width=\linewidth]{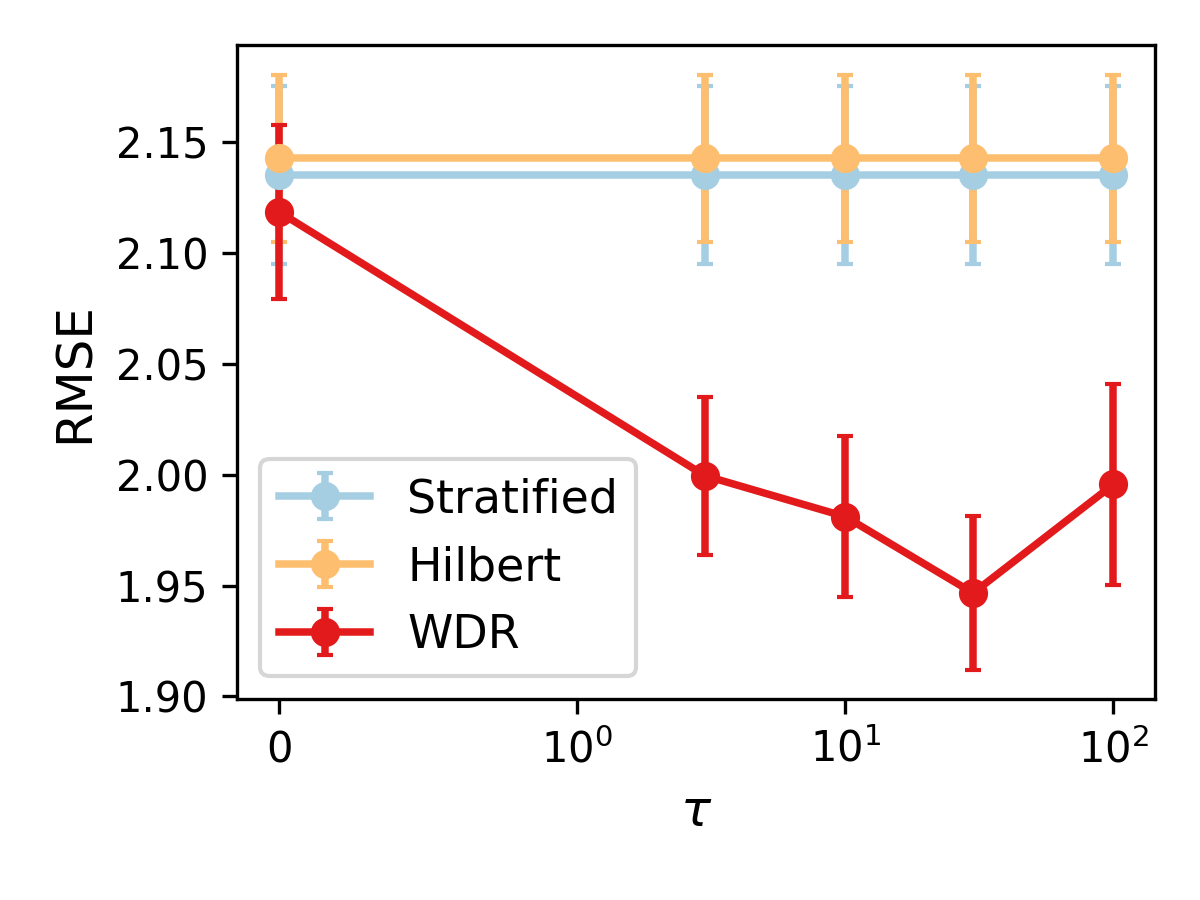}
        \caption{RMSE versus $\tau$.}
        \label{fig:l63-rmse_vs_tau}
    \end{subfigure}
    \begin{subfigure}{0.32\linewidth}
        \centering
        \includegraphics[width=\linewidth]{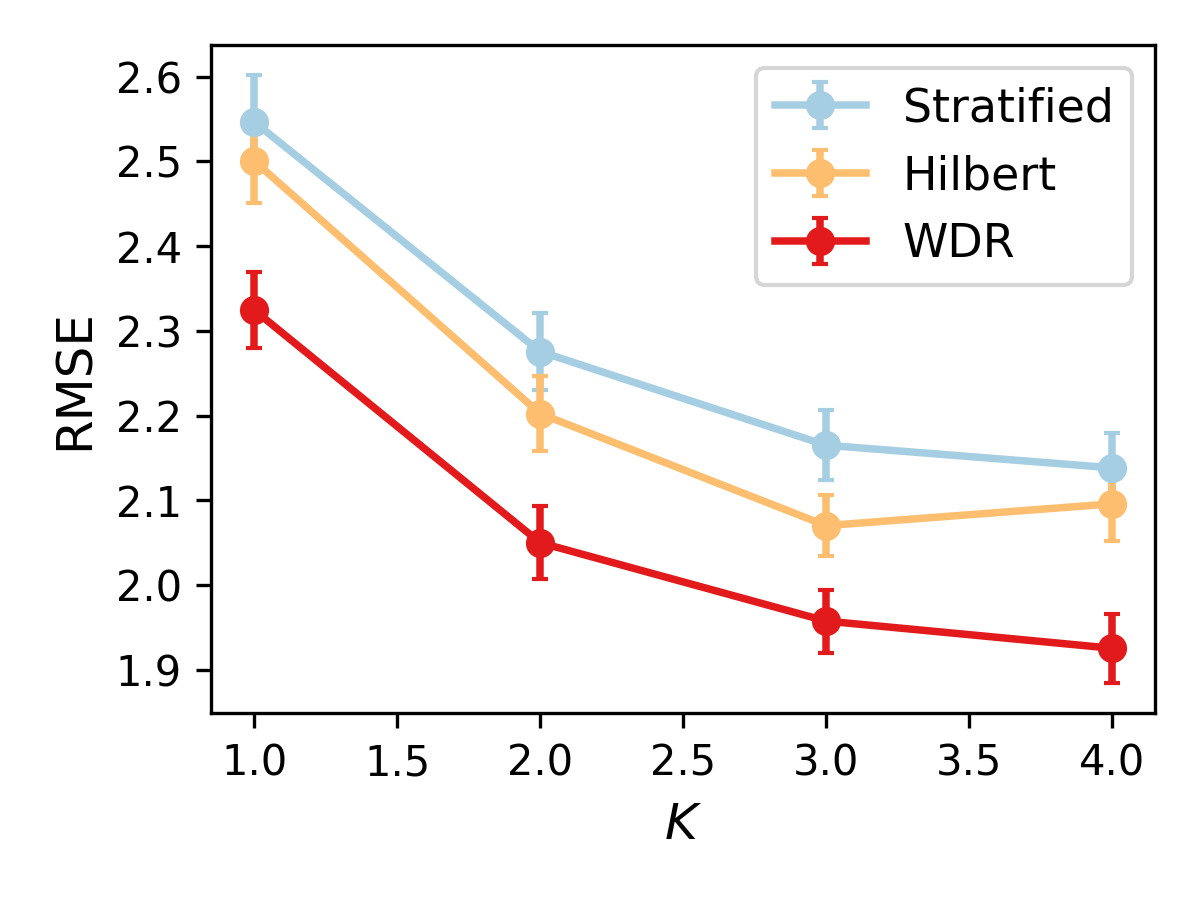}
        \caption{RMSE versus $K$.}
        \label{fig:l63-rmse_vs_K}
    \end{subfigure}
    \begin{subfigure}{0.32\linewidth}
        \centering
        \includegraphics[width=\linewidth]{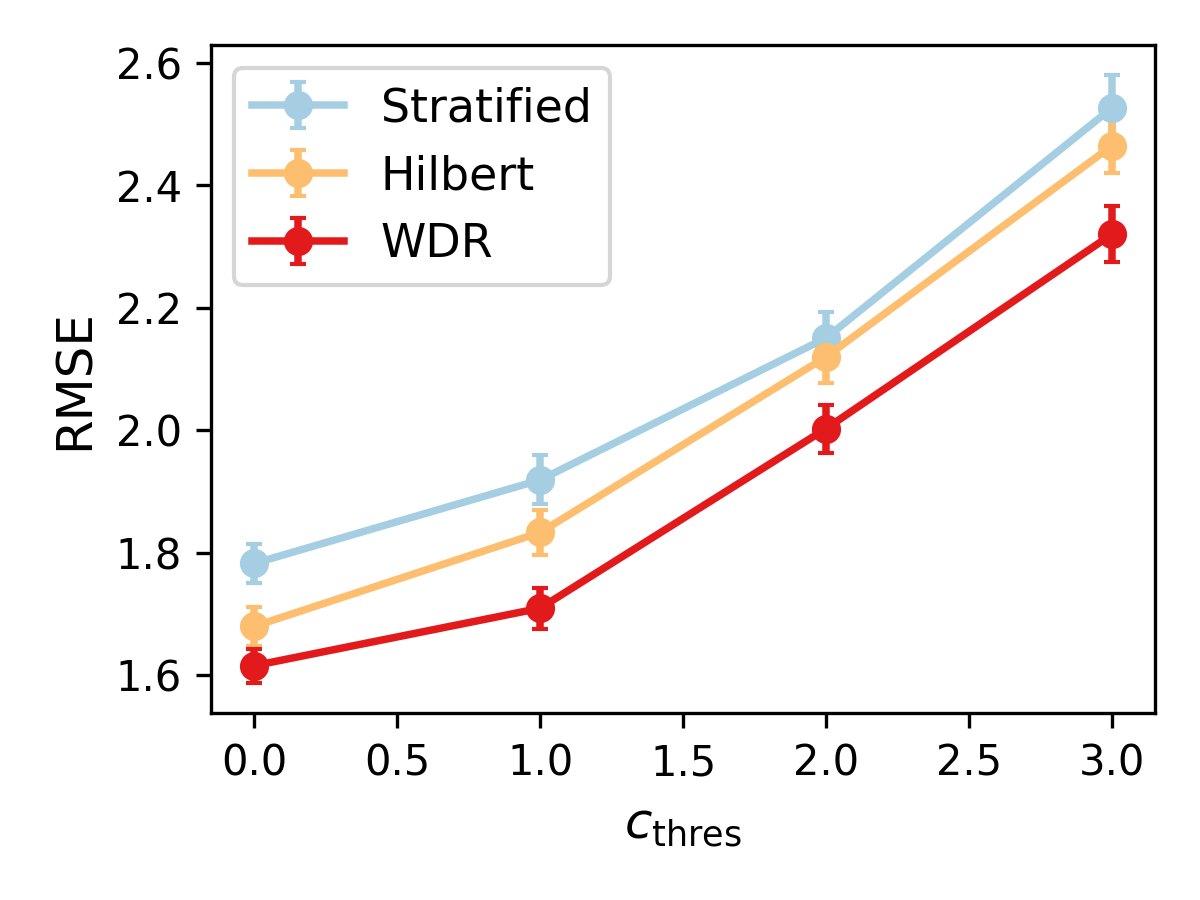}
        \caption{RMSE versus $c_{\mathrm{thres}}$.}
        \label{fig:l63-rmse_vs_cv}
    \end{subfigure}
    \caption{Time-averaged RMSE versus the Dirichlet concentration parameter $\tau$, the number of descendants $K$ in the adaptive multi-descendant proposal, and the coefficient-of-variation threshold $c_{\mathrm{thres}}$ that triggers resampling.
    Vertical bars represent standard errors based on one hundred replications.}
    \label{fig:simu-sensitivity}
\end{figure}

Figure~\ref{fig:simu-sensitivity} reports the performance of WDR as we vary the Dirichlet concentration parameter $\tau \in \{0,3,10,30,100\}$, the number of descendants $K \in \{1,2,3,4\}$ in the adaptive multi-descendant proposal, and the coefficient-of-variation threshold $c_{\mathrm{thres}} \in \{0,1,2,3\}$.

In Fig.~\ref{fig:l63-rmse_vs_tau}, the case $\tau=0$ corresponds to a fully randomized transform that does not generate new support points. Its performance is close to that of stratified and Hilbert resampling, which also only replicate existing particles. At the other extreme, very large values of $\tau$ tend to a deterministic free-support OT transform. We observe that intermediate choices, such as $\tau=10$ or $30$, outperform both limiting regimes. This pattern highlights the benefit of Dirichlet mixing. In particular, some randomness is helpful to avoid over-concentration, while a sufficiently large $\tau$ still encourages the creation of new particles and thus improves exploration of the posterior.

Figures~\ref{fig:l63-rmse_vs_K} and~\ref{fig:l63-rmse_vs_cv} display the effect of $K$ and $c_{\mathrm{thres}}$. For all resampling approaches, the RMSE decreases as $K$ increases or as $c_{\mathrm{thres}}$ is lowered. This behavior is consistent with our expectation: a larger $K$ allows a larger number of propagated particles, and a smaller $c_{\mathrm{thres}}$ triggers resampling more frequently, mitigating weight degeneracy. Overall, Wasserstein-Dirichlet resampling maintains a consistent advantage over the classical resamplers in all scenarios, indicating that the improvements are robust to the choice of tuning parameters.

\subsection{High-dimensional Gaussian spatial model}\label{subsec:supp-b3g}

Algorithm~\ref{alg:supp-bpf} states one filtering step of the block particle filter~\citep{rebeschini2015can}.
For a vector $x\in\RR^d$ and an index set $v\subseteq\{1,\dots,d\}$, let $x^v$ denote the subvector indexed by $v$.
In practice, we partition the $45\times 45$ lattice into $81$ blocks of $5 \times 5$ sites (each block is of dimension $25$).
For each block $\cB_b$, the set $\cO_b$ indexes observations at sites within Euclidean distance $5$ of at least one site in $\cB_b$.

\begin{algorithm}[ht]

  \caption{Block particle filter}
  \label{alg:supp-bpf}
  \begin{algorithmic}[1]
    \State \textbf{Input:} particles $\{\tilde X_i^{(t-1)}\}_{i=1}^{m}\subset\RR^{d}$, observation $Y^{(t)}$, blocks $\{\cB_1,\dots,\cB_L\}$, observation index sets $\{\cO_1,\dots,\cO_L\}$, multi-descendant factor $K$, resampling operator $\cR_m$
    \State Draw $X_\ell^{(t)} \sim f_t(\,\cdot\mid\tilde X_{\lceil\ell/K\rceil}^{(t-1)})$ for $\ell=1,\dots,Km$
    \For{$b=1,\dots,L$}
      \State Set $W_\ell^{(t,b)} \propto \exp\{-\|Y^{(t),\cO_b}-(H X_\ell^{(t)})^{\cO_b}\|^2/2\}$ for $\ell=1,\dots,Km$, with $\sum_{\ell=1}^{Km}W_\ell^{(t,b)}=1$
      \State $\{\tilde X_j^{(t),\cB_b}\}_{j=1}^{m}
      \sim \cR_m\bigl(\{(X_\ell^{(t),\cB_b},W_\ell^{(t,b)})\}_{\ell=1}^{Km}\bigr)$
    \EndFor
    \State Assemble $\tilde X_j^{(t)}$ from $\{\tilde X_j^{(t),\cB_b}\}_{b=1}^{L}$ for $j=1,\dots,m$
    \State \textbf{Output:} equal-weight particles $\{\tilde X_j^{(t)}\}_{j=1}^{m}$
  \end{algorithmic}
\end{algorithm}

\subsection{Energy distance to the filtering distribution}\label{subsec:supp-ed}

Both RMSE and energy score evaluate the particle set with respect to the oracle posterior mean $I^{\star(t)}$, rather than the full target distribution.
For a linear Gaussian model, the Kalman recursions provide the oracle/exact filtering distribution $Q^{(t)} = \cN(I^{\star(t)}, \Sigma^{(t)})$.
We therefore evaluate the energy distance \citep{szekely2013energy} between the particle approximation $\hat P^{(t)} = \sum_i W_i^{(t)}\delta_{X_i^{(t)}}$ and $Q^{(t)}$, that is,
\begin{equation}\label{eq:energy-distance}
\mathrm{ED}_t\bigl(\hat P^{(t)}, Q^{(t)}\bigr)
= 2\sum_i W_i^{(t)}\, \EE\bigl\|X_i^{(t)} - Y\bigr\|
- \sum_{i,j} W_i^{(t)} W_j^{(t)} \bigl\|X_i^{(t)} - X_j^{(t)}\bigr\|
- \EE\|Y - Y'\|,
\end{equation}
where $Y$ and $Y'$ are independent random variables from $Q^{(t)}$.
The energy distance is nonnegative and equals zero if and only if the two distributions coincide.
We report its time average over the simulated trajectory.

Both expectation terms in~\eqref{eq:energy-distance} reduce to $\EE\|W\|$ with $W \sim \cN(\nu, \Sigma)$.
For the cross term, $(\nu, \Sigma) = (X_i^{(t)} - I^{\star(t)}, \Sigma^{(t)})$.
For the target self term, $(\nu, \Sigma) = (0, 2\Sigma^{(t)})$ because $Y - Y' \sim \cN(0, 2\Sigma^{(t)})$.
Writing $\Sigma = U \operatorname{diag}(\lambda_1,\ldots,\lambda_d)\, U^\top$ and $v = U^\top \nu$ gives the representation:
\begin{equation*}
\EE\|W\| = \frac{1}{2\sqrt{\pi}} \int_0^\infty \Bigl[\, 1 - \prod_{j=1}^{d} (1 + 2 s \lambda_j)^{-1/2} \exp\Bigl( - s \sum_{j=1}^{d} \frac{v_j^2}{1 + 2 s \lambda_j} \Bigr) \Bigr] \, s^{-3/2}\, \mathrm{d}s .
\end{equation*}
This representation reduces the expectation terms in~\eqref{eq:energy-distance} to one-dimensional integrals, which we evaluate by quadrature.

\begin{figure}[ht]
    \centering
    \includegraphics[width=0.65\linewidth]{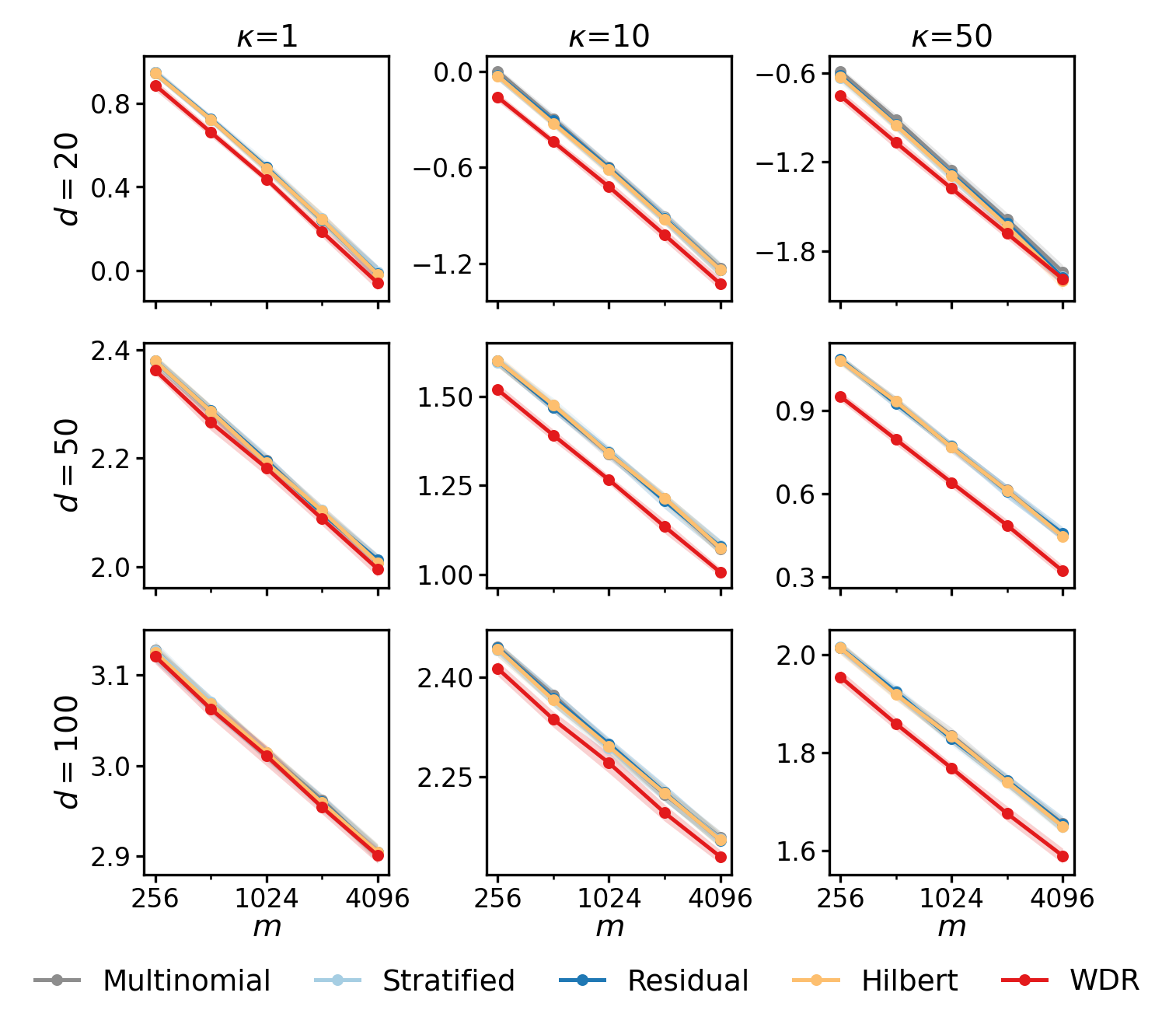}
    \caption{Time-averaged energy distance versus the particle size $m$ on the log-log scale for the linear Gaussian model.
    Rows correspond to the dimension $d \in \{20, 50, 100\}$ and columns to the condition number $\kappa \in \{1, 10, 50\}$.
    Shaded bands represent standard deviations based on twenty replications.}
    \label{fig:supp-ed-grid}
\end{figure}

Figure~\ref{fig:supp-ed-grid} shows the energy distance results for the linear Gaussian model.
WDR attains the lowest value, with a more pronounced advantage at larger $\kappa$, consistent with the RMSE and energy score results in Figure~2 of the main text.

\section{Additional details for the TDS application}\label{sec:tds-supp}

\subsection{Twisted proposals}\label{subsec:tds-proposals}
We start from the unconditional reverse diffusion transition $p_\theta(x^{(t-1)}\mid x^{(t)})$, parameterized by a score network $s_\theta(x^{(t)})$, i.e., $p_\theta (x^{(t-1)} \mid x^{(t)} ):=\mathcal{N} (x^{(t-1)} \mid x^{(t)}+\sigma_t^2 s_\theta(x^{(t)}), \sigma_t^2I)$~\citep{song2020score}.
To incorporate the conditioning variable $y$ at all reverse timesteps, TDS introduces a family of twisting potentials
$\{\tilde g_t(y\mid x^{(t)})\}_{t=0}^T$ that approximate the intractable likelihoods $p(y\mid x^{(t)})$.
Following \citet{wu2023practical}, we take
$\tilde g_t(y\mid x^{(t)})=p(y\mid \hat x_{0,\theta}(x^{(t)}))$,
the likelihood evaluated at the denoising estimate $\hat x_{0,\theta}(x^{(t)})$ produced by the diffusion model.

Using these twisting functions, TDS defines twisted proposals
\begin{equation}\label{eq:tds-proposal}
  \tilde q_t(x^{(t-1)}\mid x^{(t)},y)
  = \mathcal{N}\bigl(x^{(t-1)}\mid
    x^{(t)} + \sigma_t^2\{s_\theta(x^{(t)}) + \nabla_{x^{(t)}}\log \tilde g_t(y\mid x^{(t)})\},
    \sigma_t^2 I\bigr)
\end{equation}
and incremental importance weights
\begin{equation}\label{eq:tds-weight}
  G_t(x^{(t-1)},x^{(t)})
  \;\propto\;
  \frac{p_\theta(x^{(t-1)}\mid x^{(t)})\,\tilde g_{t-1}(y\mid x^{(t-1)})}
       {\tilde q_t(x^{(t-1)}\mid x^{(t)},y)\,\tilde g_t(y\mid x^{(t)})},
  \qquad t = T,\dots,1,
\end{equation}
so that the associated SMC targets converge to $\pi(x^{(0:T)}\mid y)$ as the number of particles grows.
In particular, at each reverse timestep TDS (i) propagates descendants via the twisted kernel $\tilde q_t(\cdot\mid X^{(t)}_i,y)$ in~\eqref{eq:tds-proposal}, (ii) updates the particle weights by multiplying by $G_t$ in~\eqref{eq:tds-weight}, and (iii) resamples particles according to these weights.

\subsection{Role of the VAE}\label{subsec:tds-vae}
We define the transport cost in a low-dimensional embedding that better reflects image semantics.
Specifically, we train a lightweight convolutional variational autoencoder (VAE) on the corresponding image domain and fix the latent dimension to 32 for both MNIST and CIFAR-10.
The VAE is used only to provide a perceptually meaningful ground cost for the resampling step: at each reverse timestep $t$, we encode the current denoised image estimate $\hat x_{0,\theta}(x^{(t)})$ of each particle through the VAE encoder to obtain a 32-dimensional embedding, solve the free-support OT problem~(4) in this embedding space, and then apply the resulting resampling outcome back to the pixel-space particles. The learned low-dimensional representation provides a practical way to define a semantically meaningful transport geometry for images, which is consistent with the general intuition that natural images are concentrated near a much lower-dimensional structure than the ambient pixel space.

\subsection{Image metrics}\label{subsec:tds-metrics}
The image generation experiments are evaluated using four metrics. Among them, FID is computed on the pooled generated samples across all classes, whereas other metrics are computed separately for each target class.

\textit{(i) Fr\'echet inception distance (FID).}
Let $\{x_i^{\mathrm{real}}\}_{i=1}^N$ and $\{x_j^{\mathrm{gen}}\}_{j=1}^M$ denote the real and generated images, pooled over all classes. Following \citet{heusel2017gans}, we pass these images through the Inception-v3 network and extract the corresponding feature vectors. Let $(\mu_r,\Sigma_r)$ and $(\mu_g,\Sigma_g)$ be the empirical means and covariance matrices of the resulting real and generated feature distributions. The FID is then defined by
$$
\mathrm{FID}
=
\|\mu_r-\mu_g\|^2
+
\mathrm{tr}\bigl(
\Sigma_r+\Sigma_g-2(\Sigma_r^{1/2}\Sigma_g\Sigma_r^{1/2})^{1/2}
\bigr).
$$
Lower values indicate that the generated distribution is closer to the real data distribution.

\textit{(ii) Classification accuracy.}
For the target class $\ell$, let $\{x_i^{(\ell)}\}_{i=1}^{N_\ell}$ denote the generated images conditioned on class $\ell$. We apply a pretrained classifier to each generated image and record whether the predicted label matches $\ell$. The classwise accuracy is
$$
\mathrm{Acc}_\ell
=
\frac{1}{N_\ell}\sum_{i=1}^{N_\ell}\mathbf{1}\{\hat y_i=\ell\}.
$$
Higher values indicate better label-conditional correctness.

\textit{(iii) Within-run LPIPS diversity.}
For the target class $\ell$, suppose run $r$ generates images $x_{r,1}^{(\ell)},\dots,x_{r,m}^{(\ell)}$. We define the within-run learned perceptual image patch similarity (LPIPS)~\citep{zhang2018unreasonable} as
$$
\mathrm{WI}_\ell
=
\mathbb E\bigl[
\mathrm{LPIPS}(x,x')
\,\big|\,
x,x' \text{ are generated in the same run for class } \ell
\bigr].
$$
Higher values indicate greater diversity within a single run.

\textit{(iv) Between-run/within-run LPIPS ratio.}
To assess whether different random seeds explore the same posterior region, we compare cross-run LPIPS distances with within-run LPIPS distances. For class $\ell$, we define
$$
\mathrm{BR/WI}_\ell
=
\frac{
\mathbb E\bigl[
\mathrm{LPIPS}(x,x')
\,\big|\,
x\in \text{run } i,\;
x'\in \text{run } j,\;
i\neq j,\;
\text{class } \ell
\bigr]
}{
\mathbb E\bigl[
\mathrm{LPIPS}(x,x')
\,\big|\,
x,x' \text{ are generated in the same run for class } \ell
\bigr]
}.
$$
Values close to one indicate that different random seeds explore essentially the same posterior region, whereas larger values suggest seed-dependent mode concentration.

\bibliographystyle{chicago}
\bibliography{refs}